\documentclass[prd,preprintnumbers,nofootinbib,twocolumn]{revtex4-1}

\usepackage{amsmath}
\usepackage{amssymb}
\usepackage{booktabs}
\usepackage{graphicx}
\usepackage{multirow}
\usepackage{listings}
\usepackage{xcolor}
\usepackage{hyperref}
\usepackage{textcomp}

\allowdisplaybreaks[1]

\numberwithin{equation}{section}
\def \refeq#1{(\ref{#1})}
\def \refsec#1{sec.~\ref{#1}}

\def \refapp#1{app.~\ref{#1}}
\def \reffig#1{fig.~\ref{#1}}
\def \reftab#1{tab.~\ref{#1}}

\newcommand\tabvsptop{\rule{0pt}{2.6ex}}

\newcommand{\wilson}[2][{}]{\mathcal{C}_{#2}^{\mathrm{#1}}}

\newcommand{\order}[1]{\mathcal{O}\left({#1}\right)}

\renewcommand{\[}{\big[}
\renewcommand{\]}{\big]}

\newcommand{\dd}{{\rm d}}
\newcommand{\para}{\parallel}

\newcommand{\GeV}{\ensuremath{\mathrm{GeV}}}
\newcommand{\msbar}{\ensuremath{\overline{\rm MS}}}

\def \order#1{ {\cal O} \left( #1 \right) }

\def \vecth{\vec{\theta}}
\def \vecnu{\vec{\nu}}

\def \cLdB1{{{\cal L}_{\Delta B = 1}^{\rm EW}}} % Delta_B = 1 effective Lagrangian
\def \BR{{\cal B}}                               % branching ratio

\def \Op{{\cal O}}

\def \One{\leavevmode\hbox{\small1\kern-3.6pt\normalsize1}} % unit matrix

\newcommand\rmdx[1]{\mbox{d} \, #1 \,}

\def \alE{\alpha_e}        % electro-magnetic coupling
\def \B0toK0ast{{ \bar{B}^0 \to K^{\ast 0}}}
\def \B0toK0mumu{{ \bar{B}^0 \to K^0 \bar{\mu} \mu}}
\def \barB0toKKpill{{ \bar{B}^0 \to \bar{K}^{\ast 0} (\to K^- \pi^+) \bar{l}l}}
\def \B0toKKpill{{ B^0 \to K^{\ast 0} (\to K^+ \pi^-) \bar{l}l}}

\def \SMnu{SM($\nu$-only)}
\def \SM{SM}
\def \SMp{SM+SM$'$}
\def \SMpNine{SM+SM$'$$(9)$}

\begin{document}

\title{Comprehensive Bayesian Analysis of Rare (Semi)leptonic and Radiative
  $\boldsymbol{B}$ Decays}

\author{Frederik Beaujean}
\affiliation{
  C2PAP, Excellence Cluster Universe,\\
  Ludwig-Maximilians-Universit\"at M\"unchen,
  Garching, Germany
}

\author{Christoph Bobeth}
\affiliation{
  Universe Cluster \& Institute for Advanced Study,
  Technische Universit\"at M\"unchen,
  Garching, Germany
}

\author{Danny van Dyk}
\affiliation{
  Theoretische Elementarteilchenphysik,
  Universit\"at Siegen,
  Siegen,
  Germany
}

\date{\today}

\preprint{EOS-2013-01, SI-HEP-2013-10, QFET-2013-07, FLAVOUR(267104)-ERC-56}

\begin{abstract}
  The available data on $|\Delta B| = |\Delta S| = 1$ decays are in good
  agreement with the Standard Model when permitting subleading power corrections
  of about $15\%$ at large hadronic recoil.  Constraining new-physics effects in
  $\wilson{7}$, $\wilson{9}$, $\wilson{10}$, the data still demand the same size
  of power corrections as in the Standard Model. In the presence of
  chirality-flipped operators, all but one of the power corrections reduce
  substantially.  The Bayes factors are in favor of the Standard Model.  Using
  new lattice inputs for $B\to K^*$ form factors, the evidence of models with
  chirality-flipped operators increases, but does not outperform the Standard
  Model. We use the data to further constrain the hadronic form factors in $B\to
  K$ and $B\to K^*$ transitions.
\end{abstract}

\maketitle

%
%
%
%--------+---------+---------+---------+---------+---------+---------+---------+
\section{Introduction}

The rare $B$ decays mediated by $b\to s \gamma$ and $b\to s \ell^+ \ell^-$
($\ell = e,\, \mu,\, \tau$) flavor-changing neutral-current transitions are
important probes of the Standard Model (SM) and provide constraints on
nonstandard effects in the flavor sector up to the TeV range. In recent years,
phenomenological analyses focused on the exclusive modes $B\to K^{*}(\to K
\pi)\, \ell^+\ell^-$, $B\to K \ell^+\ell^-$ and $B_s \to \mu^+\mu^-$.  Many
observables of these modes were recently measured at the LHC (LHCb, CMS, and
ATLAS), and previously at $B$ factories (Belle, BaBar) and the Tevatron (CDF)
\cite{Limosani:2009qg,Lees:2012ufa,Aubert:2004it,
  Iwasaki:2005sy,Aaij:2013aka,Chatrchyan:2013bka,Coan:1999kh,Aubert:2008gy,Aubert:2009ak,
  Nakao:2004th,Ushiroda:2006fi,Lees:2012tva,Wei:2009zv,CDF:2012:BKstarll,Aaij:2012vr,
  Aaij:2013iag,Chatrchyan:2013cda,ATLAS:2013ola,Aaij:2013qta}.

The goal is to measure a large number of observables accessible in the angular
analysis of the decay distributions of the three- and four-body final
states. With the CP-averaged observables one can test --- in a model-independent
fashion --- the underlying short-distance couplings of the $\Delta B = 1$
effective theory. CP-asymmetric observables also probe new sources of CP
violation beyond the SM. In $B\to K^{*}(\to K \pi)\, \ell^+\ell^-$ decays,
certain combinations of the angular observables, the ``optimized observables''
\cite{Kruger:2005ep,Egede:2008uy,Bobeth:2010wg,Bobeth:2011gi,
  Becirevic:2011bp,Matias:2012xw,Das:2012kz,DescotesGenon:2012zf,
  Bobeth:2012vn}, are free of form factors to leading order in the $1/m_b$
expansion and consequently expected to have smaller theoretical
uncertainties. The same framework also provides observables that are dominated
by ratios of $B\to K^*$ form factors
\cite{Bobeth:2010wg,Hambrock:2012dg,Beaujean:2012uj,Bobeth:2012vn,
  Hambrock:2013zya}, providing some additional data-driven control over these
hadronic quantities.

The $1/m_b$ expansions are important tools for the prediction of exclusive
decays.  At large hadronic recoil of the $K^{(*)}$ meson, QCD factorization
(QCDF) yields corrections beyond naive factorization
\cite{Beneke:2001at,Beneke:2004dp}. Effects of ($q\bar{q}$)-resonances,
dominantly from charm, as well as the chromomagnetic dipole operator can be
calculated using a light-cone operator product expansion (OPE) in combination
with dispersion relations
\cite{Ball:2006eu,Khodjamirian:2010vf,Khodjamirian:2012rm,Dimou:2012un,
  Jager:2012uw}. At low hadronic recoil, a local OPE
\cite{Grinstein:2004vb,Beylich:2011aq} can be employed.  Its prediction of
correlations between different observables can be tested experimentally through
measurement of the observables $H_T^{(1)}$ and $J_7$ in $B\to K^*\ell^+\ell^-$
\cite{Bobeth:2012vn}. Hadronic form factors are a major source of theoretical
uncertainties in the prediction of angular observables, whereas optimized
observables are sensitive to higher-order terms in the $1/m_b$ expansions,
especially at large recoil. At present, the associated uncertainties due to the
unknown $1/m_b$ contributions are estimated based on simple power-counting
arguments.

As of 2011, several global analyses of the available data --- differing in the
degree of sophistication, the statistical approach, and the estimation of theory
uncertainties --- have been performed
\cite{DescotesGenon:2011yn,Altmannshofer:2011gn,Bobeth:2011nj,
  Beaujean:2012uj,Altmannshofer:2012az,DescotesGenon:2012zf}.  Recently,
experimental updates from LHCb \cite{Aaij:2012vr, LHCb:2012kz, Aaij:2013iag,
  Aaij:2013aka, Aaij:2013dgw} and CDF \cite{CDF:2012:BKstarll} became available
as well as analogous measurements from CMS \cite{Chatrchyan:2013bka,
  Chatrchyan:2013cda} and ATLAS \cite{ATLAS:2013ola}. LHCb is the first
experiment to measure optimized observables \cite{Aaij:2013qta}. Perhaps the
most outstanding LHCb result is that the measured value of $P_{4,5}'$ is in some
tension with the SM predictions, stimulating new global fits
\cite{Descotes-Genon:2013wba, Altmannshofer:2013foa,Horgan:2013pva}.

Based on the framework developed in our previous work \cite{Beaujean:2012uj}, we
perform a global analysis using Bayesian inference, and include a total of 28
nuisance parameters to account for theory uncertainties. Apart from new-physics
parameters, our framework allows us to infer also $B\to K^{(*)}$ form factors
and the size of subleading contributions, thereby shedding some light on the
origin of tensions between data and SM predictions. Compared
to~\cite{Beaujean:2012uj}, we include the most recent measurements, add
additional observables to the fit, and account also for recent lattice
calculations of form factors \cite{Bouchard:2013eph,Horgan:2013hoa} and other
new theoretical results.

In \refsec{sec:scenarios}, the model-independent framework of $\Delta B = 1$
decays is briefly revisited, and three scenarios of new physics (NP) are
introduced. In \refsec{sec:exp-input}, we list the updated experimental
input. The results of the global analysis are presented in \refsec{sec:results}:
1) for the most important Wilson coefficients of the SM operator basis
$\wilson{7,9,10}$ and their chirality-flipped counterparts assuming them to be
real-valued, 2) for the $B\to K^{(*)}$ form factors in the SM and the two NP
scenarios, and 3) the size of subleading contributions. We also compare our
results with recent analyses that had access to the same experimental data.  In
\refapp{app:theory-predictions} we summarize the theoretical predictions of
newly included observables and changes in the treatment of form factors and
subleading contributions.

%
%
%
%--------+---------+---------+---------+---------+---------+---------+---------+
\section{Model-independent scenarios \label{sec:scenarios}}

For the global analysis of $b\to s (\gamma,\, \ell^+\ell^-)$ data we use a
model-independent approach based on the $|\Delta B| = |\Delta S| = 1$ effective
theory.  The Hamiltonian reads
\begin{align}
  \label{eq:Heff}
  {\cal{H}}_{\rm eff}=
   - \frac{4\, G_F}{\sqrt{2}}  V_{tb}^{} V_{ts}^\ast \,\frac{\alE}{4 \pi}\,
       \sum_i \wilson[]{i}(\mu)  \Op_i + \text{h.c.}
\end{align}
with dimension-six flavor-changing operators $\Op_i$ and their respective
short-distance couplings, the Wilson coefficients $\wilson{i}(\mu)$. We evaluate
the hadronic matrix elements of the operators at the scale $\mu = 4.2 \,
\mbox{GeV}$ of the order of the bottom-quark mass $m_b$. We restrict our
ana\-lysis to the set of operators present in the SM ($i = 7, 9, 10$)
\begin{equation}
\label{eq:SM:ops}
\begin{aligned}
  \Op_{7(7')} & = \frac{m_b}{e}\!\[\bar{s} \sigma^{\mu\nu} P_{R(L)} b\] F_{\mu\nu}\,,
\\
  \Op_{9(9')} & = \[\bar{s} \gamma_\mu P_{L(R)} b\]\!\[\bar{\ell} \gamma^\mu \ell\]\,,
\\[0.1cm]
  \Op_{10(10')} & = \[\bar{s} \gamma_\mu P_{L(R)} b\]\!\[\bar{\ell} \gamma^\mu \gamma_5 \ell\]
\end{aligned}
\end{equation}
and their chirality-flipped counterparts ($i = 7',9',10'$), denoted by SM$'$.
The Wilson coefficients of the four-quark and the chromomagnetic dipole
operators are set to their NNLO SM values at $\mu = 4.2 \, \mbox{GeV}$
\cite{Bobeth:2003at,Huber:2005ig}.

In principle, the scalar, pseudo-scalar, and tensor $b\to s \ell^+\ell^-$
operators can contribute to the angular distributions of $B\to K^*(\to K \pi)
\ell^+\ell^-$ and $B\to K \ell^+\ell^-$ as discussed in great detail in
\cite{Bobeth:2012vn}.  However, only a few measurements with rather large
uncertainties of sensitive observables are currently available, and many other
results by LHCb cannot be used because they have been made with the explicit
assumption that those new operators do \emph{not} contribute. We therefore
abstain from including those operators into our analysis. Instead, we focus on
comparing the SM to several new-physics scenarios with regard to their ability
to describe the data well.

In the model-independent approach, the SM and SM$'$ Wilson coefficients are the
parameters of interest; they are assumed real valued and independent \emph{a
  priori}. The nuisance parameters $\vec{\nu}$ serve to model theory
uncertainties, including CKM parameters, quark masses, and hadronic matrix
elements; see \reftab{tab:common-input} and \reftab{tab:hadronic:nuisance}. We
introduce the following scenarios of fit parameters
\begin{align}
    \nonumber
    \text{SM($\nu$-only)} & : \begin{cases}
        \wilson{7,9,10} & \text{SM values}\\
        \wilson{7',9',10'} & \text{SM values}\\
        \vec\nu & \text{free floating}
    \end{cases}\,,\\
    \label{eq:scenarios}
    \text{SM} & : \begin{cases}
        \wilson{7} & \in [-2,+2]\\
        \wilson{9,10} & \in [-15,+15]\\
    \wilson{7',9',10'} & \text{SM values}\\
        \vec\nu & \text{free floating}
    \end{cases}\,,\\
    \nonumber
    \text{\SMp{}} & : \begin{cases}
        \wilson{7,7'} & \in [-1,+1]\\
        \wilson{9,9',10,10'} & \in [-7.5,+7.5]\\
        \vec\nu & \text{free floating}
    \end{cases}\,,\\
    \nonumber
    \text{\SMpNine{}} & : \begin{cases}
        \wilson{7,7',10,10'} & \text{SM values}\\
        \wilson{9,9'} & \in [-7.5,+7.5]\\
        \vec\nu & \text{free floating}
    \end{cases}\,.
\end{align}
Expressing vague prior knowledge, we assign a flat prior distribution to the
Wilson coefficients. However, each nuisance parameter $\vec\nu$ comes with an
informative prior as discussed in detail in \refapp{app:theory-predictions}.

The SM values $\wilson{7,9,10}$ are obtained at NNLO
\cite{Bobeth:2003at,Huber:2005ig} and depend on the fundamental parameters of
the top-quark and $W$-boson masses, as well as on the sine of the weak mixing
angle. For new-physics models that fall into one of the scenarios SM and \SMp{},
the obtained fit results of the Wilson coefficients can be subsequently used to
constrain those models' fundamental parameters after accounting for the
renormalization group evolution from the high matching scale down to $\mu \sim
m_b$.

%
%
%
%--------+---------+---------+---------+---------+---------+---------+---------+
\section{Observables and experimental input \label{sec:exp-input}}

In this section we describe changes of the experimental inputs that enter our
global analysis with respect to our previous work \cite{Beaujean:2012uj}.  We
first introduce observables which are newly added to the global analysis and
refer the reader for details of their theoretical treatment to
\refapp{app:theory-predictions}. Afterward, we summarize those observables whose
measurements have been updated, or for which additional measurements have since
become available. In general we employ the full set of observables listed in
\reftab{tab:observables} except for the last row and denote this as ``full''.
Inclusion of the $B\to K^*$ lattice points from the last row is denoted as
``full (+FF)''.  For the sake of comparison with \cite{Descotes-Genon:2013wba}
we also repeat the analysis with a smaller subset called ``selection'' as
specified in the same table.  Generally, we model the probability distributions
of experimental measurements as (multivariate) Gaussian distributions. In
practice, however, experiments do not yet provide correlations, except for $S$
and $C$ in $B\to K^*\gamma$.  For measurements with asymmetric uncertainties, we
model the probability distribution as a split Gaussian with two different
widths.  For the measurement of $\BR(B_s \to \mu^+\mu^-)$, we use the Amoroso
distribution to avoid the unphysical region $\BR < 0$ as described in
\cite{Beaujean:2012uj}.

In the following, all observables are understood to be CP-averaged unless noted
otherwise. The dilepton invariant mass in inclusive and exclusive $b\to
s\ell^+\ell^-$ decays is denoted by $q^2$ throughout.

\begin{table*}
\begin{center}
\renewcommand{\arraystretch}{1.4}
\begin{tabular}{lcccc}
\hline
Channel & Constraints & Kinematics & Source & Selection
\\
\hline
$B\to X_s \gamma$
    & $\BR$ &  $1.8\, \mbox{GeV} < E_\gamma$  & \cite{Limosani:2009qg,Lees:2012ufa} & \checkmark
\\
\hline %\hline
$B\to X_s \ell^+\ell^-$
    & $\BR$ &  $q^2\in [1,\, 6]$ GeV$^2$  & \cite{Aubert:2004it,Iwasaki:2005sy} & \checkmark
\\
\hline %\hline
$B_s\to \mu^+\mu^-$
& $\int\dd\tau \BR(\tau)$ &  --  & \cite{Aaij:2013aka,Chatrchyan:2013bka} & \checkmark
\\
\hline %\hline
$B\to K^*\gamma$
    & $\BR$, $S$, $C$ &  --  & \cite{Coan:1999kh,Aubert:2008gy,Aubert:2009ak,Nakao:2004th,Ushiroda:2006fi} & \checkmark
\\
\hline %\hline
\multirow{2}{*}{${B\to K \ell^+\ell^-}$}
    & \multirow{2}{*}{$\BR$}
    & $q^2\in [1,\, 6],\, [14.18,\, 16],\, [> 16]$ GeV$^2$  & \cite{Lees:2012tva,Wei:2009zv,CDF:2012:BKstarll} & ---\\
    &
    &  $\,\,\,\,\, q^2 \in [1,\, 6],\, [14.18,\, 16],\, [16,\, 18],\, [18, \, 22]$ GeV$^2$  & \cite{Aaij:2012vr} & ---
\\
\hline %\hline
\multirow{5}{*}{${B\to K^* \ell^+\ell^-}$}
    & $\BR$ &  $q^2\in [1,\, 6],\, [14.18,\, 16],\, [> 16]$ GeV$^2$  & \cite{Lees:2012tva,Wei:2009zv,CDF:2012:BKstarll,Aaij:2013iag,Chatrchyan:2013cda} & ---\\
    & $F_L$   &  $-$\textquotestraightdblbase$-$  &  \cite{Lees:2012tva,Wei:2009zv,CDF:2012:BKstarll,Aaij:2013iag,Chatrchyan:2013cda,ATLAS:2013ola} & ---\\
    & $A_{\rm FB}$   &  $-$\textquotestraightdblbase$-$  &  \cite{Lees:2012tva,Wei:2009zv,CDF:2012:BKstarll,Aaij:2013iag,Chatrchyan:2013cda,ATLAS:2013ola} & $\dagger$\\
    & $A_T^{(2)}$ &  $-$\textquotestraightdblbase $-$  &  \cite{CDF:2012:BKstarll,Aaij:2013iag} & $\dagger$\\
    & $A_T^{\rm re}$, $P'_{4,5,6}$ &  $-$\textquotestraightdblbase$-$  &  \cite{Aaij:2013iag,Aaij:2013qta} & $\dagger$\\
\hline %\hline
$B$ properties
    & $M_{B^*} - M_{B^{}}$ &  --  & \cite{Beringer:1900zz} & \checkmark\\
\hline %\hline
${B\to K}$ form factor
    & $f_+$   &  $q^2 = 17,\, 20,\, 23$ GeV$^2$  & \cite{Bouchard:2013eph} & ---\\
\hline %\hline
\multirow{3}{*}{${B\to K^*}$ {form factors}}
    & $V/A_1$ &  $q^2 = 0$ GeV$^2$  & \cite{Hambrock:2013zya} & \checkmark\\
    & $A_0$   &  $q^2 = 0$ GeV$^2$  & \cite{Khodjamirian:2010vf} & \checkmark\\
    & $V$, $A_1$, $A_{12}$
              &  $q^2 = 15,\, 19.21$ GeV$^2$     & \cite{Horgan:2013hoa}   & ---\\
\hline
\end{tabular}
\renewcommand{\arraystretch}{1.0}
\caption{\label{tab:observables} List of all observables in the various
  inclusive and exclusive $b\to s (\gamma,\, \ell^+\ell^-)$ decays that enter
  the global fits with their respective kinematics and experiments that provide
  the measurements. The $B^*$--$B^{}$ mass splitting is used to constrain matrix
  elements of dimension five operators. Lattice results of $B\to K$ form factors
  are used to constrain their parameters, and theoretical constraints on $B\to
  K^*$ form factors are included. For more details we refer to
  \refsec{sec:exp-input} and \refapp{app:theory-predictions}.  $\dagger$: Note
  that we include only the LHCb measurements in the $[1,\,6]$ GeV$^2$ bin as
  part of the ``selection'' data set, but not the low recoil bins.
}
\end{center}
\end{table*}

%
%--------+---------+---------+---------+---------+---------+---------+---------+
\subsection{New observables}

Measurements of the branching ratio of the inclusive radiative decay $B\to X_s
\gamma$
\begin{align}
  \BR_{1.8\, {\rm GeV}} &
  = (3.36 \pm 0.13 \pm 0.25) \cdot 10^{-4}\,, &  \mbox{\cite{Limosani:2009qg}}
\\[0.1cm]
  \BR_{1.8\, {\rm GeV}} &
  = (3.21 \pm 0.15 \pm 0.29) \cdot 10^{-4}\,, &  \mbox{\cite{Lees:2012ufa}}
\end{align}
are included with a lower cut on the photon energy $E_\gamma > 1.8$~GeV. For the
branching ratio of the inclusive semileptonic decay $B\to X_s \ell^+\ell^-$
integrated over the low-$q^2$ region $q^2\in [1,\, 6]$ GeV$^2$, we use
\begin{align}
    \langle \BR\rangle_{[1, 6]} &
  = (1.8 \pm 0.7 \pm 0.5) \cdot 10^{-6}\,, &  \mbox{\cite{Aubert:2004it}}
\\[0.1cm]
    \langle\BR\rangle_{[1, 6]} &
  = (1.493 \pm 0.504 {}^{+0.411}_{-0.321}) \cdot 10^{-6}\,. &  \mbox{\cite{Iwasaki:2005sy}}
\end{align}
A source of parametric uncertainty in inclusive decays arises from matrix
elements of dimension-five operators, $\mu^2_{\pi}$ and $\mu^2_{G}$, discussed
in more detail in \refapp{sec:inclusive-decays}. They appear in the heavy-quark
expansion at order $\Lambda_{\rm QCD}^2/m_b^2$, and $\mu^2_{G}$ enters also the
$B^*$--$B$ mass splitting~\cite{Beringer:1900zz}
\begin{align}
  \label{eq:Bmass:splitting:exp}
  M_{B^*} - M_{B^{}} & = (4.578 \pm 0.035) \cdot 10^{-2}\, \mbox{GeV} \,,
\end{align}
incorporated as an additional experimental constraint.

Previous experimental angular analyses of $B\to K^{*} (\to K\pi)\, \ell^+
\ell^-$ were restricted to the measurements of the longitudinal
$K^*$-polarization fraction, $F_L$, and the lepton forward-backward asymmetry,
$A_{\rm FB}$. The CDF collaboration was the first to measure $A_T^{(2)}$ and the
CP-asymmetry $A_9 = A_{im}$~\cite{Aaltonen:2011ja}. Most recently, LHCb extended
the angular analysis to measure $A_T^{\rm re}$ \cite{Aaij:2013iag} as well as
$S_{4,5,7,8}$ and their optimized analogues $P_{4,5,6,8}'$ \cite{Aaij:2013qta}
in addition to the previously published results on $A_T^{(2)}$, $S_3$, $A_9$,
and $S_9$. The original definitions of the observables $S_i$ and $P_i'$ can be
found in \cite{Altmannshofer:2008dz} and \cite{DescotesGenon:2012zf},
respectively. Here we include the measurements of $A_T^{\rm re}$ and
$P_{4,5,6}'$ in the $q^2$-bins $[1,\,6]$, $[14.18,\,16.0]$, and
$[>16.0]$~GeV$^2$.  Compared to \cite{Beaujean:2012uj}, we replace $S_3$ data
from LHCb by the corresponding $A_T^{(2)}$ results.

%
%--------+---------+---------+---------+---------+---------+---------+---------+
\subsection{Updated experimental input}

We use the same experimental input as in our previous analysis
\cite{Beaujean:2012uj}, unless the experimental collaborations provide updated
measurements. For some of the observables, additional measurements by further
experimental collaborations have become available; they are added to the
previous ones.  Both types of updates are listed below.

The measurement of the time-integrated and CP-averaged branching ratio of the
leptonic decay $B_s \to \mu^+\mu^-$ has been recently updated by LHCb and
measured for the first time by CMS
\begin{align}
  \BR &
  = \left(2.9 {}^{+1.1}_{-1.0} {}^{+0.3}_{-0.1}\right) \cdot 10^{-9}\,, &  \mbox{\cite{Aaij:2013aka}}
\\[0.1cm]
  \BR &
  = \left(3.0 {}^{+1.0}_{-0.9}\right) \cdot 10^{-9}\,, &  \mbox{\cite{Chatrchyan:2013bka}}
\end{align}
with $4.0\,\sigma$ and $4.3\,\sigma$ signal significance, respectively. To
faithfully model the physical constraint $\BR \ge 0$ and the reported asymmetric
uncertainties of the experimental probability distribution (PDF), we use the
Amoroso distribution \cite{Beaujean:2012uj}.

The LHCb measurement of the $B^+ \to K^+ \ell^+\ell^-$ branching ratio
(CP-averaged) \cite{Aaij:2012vr}, based on $1$ fb$^{-1}$ integrated luminosity,
is used in the $q^2$ bins $[1,\,6]$, $[14.18,\,16.0]$, $[16.0,\,18.0]$, and
$[18.0,\,22.0]$~GeV$^2$, in addition to the previous results from Belle and
BaBar. The new CDF results are now based on the full~\cite{CDF:2012:BKstarll}
rather than a partial data set~\cite{Aaltonen:2011qs}.  Very recently, LHCb
reported a broad peaking structure in the branching ratio at high $q^2$
compatible with $\psi(4160)$ \cite{Aaij:2013pta} using the larger data set of
$3$ fb$^{-1}$. Within the picture of quark-hadron duality, an adapted larger bin
$[\geq 15]$ GeV$^2$ around the peak should satisfy the necessary conditions
required by the theoretical framework \cite{Grinstein:2004vb, Beylich:2011aq} in
the future. For the moment, we continue to use the binning provided by
experiments and do not enlarge theoretical uncertainties in $B \to K^{(*)}
\ell^+\ell^-$ at high~$q^2$.

There are numerous updates on $B\to K^* \ell^+ \ell^-$ for which we use the
three $q^2$ bins $[1,\,6]$, $[14.18,\,16.0]$, and $[> 16.0]$~GeV$^2$. We add
recent measurements of the branching ratio from CMS \cite{Chatrchyan:2013cda} as
well as for $F_L$ and $A_{\rm FB}$ from CMS \cite{Chatrchyan:2013cda} and
ATLAS~\cite{ATLAS:2013ola}.  For the branching ratio, $F_L$ and $A_{\rm FB}$ we
now use the updated values \cite{Aaij:2013iag} instead of
\cite{LHCb:2012:BKstarll} in the case of LHCb, and \cite{CDF:2012:BKstarll}
instead of \cite{Aaltonen:2011qs, Aaltonen:2011ja} in the case of CDF.

For $B^0\to K^{*0}(\to K \pi)\, \ell^+\ell^-$, LHCb provides results of
``optimized observables''. Combining these with both the observables $F_L$ and
$A_{\rm FB}$ can lead to double counting \cite{Matias:2012xw} in the strict
limit of vanishing lepton masses that is well justified in the NP scenarios
considered in this work\footnote{This does not apply to NP scenarios with
  additional (pseudo-) scalar and tensor operators, however, the present LHCb
  analysis~\cite{Aaij:2013qta} can not be applied to such scenarios.}.  For this
reason, we replace the LHCb measurements of $\langle A_{\rm FB}\rangle_{[1,6]}$
by $\langle A_T^{\rm re}\rangle_{[1,6]}$.  However, we continue to use $\langle
A_{\rm FB}\rangle$ rather than $\langle A_T^{\rm re}\rangle$ for the low recoil
bins because neither is an optimized observable at high $q^2$. It is difficult
to include the LHCb measurement of $\langle A_T^{\rm re} \rangle_{[14.18,
  16.0]}$, since it implies that $A_T^{\rm re}(q^2)$ is constant --- against the
generic theory prediction --- and it attains the maximum value allowed by the
theory.

Overall, the measured $q^2$ dependence of individual observables is in quite
good agreement with the SM predictions. The largest deviation of $3.7\sigma$ is
reported for the optimized observable $\langle P_5' \rangle_{[4.3,8.68]}$
\cite{Aaij:2013qta} when compared to the SM prediction
\cite{Descotes-Genon:2013wba}.  We do not use any of the $[4.3,8.68]$ GeV$^2$
bins since their theory predictions receive large contributions from
$c\bar{c}$-loops \cite{Khodjamirian:2010vf}. The $[1,6]$~GeV$^2$ bins are less
affected by these effects. The remaining uncertainty is accounted for by the
parameters of subleading contributions at the level of the decay amplitudes
\cite{Beaujean:2012uj} in our predictions. The SM predictions available in the
literature for this bin (and the low-recoil bins) are compared with our results
in \reftab{tab:prediction-sl}.  Based on our prior input, we obtain central
values as in \cite{Descotes-Genon:2013wba} deviating by $2.5\sigma$ from the
measurement whereas the analysis \cite{Jager:2012uw} has a different central
value and larger errors with a $1.0\sigma$ deviation from experiment.

A second interesting deviation appears in the optimized observable $\langle P_4'
\rangle_{[14.18,16.0]}$. In the SM operator basis it is given by a ratio of form
factors \cite{Hambrock:2013zya} up to strongly suppressed subleading
corrections. The extrapolation of LCSR form-factor results
\cite{Khodjamirian:2010vf} from low to high $q^2$ yields a much larger value
compared to the measurement. This is also observed \cite{Horgan:2013pva} with
recent lattice QCD determinations of $B\to K^*$ form factors at high $q^2$
\cite{Horgan:2013hoa}. They allow us to further constrain the form-factor
parameters \emph{a priori}. However, these determinations do not reliably
consider systematic uncertainties due to finite width effects
\cite{Horgan:2013hoa} of the $K^*$. For this reason, we provide our fit results
for analyses with and without the $B\to K^*$ lattice inputs. Note, however, that
this issue does not affect the lattice determinations of $B\to K$ form factors.

A third deviation from the SM prediction is seen in the preliminary measurements
of $\langle F_L \rangle_{[1,6]}$ from BaBar and ATLAS that are both too low by
more than $3\sigma$ and $2\sigma$, respectively. This stands in contrast to the
published results of Belle, CDF, CMS, and LHCb that are all in good agreement
with the SM at low $q^2$.  The BaBar results are an average of $B^0 \to
K^{*0}Ê\ell^+\ell^-$ and $B^+ \to K^{*+}Ê\ell^+\ell^-$. While the neutral mode
yields $F_L$ consistent or close to the SM, the charged mode deviates strongly
in the low $q^2$ region and points in principle to a large isospin asymmetry for
the longitudinally polarized $K^*$ branching fraction \cite{Poireau:2012by}. The
ATLAS measurement has been performed only for the neutral mode. Although
preliminary and despite the isospin average in the case of BaBar, we include
both measurements in the fit.

%
%
%
%--------+---------+---------+---------+---------+---------+---------+---------+
\section{Results \label{sec:results}}

In this section, we review briefly our statistical approach and summarize our
fit results in several subsections, providing measures for the goodness of
fit. We describe several solutions in the subspace of the Wilson coefficients
for all four scenarios introduced in \refsec{sec:scenarios}. Using Bayes
factors, we compare models with non-SM Wilson coefficients to the \SMnu{} fit.
Finally, we present results for the nuisance parameters of the form factors and
subleading corrections to the $B\to K^*\ell^+\ell^-$ transition amplitudes in
each of the scenarios. Throughout we will compare with recent similar analyses
in the literature.

%
%
%--------+---------+---------+---------+---------+---------+---------+---------+
\subsection{Statistical Approach}

Our results are obtained from a Bayesian fit, similar to our previous work
\cite{Beaujean:2012uj}.  The main outputs are samples drawn from the posterior
distribution using the EOS flavor program \cite{EOS}. The samples are obtained
using an algorithm that employs Markov chains, hierarchical clustering, and
adaptive importance sampling (for a detailed description we refer to
\cite{Beaujean:2013:PMC}).

Throughout we denote the posterior probability density by $P(\vecth \,| D,M)$,
where $D\in\lbrace$full, full (+FF), selection$\rbrace$ represents the data set,
and $\vecth$ all parameters ($\wilson{i}$ and nuisance parameters $\vecnu$) of
the model $M \in \lbrace$\SMnu{}, \SM{}, \SMp{}, \SMpNine{}$\rbrace$ as defined
in \refsec{sec:scenarios}. The weighted posterior samples provide access to all
marginal distributions and to the evidence
\begin{equation}
  \label{eq:evidence}
  P(D|M) = \int_{V_0} \rmdx{\vecth}\, P(D|\vecth,M)\, P_0(\vecth\, | M )\,,
\end{equation}
where the integration extends over the whole prior volume $V_0$ spanned by the
parameters $\vecth$. The likelihood and prior distribution are denoted by
$P(D|\vecth,M)$ and $P_0(\vecth\,|M)$, respectively.  For $M\in\lbrace$\SM{},
\SMp{}$\rbrace$, the posterior has numerous well separated local maxima.  Most
of these have a negligible impact, and we consider only those solutions with
significant posterior mass. We require the ratio $R$ of the local evidence ---
integration volume $V_0$ restricted to contain only a single solution --- to the
global evidence \refeq{eq:evidence} exceeds $0.001$.  We label the individual
solutions as $A$ in the \SMnu{} scenario, $A$ and $B$ in the \SM{} scenario, and
$A'$ through $D'$ in the \SMp{} scenario, whereas in \SMpNine{} only $A'$
appears. For each model, $A^{(\prime)}$ denotes the solution in which the
signature of $(\wilson{7}, \wilson{9}, \wilson{10})$ is $(-,+,-)$ as predicted
in the \SMnu{}, and $B^{(\prime)}$ indicates flipped signs; i.e, $(+,-,+)$.

To determine the goodness of fit, we first find the best-fit point,
$\vecth^{*}$, in each solution by running the two local gradient-free optimizers
BOBYQA~\cite{Powell:2009} and COBYLA~\cite{Powell:1994} via NLopt~\cite{NLopt}
with the same initial point; usually the two results differ only slightly, and
we accept the point with the higher posterior. Next, we calculate the pull value
as in~\cite{Beaujean:2012uj} for fixed $M$,~$\vecth^{*}$ for each constraint,
and finally $\chi^2$ as the quadratic sum of pulls. From $\chi^2$, a $p$ value
follows assuming $N_{\rm dof}$ degrees of freedom. Note that there are $N$
experimental constraints and $\dim \vecnu$ informative priors. For the goodness
of fit, we consider each informative prior as \emph{one} constraint. With $K$
Wilson coefficients varied in $M$, we have
\begin{equation}
  \label{eq:n-dof}
  N_{\rm dof} = (N + \dim \vecnu) - (K + \dim \vecnu) = N - K
\end{equation}
degrees of freedom. For the full data set we have $\dim \vecnu = 28$ nuisance
parameters, and include $N=93$ constraints. In addition there are eleven or five
theory constraints on the form factors, depending on whether we include the
lattice results of the $B\to K^*$ form factors or not; see
\reftab{tab:btok-lattice} and \reftab{tab:btokStar-lattice}.  We consider those
constraints part of the prior, and do \emph{not} include them in $N_{\rm dof}$.
Below we denote both setups as ``full (+FF)'' and ``full'', respectively.  For
the ``selection'' data set, we have $N=20$ experimental inputs, two theory
constraints, and $\dim \vecnu = 24$.  For this data set we do not include the
lattice form-factor results since their theory uncertainty, when extrapolated to
low $q^2$, is comparable to the uncertainty of the LCSR results.

We calculate the Bayes factor between two statistical models $M_1$ and $M_2$ and
for a common data set $D$,
\begin{equation}
    B(D|M_1,M_2 ) \equiv \frac{P(D|M_1)}{P(D|M_2)}\,.
\end{equation}
The standard quantity to compare two models, the posterior odds, are defined as
\begin{equation}
  \label{eq:posterior-odds}
  \frac{P(M_1 | D)}{P(M_2|D)} = B(D|M_1, M_2) \frac{P_0(M_1)}{P_0(M_2)} \, ;
\end{equation}
i.e., the product of the Bayes factor and the prior odds $P_0(M_1) /
P_0(M_2)$. It is important to note that the prior of the model parameters,
$P_0(\vecth | M)$, is an integral part of the model $M$. Therefore, the Bayes
factor penalizes $M_2$ versus $M_1$ if $M_2$ contains extra parameters because
the evidence (\ref{eq:evidence}) is just the likelihood weighted by the prior,
and the average typically decreases when the same unit probability mass is
smeared over a larger volume $V_0$. This occurs, e.g., in the present analysis
for $M_1=$\,\SMnu{} and $M_2=$\,\SM{} as the Wilson coefficients are fixed in
$M_1$ but variable with flat priors over a large volume covering multiple
solutions in $M_2$. With the evidence given separately for each solution in
\reftab{tab:goodness-of-fit}, the reader can, for example, compute the Bayes
factor between $M_1$ and $M_2$ as though only one of the solutions had been
allowed a priori by reducing the (flat) prior ranges of the Wilson coefficients
and scaling the evidence accordingly. We focus on the SM-like solution of each
scenario that is fully contained in a hyperrectangle with edge lengths
\begin{equation}
  \label{eq:reduced:prior}
\begin{aligned}
  0.4 & & \mbox{for} & & & \wilson{7,7'} \,,
\\
  4.0 & & \mbox{for} & & & \wilson{9,9, 10,10'}\,.
\end{aligned}
\end{equation}
These reduced ranges yield Bayes factors that minimally penalize the NP
models. In other words, if $B(D|\text{SM}(\nu\text{-only}), M_2) > 1$, then any
larger range would increase $B(D|\text{SM}(\nu\text{-only}), M_2)$ even more.
The penalty due to extra parameters can be overcome if $M_2$ provides a
significantly better description of the data; i.e., higher likelihood values. In
conclusion, $B(D|M_1, M_2)>1$ implies that the data favor $M_1$.

We stress that the evidence by itself is not meaningful because of the arbitrary
likelihood normalization due to the fact that we do not have the actual events
seen by an experiment but only a concise summary usually in the form of an
observable's value maximizing the likelihood value plus uncertainties. Using a
consistent normalization, at least ratios of evidences for identical data $D$
--- as in the Bayes factor --- have a well defined interpretation.

\begin{table}[floatfix]
    \begin{center}
    \renewcommand{\arraystretch}{1.4}
    \resizebox{\columnwidth}{!}{
    \begin{tabular}{ccccccc}
        \hline
        Scenario & Data set & Solution & $\chi^2$ & $p$ value & $\ln P(D|M)$& $R$
        \\
        \hline
        \multirow{3}{*}{\SMnu{}}
        & full                       & $A$ & 109.4 & 0.12 & 572.3 & 1\\
        \cline{2-7}
        & full (+FF)                 & $A$ & 114.5 & 0.06 & 580.2 & 1\\
        \cline{2-7}
        & selection                  & $A$ & 12.4  & 0.90 & 118.0 & 1\\
        \hline
        % SM-like solution
        \multirow{6}{*}{\SM{}}
        & \multirow{2}{*}{full}      & $A$ & 106.0 & 0.12 & 562.1 & 0.82\\
                                   & & $B$ & 110.4 & 0.07 & 560.6 & 0.18\\
        \cline{2-7}
        & \multirow{2}{*}{full (+FF)}& $A$ & 109.7 & 0.08 & 570.1 & 0.75\\
                                   & & $B$ & 111.8 & 0.06 & 569.0 & 0.25\\
        \cline{2-7}

        & selection                  & $A$ & 6.2   & 0.99 & 112.1 & 1\\
        \hline
        \multirow{8}{*}{\SMp{}}
        & \multirow{4}{*}{full}      & $A'$& 107.0 & 0.07 & 557.8 & 0.37\\
                              &      & $B'$& 106.9 & 0.07 & 556.8 & 0.14\\
                              &      & $C'$& 106.2 & 0.08 & 556.9 & 0.15\\
                              &      & $D'$& 105.4 & 0.09 & 557.7 & 0.34\\
        \cline{2-7}
        & \multirow{4}{*}{full (+FF)}& $A'$& 109.7 & 0.05 & 566.7 & 0.35\\
                              &      & $B'$& 106.9 & 0.07 & 565.9 & 0.16\\
                              &      & $C'$& 107.6 & 0.07 & 566.0 & 0.17\\
                              &      & $D'$& 105.5 & 0.09 & 566.6 & 0.32\\
        \hline
        \multirow{2}{*}{\SMpNine{}}
        & full                       & $A'$& 105.7 & 0.14 & 568.7 & 1\\
        \cline{2-7}
        & full (+FF)                 & $A'$& 110.1 & 0.08 & 577.6 & 1\\
    \hline
    \end{tabular}
    }
    \renewcommand{\arraystretch}{1.0}
    \end{center}
    \caption{
        Goodness of fit and posterior evidence (ratio) for various combinations
        of constraints and fit models. Individual solutions are labeled as $A$
        and $B$ in the \SMnu{} and the \SM{}, and $A'$ through $D'$ in the
        \SMp{} and $A'$ in the \SMpNine{}. The solutions with SM-like and flipped
        signs of $\wilson{i}$ are $A^{(\prime)}$ and $B^{(\prime)}$, respectively.
        For the definitions of $P(D|M)$ and $R$ see \refeq{eq:evidence} and below.
    }
    \label{tab:goodness-of-fit}
\end{table}

The fit results regarding model comparison and goodness of fit are listed in
\reftab{tab:goodness-of-fit}, in which the local evidence is computed at a
relative precision of 1\% on the linear scale. For an overview of all of the
following results on the Wilson coefficients, we refer to
\reftab{tab:wilson:coeff:1-dimCLs}.  The pull values entering the $p$ value are
compiled for each experimental constraint in \reftab{tab:p-value-sm-full} for
the solution $A$ of \SM{} and $A'$ of the \SMp{}.

%
%--------+---------+---------+---------+---------+---------+---------+---------+
\subsection{Fitting the Nuisance Parameters}
\label{sec:fit:sm-nu-only}

\begin{figure}
  \includegraphics[width=.45\textwidth]{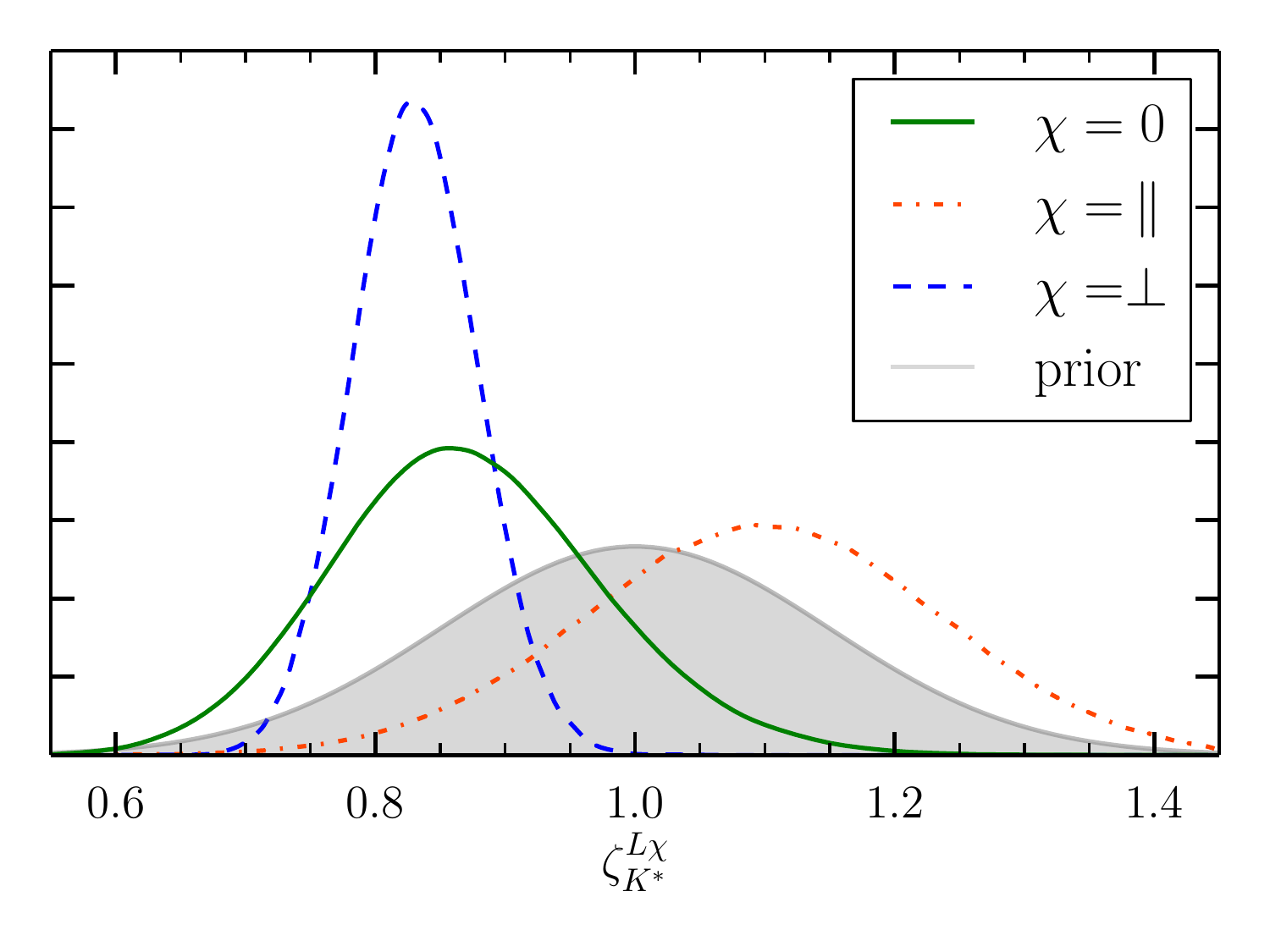}\\
  \includegraphics[width=.45\textwidth]{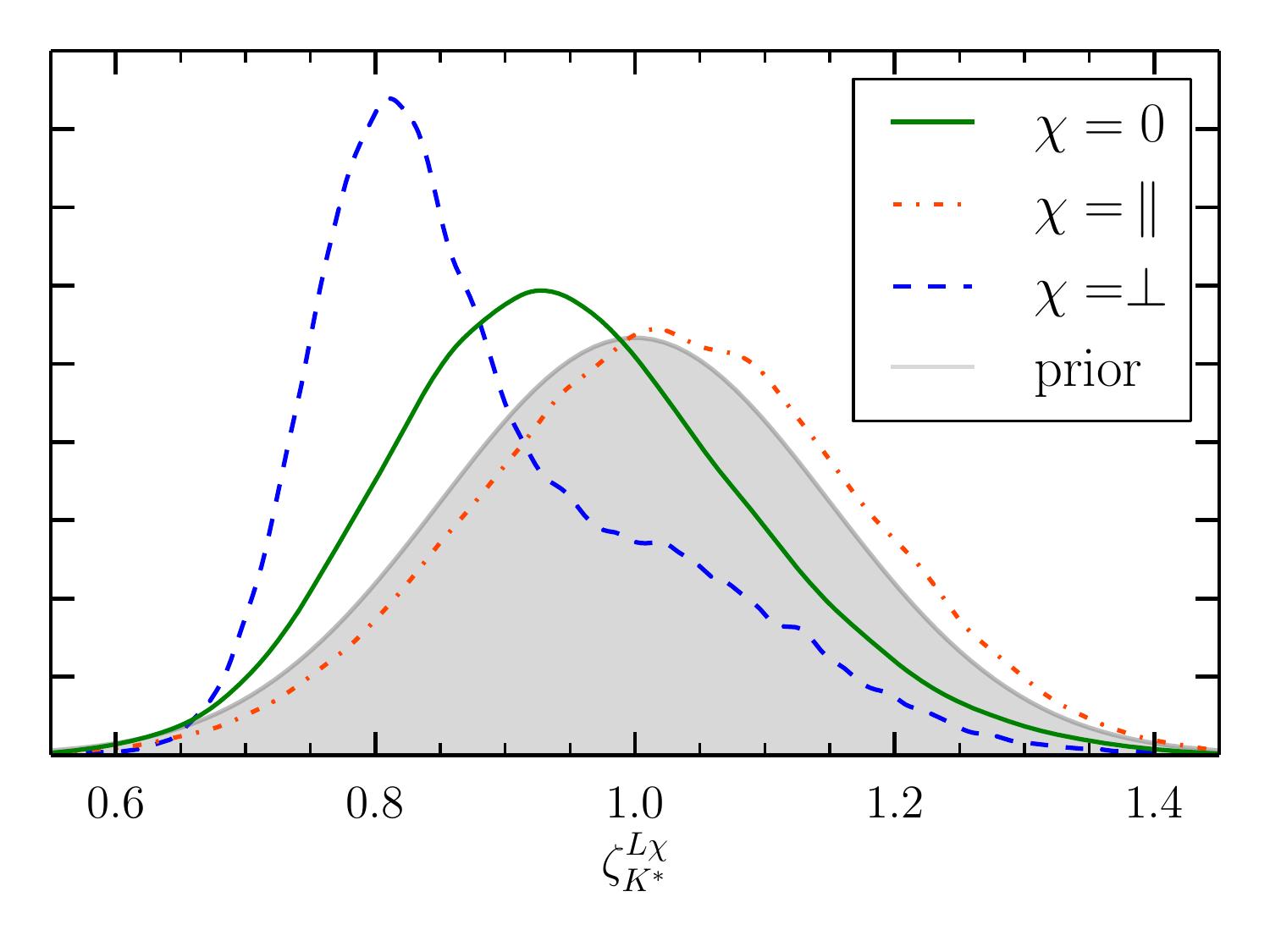}
  \caption{
    Comparison of the prior (gray shaded) and marginalized posterior distributions,
    using the ``full'' data set, for the parameters $\zeta^{L\chi}_{K^*}$,
    $\chi=\perp,\para,0$, describing the unknown $1/m_b$ contributions to the
    $B\to K^*\ell^+\ell^-$ transversity amplitudes $A_\chi^{L}$ at large recoil.
    Upper panel: \SMnu{}; in the \SM{}, the results are very similar (not shown).
    Lower panel: \SMp{}. Note that the tail for $\chi = \perp$ is a consequence
    of the suppressed solutions $C'$ and $D'$.
  }
  \label{fig:subleading-btokstar}
\end{figure}

\begin{table*}
  \renewcommand{\arraystretch}{1.4}
  \setlength{\tabcolsep}{5pt}
  \begin{tabular}{c|cccc|cccc}
  \hline
                    & \multicolumn{4}{c|}{no $B\to K^*$ lattice} & \multicolumn{4}{c}{$B\to K^*$ lattice} \\
                    & prior                   &  \SMnu{}                &  \SM{}                  &  \SMp{}                 & prior                   & \SMnu{}                 & \SM{}                   &  \SMp{}                 \\
    \hline
    $V(0)$          & $ 0.35^{+0.14}_{-0.09}$ & $ 0.40^{+0.03}_{-0.03}$ & $ 0.40^{+0.03}_{-0.03}$ & $ 0.39^{+0.03}_{-0.03}$ & $ 0.36^{+0.03}_{-0.03}$ & $ 0.38^{+0.03}_{-0.02}$ & $ 0.38^{+0.03}_{-0.02}$ & $ 0.37^{+0.02}_{-0.02}$ \\
    $b_1^V$         & $-4.8^{+0.7}_{-0.5}$    & $-4.7^{+0.7}_{-0.5}$    & $-4.8^{+0.5}_{-0.4}$    & $-4.9^{+0.5}_{-0.3}$    & $-4.8^{+0.7}_{-0.4}$    & $-4.6^{+0.8}_{-0.4}$    & $-4.8^{+0.7}_{-0.4}$    & $-4.9^{+0.6}_{-0.3}$    \\
    $A_1(0)$        & $ 0.28^{+0.08}_{-0.07}$ & $ 0.24^{+0.03}_{-0.02}$ & $ 0.25^{+0.03}_{-0.02}$ & $ 0.26^{+0.03}_{-0.03}$ & $ 0.28^{+0.04}_{-0.03}$ & $ 0.26^{+0.03}_{-0.02}$ & $ 0.26^{+0.03}_{-0.02}$ & $ 0.27^{+0.03}_{-0.03}$ \\
    $b_1^{A_1}$     & $ 0.4^{+0.7}_{-1.0}$    & $ 0.4^{+0.6}_{-0.6}$    & $ 0.5^{+0.6}_{-0.6}$    & $ 0.5^{+0.6}_{-0.7}$    & $ 0.5^{+0.5}_{-0.7}$    & $ 0.3^{+0.5}_{-0.6}$    & $ 0.4^{+0.5}_{-0.6}$    & $ 0.2^{+0.6}_{-0.5}$    \\
    $A_2(0)$        & $ 0.24^{+0.13}_{-0.07}$ & $ 0.23^{+0.04}_{-0.04}$ & $ 0.24^{+0.04}_{-0.04}$ & $ 0.24^{+0.05}_{-0.04}$ & $ 0.28^{+0.05}_{-0.05}$ & $ 0.25^{+0.04}_{-0.03}$ & $ 0.26^{+0.04}_{-0.04}$ & $ 0.27^{+0.04}_{-0.04}$ \\
    $b_1^{A_2}$     & $-0.5^{+2.1}_{-1.7}$    & $-0.6^{+1.5}_{-1.3}$    & $-0.9^{+1.6}_{-1.1}$    & $-0.8^{+1.4}_{-1.2}$    & $-1.4^{+1.3}_{-0.9}$    & $-1.4^{+1.0}_{-0.9}$    & $-1.5^{+1.1}_{-0.7}$    & $-1.4^{+1.2}_{-0.8}$    \\
    \hline
    $f_+(0)$        & $ 0.33^{+0.04}_{-0.03}$ & $ 0.30^{+0.02}_{-0.02}$ & $ 0.30^{+0.02}_{-0.02}$ & $ 0.29^{+0.02}_{-0.02}$ & $ 0.33^{+0.04}_{-0.03}$ & $ 0.30^{+0.02}_{-0.02}$ & $ 0.31^{+0.02}_{-0.02}$ & $ 0.29^{+0.02}_{-0.02}$ \\
    $b_1^{f_+}$     & $-2.3^{+0.6}_{-0.8}$    & $-3.1^{+0.5}_{-0.5}$    & $-3.1^{+0.5}_{-0.5}$    & $-3.2^{+0.4}_{-0.5}$    & $-2.3^{+0.6}_{-0.8}$    & $-3.1^{+0.5}_{-0.5}$    & $-2.9^{+0.4}_{-0.6}$    & $-3.4^{+0.6}_{-0.5}$    \\
    \hline
    $V(0)/A_1(0)$   & $1.3^{+0.3}_{-0.3}$     & $ 1.6^{+0.2}_{-0.1}$    & $ 1.6^{+0.2}_{-0.2}$    & $ 1.5^{+0.2}_{-0.2}$    & $1.2^{+0.2}_{-0.1}$     & $ 1.5^{+0.2}_{-0.1}$    & $ 1.4^{+0.2}_{-0.2}$    & $ 1.4^{+0.2}_{-0.2}$    \\
    $A_2(0)/A_1(0)$ & $0.99^{+0.10}_{-0.15}$  & $ 0.95^{+0.08}_{-0.08}$ & $ 0.96^{+0.07}_{-0.08}$ & $ 0.96^{+0.08}_{-0.08}$ & $0.98^{+0.09}_{-0.10}$  & $ 0.98^{+0.07}_{-0.07}$ & $ 0.99^{+0.07}_{-0.08}$ & $ 0.98^{+0.07}_{-0.07}$ \\
    \hline
  \end{tabular}
  \renewcommand{\arraystretch}{1.0}
  \setlength{\tabcolsep}{6pt}
  \caption{
    1D-marginalized posterior results at 68\% probability in comparison to the
    prior inputs for the various $B\to K^*$ (upper rows) and $B\to K$ (middle
    two rows) form-factor parameters. The results are shown for the ``full''
    (left) and ``full (+FF)'' (right) data set in various scenarios. The priors
    for the ``full'' data set comprise LCSR \cite{Khodjamirian:2010vf} inputs
    combined with the additional constraints \refeq{eq:FF-constr:VoverA1} --
    \refeq{eq:FF-constr:A0:LCSR} and $B\to K$ lattice results \cite{Bouchard:2013eph},
    whereas for ``full (+FF)'' the $B\to K^*$ lattice results \cite{Horgan:2013hoa}
    are added. Note that the marginalization has been performed over all solutions
    $A,B$ in the case of \SM{} and $A'-D'$ in the case of \SMp{}.
  }
  \label{tab:FF:fit-results}
\end{table*}

Let us begin the summary of our results with the fit of the scenario \SMnu{};
i.e., the fit of nuisance parameters by fixing Wilson coefficients to their
values in the SM at the scale $\mu = 4.2$ GeV
\begin{align}
  \wilson[SM]{7}  & = -0.34\,, &
  \wilson[SM]{9}  & =  4.27\,, &
  \wilson[SM]{10} & = -4.17\,.
\end{align}
The main purpose is to check whether the \SMnu{}, including all theory
uncertainties, provides a good description of the available data.  This scenario
serves as the reference point to compare with scenarios \SM{}, \SMpNine{}, and
\SMp{} later on.

Within the \SMnu{} scenario, we perform three fits to the data sets ``full'',
``full (+FF)'', and ``selection''. All fits exhibit satisfactory or excellent
$p$ values of $0.12$, $0.06$ and $0.90$, respectively, and show that the
posterior has only one relevant mode. For the values of the evidence and
$\chi^2$, we refer to \reftab{tab:goodness-of-fit}. The large $p$ value for the
data set ``selection'' can be explained through the smaller overall number of
measurements relative to $\dim \vecnu$, and especially the absence of
measurements deviating substantially from their SM predictions.  The
measurements with the largest pull values above $2\sigma$ are all in $B\to
K^*\ell^+\ell^-$ as can be seen in \reftab{tab:p-value-sm-full}: the
aforementioned $\langle F_L \rangle_{[1,6]}$ from BaBar and ATLAS, $\langle \BR
\rangle_{[16,19]}$ from Belle, $\langle A_{\rm FB} \rangle_{[16,19]}$ from ATLAS
and the two optimized observables $\langle P_4' \rangle_{[14.18,16]}$ and
$\langle P_5' \rangle_{[1,6]}$ from LHCb (see also the caption of
\reftab{tab:p-value-sm-full}).

We note that removing the ATLAS and BaBar measurements of $\langle
F_L\rangle_{[1,6]}$ increases the $p$ value substantially, from $0.12$ to $0.63$
and from $0.06$ to $0.55$ for the ``full'' and ``full (+FF)'' data sets,
respectively. The smaller $p$ value of the ``full (+FF)'' data set compared to
the one of the ``full'' data set arises in part from a known tension of the
$\BR(B\to K^*\ell^+\ell^-)$ data at high $q^2$ with predictions based on $B\to
K^*$ lattice form factors \cite{Horgan:2013hoa,Horgan:2013pva}.

We find that the impact of experimental measurements published after our
previous analysis \cite{Beaujean:2012uj} does not change the main outcome; i.e.,
the data can be accurately described without resort to new physics beyond the
SM. This result may appear surprising given the large tensions that were seen in
\cite{Descotes-Genon:2013wba}.  Within our approach, however, the tension
between SM prediction and measurement of $\langle P'_5\rangle_{[1,6]}$ can be
eased by shifts in the parameters $\zeta^{L \chi}_{K^*}$, $\chi = \perp,\para,0$
that parametrize the size of subleading contributions at large recoil in $B\to
K^*\ell^+\ell^-$, see \refapp{sec:subleading} for their definition. The shifts
of about $-(15$--$20)\%$ to $\zeta^{L \chi}_{K^*}$, $\chi=\perp,0$ and about
$+10\%$ to $\zeta^{L \para}_{K^*}$ are compatible with the power-counting
expectation $\Lambda_\text{QCD}/m_b$. They suffice to increase the most likely
value of $\langle P'_5\rangle_{[1,6]}$ from $-0.34$ (nominal prior) to $-0.27$
(see \reftab{tab:prediction-sl}), thereby slightly reducing the tension of the
$B\to K^*\ell^+\ell^-$ ``anomaly''. The prior and posterior distributions of
$\zeta^{L \chi}_{K^*}$ are shown in \reffig{fig:subleading-btokstar}.  We do not
find any shifts in the parameters $\zeta^{R \chi}_{K^*}$, $\chi = \perp,\para,0$
exceeding a few percent. There are no significant differences between the
``full'' and ``full (+FF)'' fits, except for two shifts in subleading
parameters. At low~$q^2$, $\zeta^{L 0}_{K^*}$ reduces from $-15\%$ to about
$-10\%$, and at high $q^2$, $\Lambda_\parallel$ changes from $0\%$ to about
$-3\%$ due to the tension between the data and predictions using lattice $B\to
K^*$ form factors for the branching ratio $\BR$ (see also the $\BR$ predictions
in \cite{Horgan:2013pva}).  The large shift in $\zeta^{L\perp}_{K^*}$ is mostly
due to $B\to K^* \gamma$ and reduces to a few percent once removing measurements
of this decay from the fit. Choosing a $q^2$-dependent parametrization of
subleading corrections should alleviate this tension.  We also find a negative
subleading contribution to the $B\to K\ell^+\ell^-$ amplitude at low $q^2$,
which lowers $\langle\mathcal{B}\rangle_{[1,6]}$ to accommodate the measurement
by LHCb.  This is in agreement with the observations of
\cite{Khodjamirian:2012rm}.

Beyond inference of the size of subleading contributions to the amplitudes, we
extract information on the $B\to K$ and $B\to K^*$ hadronic form factors. We
confront our results for the various fit scenarios with the prior values in
\reftab{tab:FF:fit-results}. Within the ``full'' data set, the ratio
$V(0)/A_1(0)$ is generally higher than the results labeled ``SE2 LCSR'' of
\cite{Hambrock:2013zya}, with less than $1\sigma$ tension for our scenarios
\SM{} and \SMp{}, and $\simeq 1\sigma$ deviation for our scenario \SMnu{}. Our
results for the ratio $A_2(0)/A_1(0)$ are consistently higher than those of
\cite{Hambrock:2013zya}, with a $2\sigma$ variance, which can be attributed to
our usage of the constraint on $A_0(0)$, see \refeq{eq:FF-constr:A0:LCSR}. In
the case of the ``full (+FF)'' data set, the prior values of $B\to K^*$ form
factors at $q^2 = 0$ show two to four times smaller uncertainties when taking
into account the lattice results \cite{Horgan:2013hoa}, whereas the shape
parameters $b_1$ are not affected except for the form factor $A_2$. Although
lattice results provide a reduction of the prior uncertainty, there is still
agreement with \cite{Hambrock:2013zya} within 1$\sigma$ for the ``full (+FF)''
data set.  Except for a slight increase of the variance of $V(0)/A_1(0)$ in
scenarios \SM{} and \SMp{}, the ratios do not change qualitatively.

\begin{figure*}[t]
    \begin{center}
        \includegraphics[width=.32\textwidth]{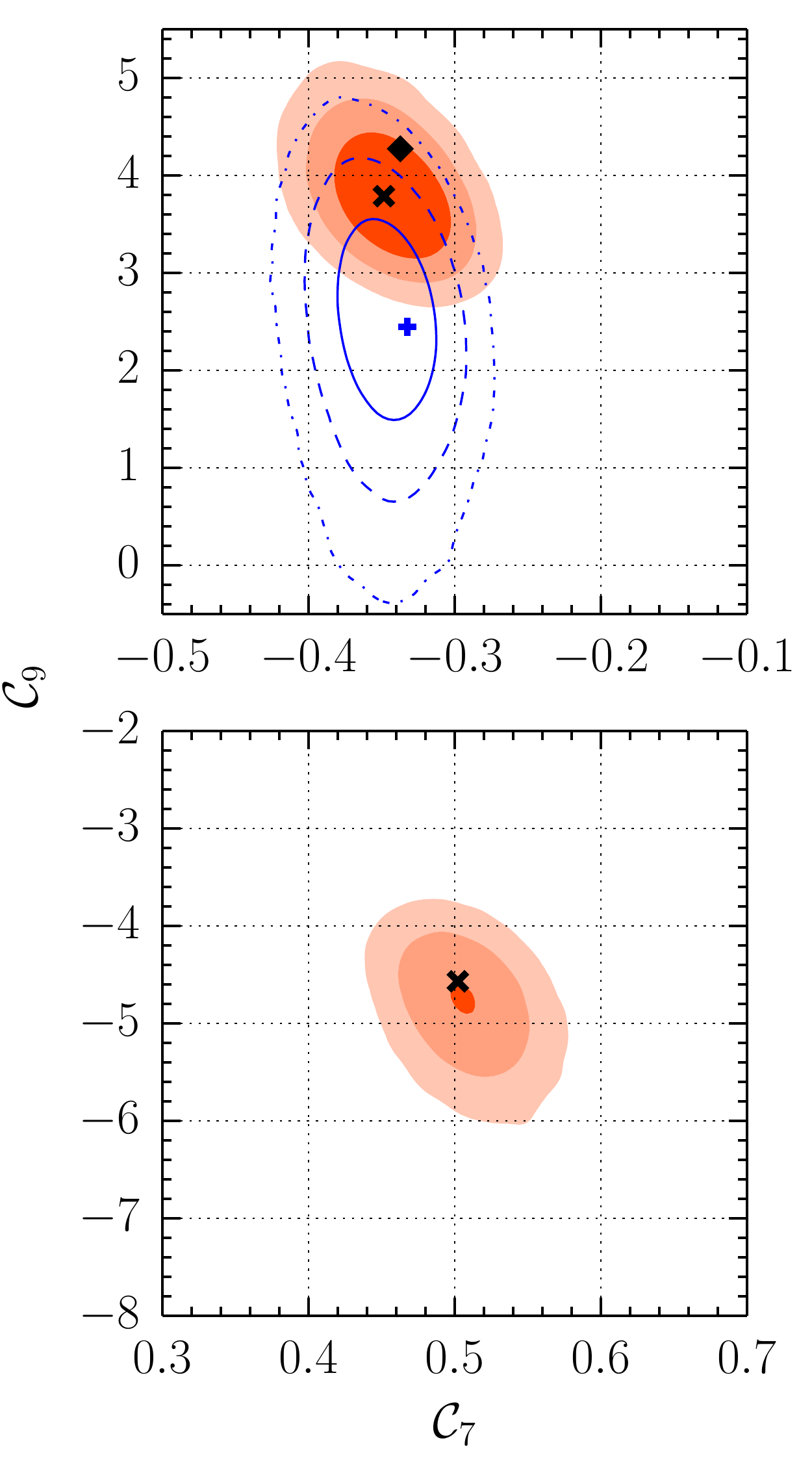}
        \includegraphics[width=.32\textwidth]{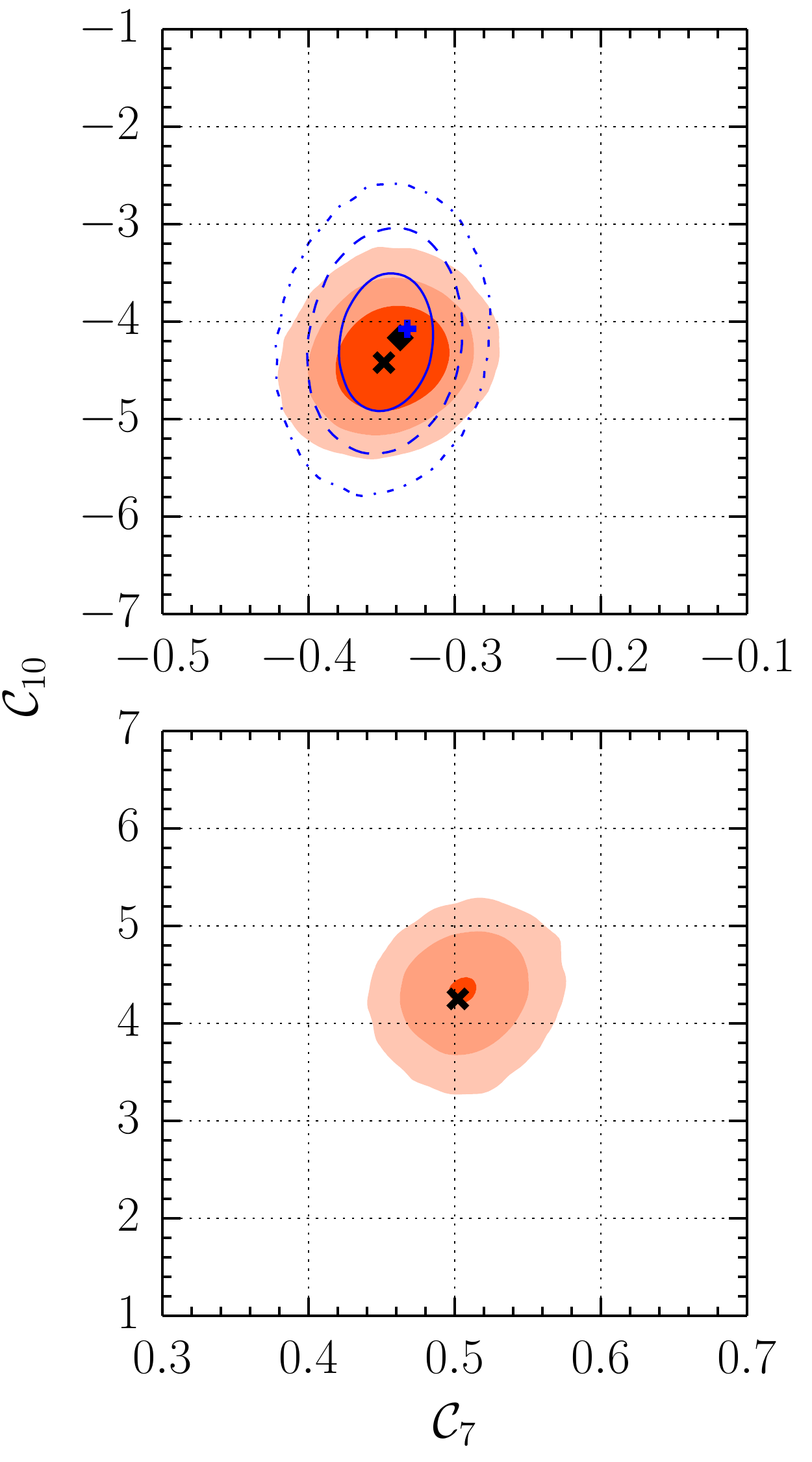}
        \includegraphics[width=.32\textwidth]{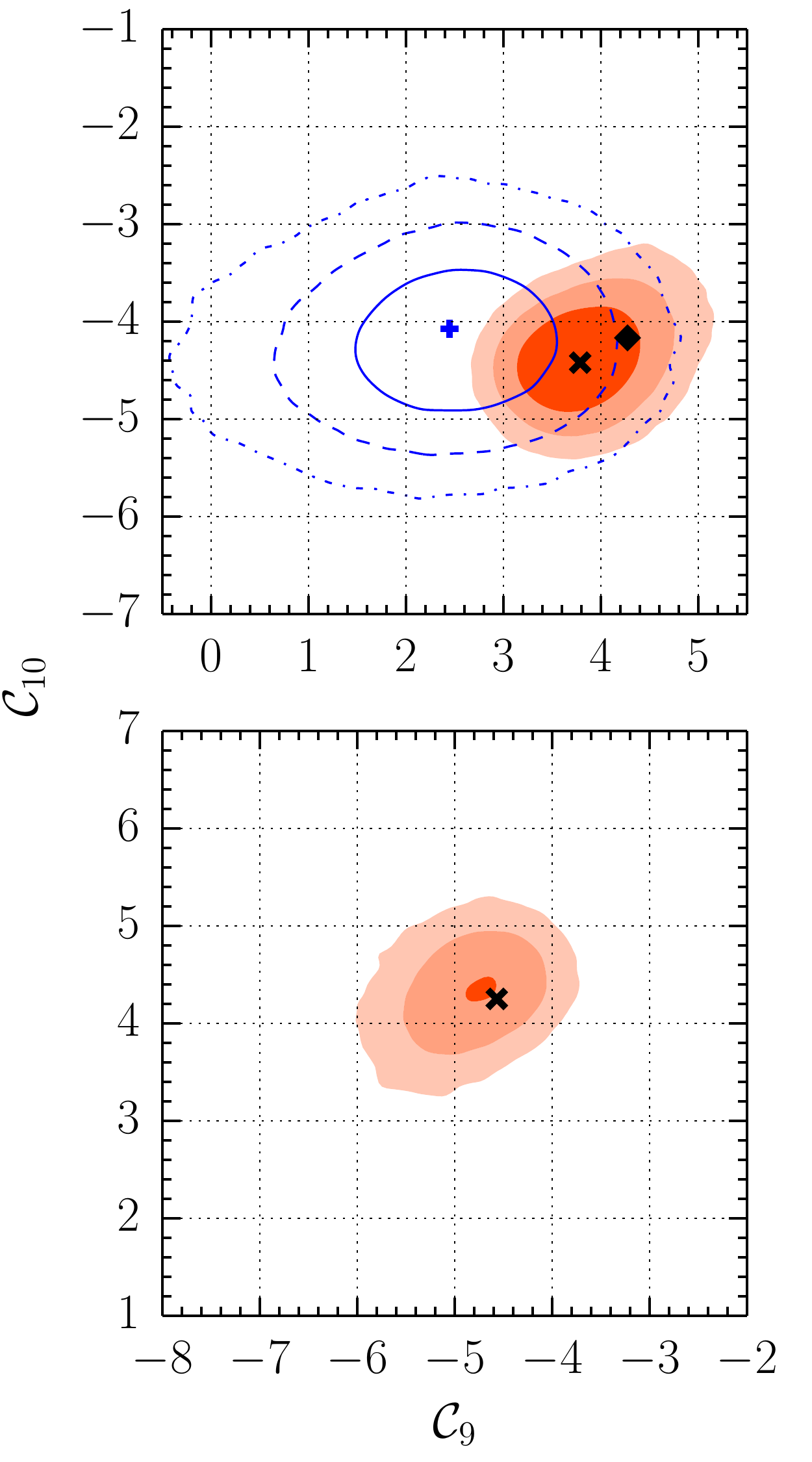}
    \end{center}
    \caption{
      Credibility regions of the Wilson coefficients $\wilson{7,9,10}$
      obtained from the fit of the ``full'' data set after the EPSHEP 2013 conference at
      $68.3\%$, $95.4\%$, $99.7\%$ (dark, normal, and light red) probability.
      The SM-like solution $A$ (upper row) and the flipped-sign
      solution $B$ (lower row) are magnified.  Overlaid are the results of the
      fit to the ``selection'' data set at $68.3\%$, $95.4\%$, $99.7\%$ (solid, dashed,
      and dash-dotted blue lines) probability. The projection of the SM point is represented
      by the black diamond, whereas the black and blue crosses mark the best-fit points.
      \label{fig:results-wilson-coeff-SM}
    }
\end{figure*}

%
%--------+---------+---------+---------+---------+---------+---------+---------+
\subsection{Fit in the SM basis}
\label{sec:fit:sm}

Although the \SMnu{} fit shows that the SM provides a reasonable description of
the available data, we still extend our analysis to obtain model-independent
constraints on NP couplings. In the SM scenario we fit the real-valued Wilson
coefficients $\wilson{7,9,10}$ in addition to the nuisance parameters (see
eq. \refeq{eq:scenarios}) to the data sets ``full'', ``full (+FF)'' and
``selection''.  For all data sets we obtain two dominant solutions $A$ and $B$
with SM-like and flipped signs of the Wilson coefficients, and many more
solutions with negligible posterior mass.

In the case of the ``selection'' data set, a $p$ value of $0.99$ is obtained for
solution $A$, depicted in \reffig{fig:results-wilson-coeff-SM}.  It is larger
compared to $0.90$ obtained in the \SMnu{} scenario, indicating that the
additional parameters allow to further reduce the tension with the data by
$\Delta \chi^2 \simeq -6$. Within solution $A$, the fit yields a deviation from
the SM value of $\wilson{9}$
\begin{equation}
  \Delta_9 = \wilson{9} - \wilson[SM]{9} \simeq -1.7 \pm 0.7\,,
\end{equation}
with $68\%$ probability; see \reftab{tab:wilson:coeff:1-dimCLs}. However, we
find no significant deviations in either $\wilson{7}$ or $\wilson{10}$. This
observation is compatible with the findings of \cite{Descotes-Genon:2013wba},
where only solution $A$ had been kept. However, in our results the
2D-marginalized posterior shows a $\simeq 2.5\sigma$ deviation in the
$(\wilson{7}-\wilson{9})$ plane from the SM, in contrast to $3.2\sigma$ as in
\cite{Descotes-Genon:2013wba}.

As in the case of \SMnu, the $p$ values are much smaller for the ``full'' data
set compared to the ``selection'' data set, but still indicate a decent fit at
$0.12$ and $0.07$ for both solutions $A$ and $B$, respectively. Removing the
$\langle F_L \rangle_{[1,6]}$ measurement from BaBar and ATLAS increases the $p$
values to $0.59$ and $0.53$ for solutions $A$ and $B$, respectively. Contrary to
the ``selection'', solution $A$ is now strongly favored over solution $B$: $R_A
: R_B = 82\% : 18\%$. This underlines the importance of a combined analysis of
all available experimental data rather than a selected subset.

As can be seen in \reffig{fig:results-wilson-coeff-SM}, the SM lies within the
$1\sigma$ credibility regions of all 2D-marginalized posterior distributions.
With the updated experimental data, the credibility regions are reduced in size
by roughly a factor of two when compared to our previous results
\cite{Beaujean:2012uj}. For the 1D credibility regions with both solutions $A$
and $B$ we refer to \reftab{tab:wilson:coeff:1-dimCLs}, whereas for the single
solution~$A$ we find smaller 68\% probability regions of
\begin{align*}
  \Delta_7    & =  0.0 \pm 0.02 , &
  \Delta_9    & = -0.5 \pm 0.3 ,  &
  \Delta_{10} & = -0.2 \pm 0.3 .
\end{align*}
This value $\Delta_9$ seems to contradict \reftab{tab:wilson:coeff:1-dimCLs} and
\reffig{fig:results-wilson-coeff-SM} as the SM value of $\wilson{9}$ is within
the 68\% region in both cases. But the 68\% regions would only agree if solution
$B$ had either negligible or equal weight compared to solution $A$.

The authors of \cite{Altmannshofer:2013foa} do not consider a scenario of
simultaneous NP contributions to $\wilson{7,9,10}$, but only
single-Wilson-coefficient scenarios $\wilson{7}$ and $\wilson{9}$, the
two-Wilson-coefficient scenario $\wilson{7,9}$ and the full set of Wilson
coefficients of \SMp{}. Their results show a decrease of $|\Delta_7|$ once
allowing NP contributions to $\wilson{9,10}$ similar to our findings\footnote{
  Note that in \cite{Altmannshofer:2013foa} Wilson coefficients are determined
  at the scale $\mu = 160$ GeV but RGE effects are only of concern for
  $\Delta_7$.}.  The NP contributions $\Delta_9$ and $\Delta_{10}$ are also
found to be preferentially negative.

The situation of the $P'_5$ anomaly is the same as in the \SMnu{} fit, and the
modifications to the posterior distributions of $\zeta^{L(R) \chi}_{K^*}$,
$\chi=\perp,\para,0$ are of the same type and similar size for both data
sets. The same applies to the postdiction $\langle P_5'\rangle_{[1,6]}$ given in
\reftab{tab:prediction-sl}. The pull value of $\langle P'_5\rangle_{[1,6]}$
decreases only little from $2.3\sigma$ in the \SMnu{} fit to $1.4\sigma$ in the
\SM{} fit when allowing NP contributions to $\wilson{7,9,10}$. However, the
tensions in other measurements are not eased, see \reftab{tab:p-value-sm-full}.

Focusing on solution $A$ (cf. \refeq{eq:reduced:prior}), the fit yields a Bayes
factor of
\begin{equation}
    \label{eq:bayes:solA:SM}
    \frac{P(\text{full} | \text{SM})}{P(\text{full} | \text{\SMnu})} \Big|_{A}
    = 1:48 \,.
\end{equation}
In the absence of substantial improvements in the handling of subleading
contributions to the $B\to K^{(*)}\ell^+\ell^-$ amplitudes, we are forced to
conclude that the SM interpretation of the data is more economical than the
hypothesis of New Physics contributions to the SM Wilson coefficients.

The inclusion of lattice $B\to K^*$ form factors in the ``full (+FF)'' analysis
shifts the ratio of the probability mass of solutions $A$ and $B$ towards $B$,
$R_A :R_B = 0.75 : 0.25$. The $p$ values at the best fit points decrease to
$0.08$ and $0.06$, respectively. Omitting $F_L$ from BaBar and ATLAS, the $p$
values jump to $0.53$ and $0.54$ in $A$ and $B$, respectively.  The
1D-marginalized 68\% probability regions of the Wilson coefficients in solution
$A$ remain the same
\begin{align*}
  \Delta_7    & =  0.0 \pm 0.02 , &
  \Delta_9    & = -0.5 \pm 0.3 , &
  \Delta_{10} & = -0.1 \pm 0.3 ,
\end{align*}
when compared to the ``full'' data set, except for the central value of
$\Delta_{10}$, which moves even closer to the SM. Finally, the Bayes factor of
solution $A$ is $1:43$.

\begin{figure}%[t]
  \begin{center}
        \includegraphics[width=.38\textwidth]{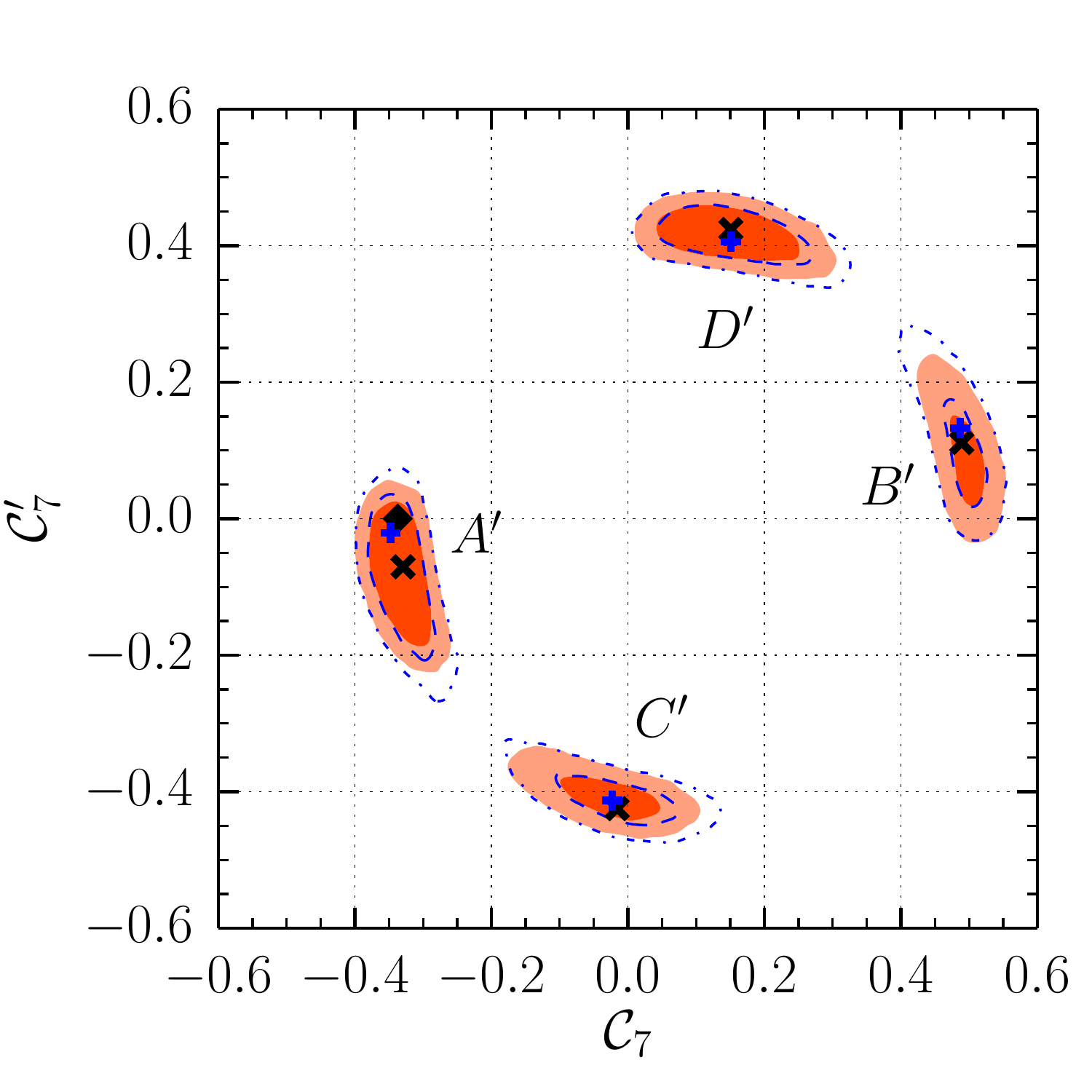} \\[-0.2cm]
        \includegraphics[width=.38\textwidth]{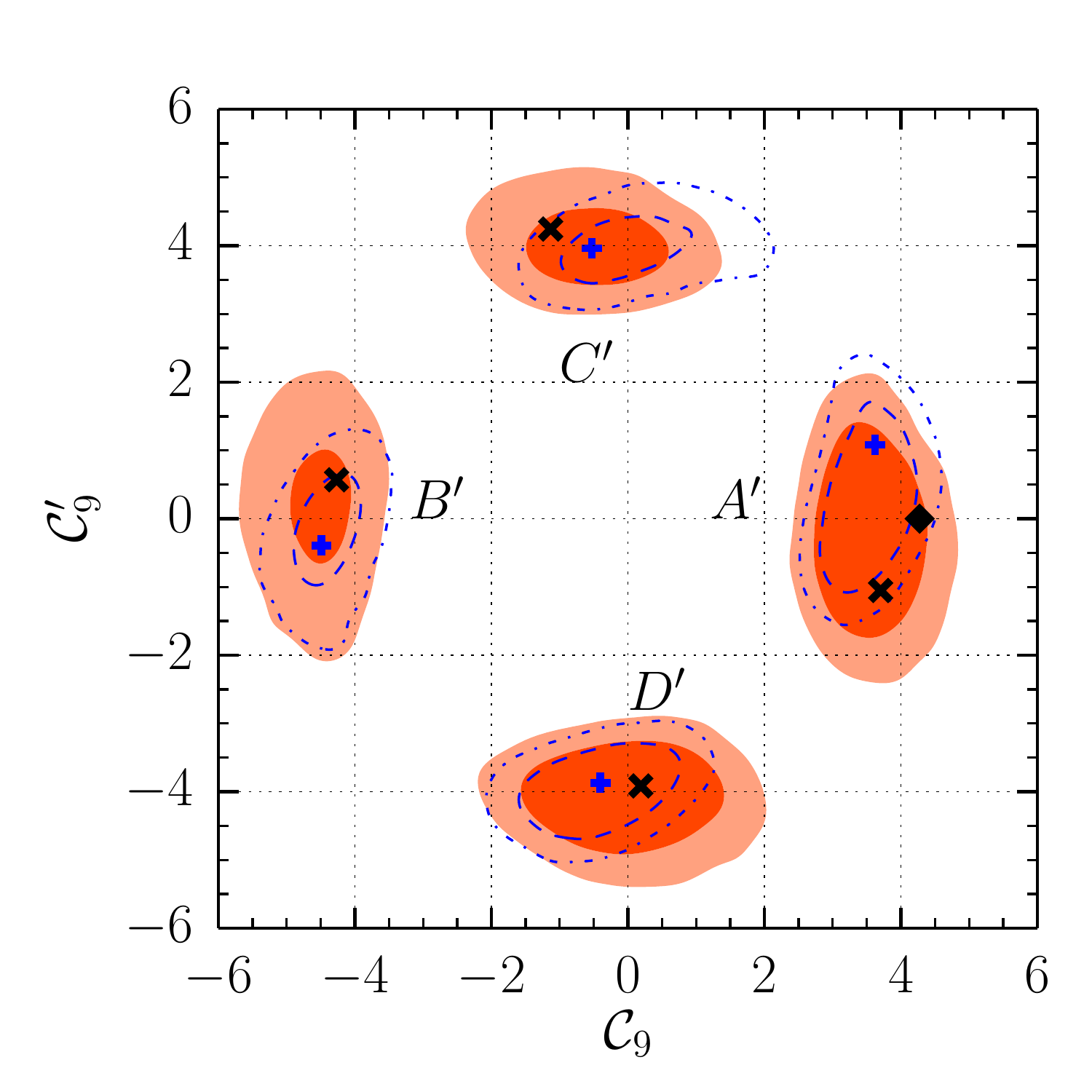} \\[-0.2cm]
        \includegraphics[width=.38\textwidth]{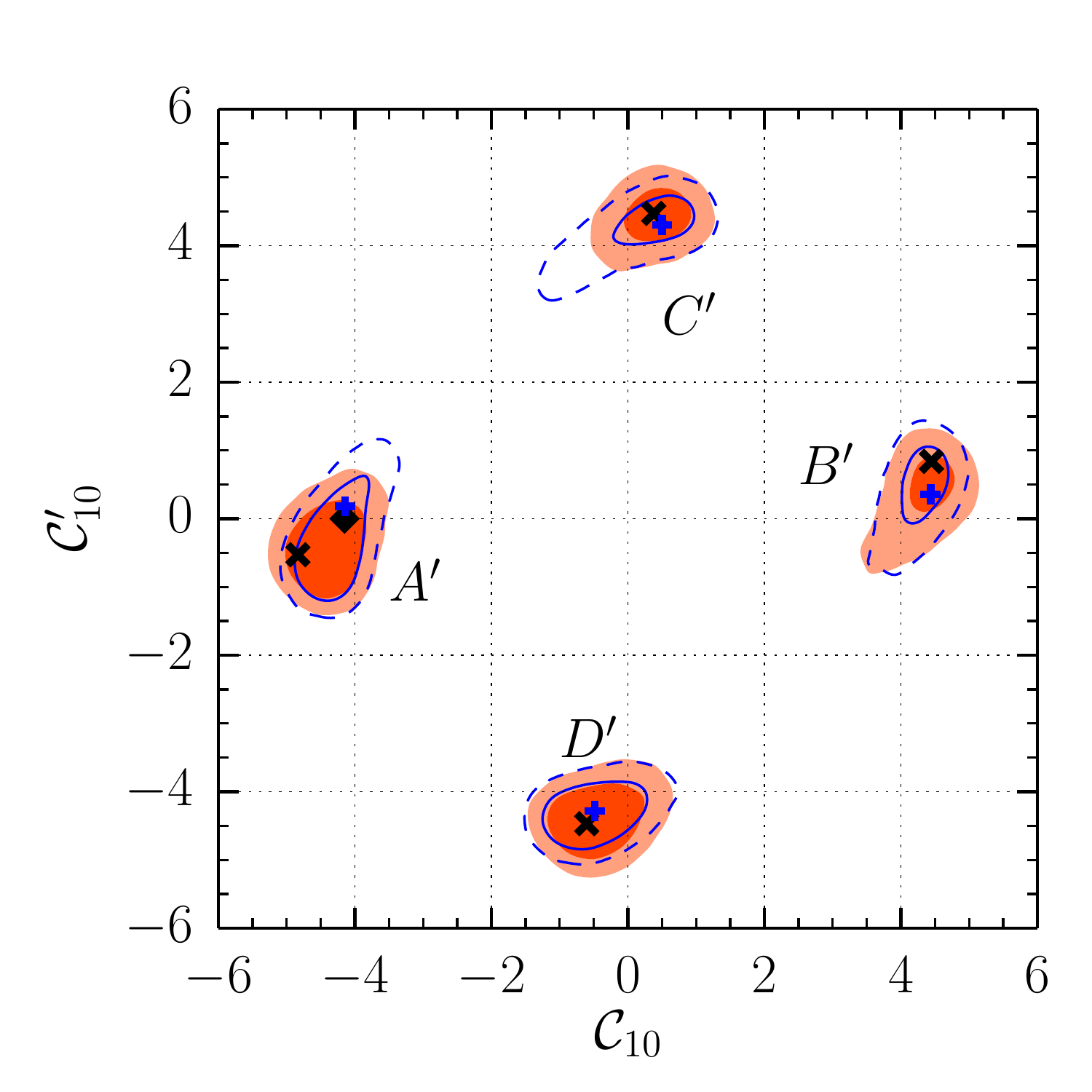}
  \end{center}
  \caption{
        Credibility regions obtained from the fit in the \SMp{} model. We show
        the results of the ``full'' data set after the EPSHEP 2013 conference at
        $68.3\%$ (dark red) and $95.4\%$ (light red) probability. The regions from the fit
        of the ``full (+FF)'' data set are overlaid by blue lines at $68.3\%$ (solid)
        and $95.4\%$ (dashed) probability. The projection of the SM point is shown by
        the black diamond and the black and blue crosses mark the
        best-fit point in the respective 2D plane.
  }
  \label{fig:results-wilson-coeff-SMprime}
\end{figure}

%
%--------+---------+---------+---------+---------+---------+---------+---------+
\subsection{Fit in the extended \SMp{} basis}

We proceed with fitting the SM-like and chirality-flipped Wilson coefficients in
the \SMp{} scenario. Using the ``full'' data set we obtain a good fit with $p$
values between $0.07$ and $0.09$ in four well separated solutions $A'$ through
$D'$, as can be seen in the 2D-marginalized planes in
\reffig{fig:results-wilson-coeff-SMprime}. Here $A'$ and $B'$ denote solutions
that show the same signs of the Wilson coefficients $\wilson{7,9,10}$ of the SM
operator basis as the solutions $A$ and $B$ in the previous section, and $C'$
and $D'$ denote further solutions. The corresponding $p$ values of the ``full
(+FF)'' data set are slightly smaller for $A'$ and $C'$, ranging from $0.05$ to
$0.09$.  Of all four solutions, $A'$ and $D'$ dominate over $B'$ and $C'$ in
terms of the posterior mass:
\begin{equation*}
    R_{A'} : R_{B'} : R_{C'} : R_{D'} =
    37\% : 14\% : 15\% : 34\% \,.
\end{equation*}
The use of the lattice results of the $B\to K^*$ form factors decreases the
posterior mass of solution $A'$ and $D'$ by $2$\% in favor of $C'$ and $B'$, see
table \reftab{tab:goodness-of-fit}.

The posterior distributions marginalized to the 2D $(\wilson{i} - \wilson{i'})$
planes ($i=7,9,10$) are shown in \reffig{fig:results-wilson-coeff-SMprime} with
the SM point and the projection of the best-fit points in each solution $A'$
through $D'$. Note that the projection of the best-fit point can deviate from
the position of the modes of the marginalized distributions; compare to the 1D
intervals in \reftab{tab:wilson:coeff:1-dimCLs}. Unlike in the \SM{} scenario,
it is not possible to disentangle the individual solutions $A'$ through $D'$
within the 1D marginalized posterior distributions. In order to compare our
findings with \cite{Altmannshofer:2013foa} we choose those intervals that
contain the SM-like solution of $\wilson{7,9,10}$ and find with 68\% probability
\begin{equation*}
\begin{aligned}
  \Delta_{7}  & =  0.01 \pm 0.02, &
  \Delta_{9}  & = -0.8 \pm 0.4  , &
  \Delta_{10} & = -0.2 \pm 0.3  ,
\end{aligned}
\end{equation*}
in agreement with \cite{Altmannshofer:2013foa}. Here $\Delta_9$ represents a
$1.8\sigma$ deviation from the SM. These results hardly change when taking $B\to
K^*$ lattice results into account, the deviation of $\Delta_9$ from the SM
increases only slightly to $2.0\sigma$.  The best-fit points for
$\wilson{7',9',10'}$ of \cite{Altmannshofer:2013foa} fall into the intervals
given in \reftab{tab:wilson:coeff:1-dimCLs}, with larger deviations from the
modes of the 1D posterior distributions. The SM prediction $\wilson[SM]{7'} =
-0.01$ and $\wilson[SM]{9',10'} = 0$ is contained in the smallest 68\% region of
the 1D-marginalized posterior distributions. The largest deviation of the SM
point in any 2D marginalized distribution is $1.6\sigma$ in the ($\wilson{9}$ --
$\wilson{7'}$) plane, dominantly due to a shift in $\wilson{9}$.

The additional NP contributions in chirality-flipped operators in scenario
\SMp{} do not help in reducing the tension in the measurement of $\langle P_5'
\rangle_{[1,6]}$. The previously mentioned large pull values for $\langle F_L
\rangle_{[1,6]}$, $\langle \BR \rangle_{[16,19]}$, $\langle P_4'
\rangle_{[14.18,16]}$ and $\langle A_{\rm FB} \rangle_{[16,19]}$ remain almost
unchanged (see \reftab{tab:p-value-sm-full}). This corroborates the findings of
\cite{Hambrock:2013zya,Altmannshofer:2013foa} that the pull value of $\langle
P_4'\rangle_{[14.18,16]}$ can not be pushed below $2\sigma$.

Focusing on the solution $A'$ (cf. \refeq{eq:reduced:prior}), the Bayes factor
becomes
\begin{equation}
    \label{eq:bayes:solA:SMp}
    \frac{P(\text{full} | \text{\SMp})}
         {P(\text{full} | \text{\SMnu})}\Big|_{A'}
    = 1:401 \,.
\end{equation}
Thus the NP hypothesis with chirality-flipped Wilson coefficients is disfavored
in comparison to the SM($\nu$-only) hypothesis. The data favor \SM{} over \SMp{}
with roughly $8:1$. Taking the lattice form factor results into account, we find
\begin{equation*}
\begin{aligned}
  \Delta_{7}  & =  0.00 ^{+0.03}_{-0.02}, &
  \Delta_{9}  & = -0.8 ^{+0.4}_{-0.3}, &
  \Delta_{10} & = -0.1 \pm 0.3
\end{aligned}
\end{equation*}
and
\begin{equation}
    \label{eq:bayes:solA:SMpFF}
    \frac{P(\text{full (+FF)} | \text{\SMp})}
         {P(\text{full (+FF)} | \text{\SMnu})}\Big|_{A'}
    = 1:148\,,
\end{equation}
an increase by a factor of about 3 compared to \refeq{eq:bayes:solA:SMp}.

In the \SMp{}, the size of subleading contributions to transversity amplitudes
$\chi = 0\, (\para)$ reduces to about $-5\%$ ($+5\%$) for $\zeta_{K^*}^{L
  \chi}$, in contrast to $\zeta_{K^*}^{L \perp}$, which remains large, see
\reffig{fig:subleading-btokstar}.  The subleading parameter $\Lambda_\parallel$
at high $q^2$ decreases back to 0\% for the ``full (+FF)'' data set. There are
no differences between the ``full'' and ``full (+FF)'' data sets.  Similarly,
the subleading contributions to the $B\to K\ell^+\ell^-$ amplitude disappear.
The small shifts we observe between the SM and \SMp{} scenarios suggest that
both $\zeta^{L\chi}_{K^*}$ and $\wilson[\prime]{i}$ can ease the tensions
between predictions and data.  It is therefore desirable to better understand
size, chirality structure, and $q^2$-dependence of the power corrections.

\begin{figure}[t]
    \begin{center}
        \includegraphics[width=0.38\textwidth]{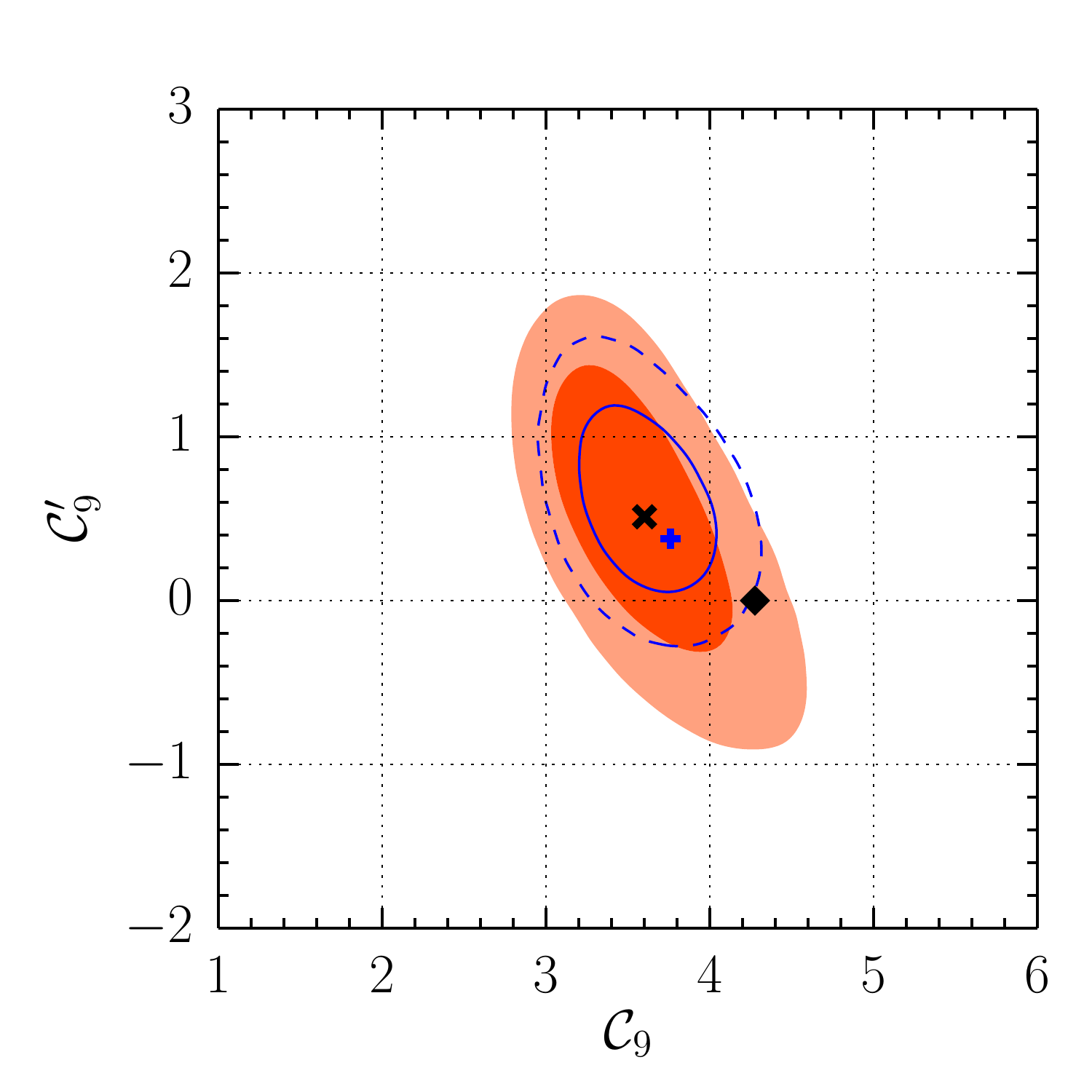}
    \end{center}
    \caption{
        \label{fig:res-wilson-coeff-SMprimeNine}
        Credibility regions obtained from the fit in the \SMpNine{} model.
        We show the results of the ``full'' data set at $68.3\%$ (dark red) and
        $95.4\%$ (red) probability. The regions from the fit of the
        ``full (+FF)'' data set are overlaid by blue lines at $68.3\%$ (solid) and
        $95.4\%$ (dashed) probability. The projection of the SM point is represented
        by the black diamond, whereas the black and blue crosses mark the best-fit points.
    }
\end{figure}

For comparison with \cite{Descotes-Genon:2013wba, Altmannshofer:2013foa,
  Horgan:2013pva}, we investigate also the variant \SMpNine{}. This scenario has
only two additional new physics parameters $\wilson{9,9'}$ compared to six in
the \SMp{} scenario, implying a ``smaller punishment'' due to the larger prior
volume w.r.t. the \SMnu{} case. The 2D-marginalized $(\wilson{9}-\wilson{9'})$
plane in \reffig{fig:res-wilson-coeff-SMprimeNine} shows the SM point on the
border of the $2.0\sigma$ region for the only solution $A'$ when using the
``full (+FF)'' data set. However, this changes to $1.4\sigma$ when discarding
the $B\to K^*$ lattice form factor results. The $1\sigma$ probability regions
from the 1D marginalized posterior distributions for ``full'' data set are
\begin{equation*}
\begin{aligned}
  \Delta_{9}  & = -0.8^{+0.4}_{-0.3}\,, &
  \Delta_{9'} & = +0.5 \pm 0.6 \,.
\end{aligned}
\end{equation*}
which is compatible with the results of \cite{Altmannshofer:2013foa}.
For the ``full (+FF)'' data set, we find
\begin{equation*}
\begin{aligned}
  \Delta_{9}  & = -0.7 \pm 0.3 \,, &
  \Delta_{9'} & = +0.6 \pm 0.4 \,.
\end{aligned}
\end{equation*}
With the same parameter ranges as in \refeq{eq:bayes:solA:SMp}, the Bayes factor
with the ``full'' data set comparing the SM-like solution $A'$ to \SMnu{} is
\begin{equation}
  \label{eq:bayes:solA:SMpNine}
  \frac{P(\text{full} | \text{\SMpNine})}
       {P(\text{full} | \text{\SMnu})}\Big|_{A'}
       = 1:3 \,,
\end{equation}
slightly favoring the \SM{} over the \SMpNine{}. With the ``full (+FF)'' data
set, the Bayes factor favors neither model
\begin{equation}
  \label{eq:bayes:solA:SMpNineFF}
  \frac{P(\text{full (+FF)} | \text{\SMpNine})}
       {P(\text{full (+FF)} | \text{\SMnu})}\Big|_{A'}
       = 1:1 \,.
\end{equation}

%
%
%--------+---------+---------+---------+---------+---------+---------+---------+
\subsection{Interpretation}\label{sec:interpretation}

We conclude this section with a remark on the model comparison.  We emphasize
again that, to derive the posterior odds, the Bayes factor has to be multiplied
by the appropriate prior odds (see \refeq{eq:posterior-odds}), the determination
of which is beyond the scope of the present work. Given that the standard model
successfully describes the vast majority of particle physics data, our prior
odds are in strong favor of the \SMnu{}. This is true even for the considered
rare decays despite their particular sensitivity to new physics because the
majority of standard model predictions are close to the other experimental
measurements of $\Delta B = 1, 2$ observables.

Currently the lack of experimental evidence for right-handed weak interactions
significantly reduces our prior probabilities of scenarios with
chirality-flipped Wilson coefficients \SMp{} and \SMpNine{} compared to
\SMnu{}. Among the scenarios with chirality-flipped Wilson coefficients we would
set prior odds in favor of \SMp{} because \SMpNine{} is only a restricted subset
of \SMp{}.

It remains to be noted that the \SMpNine{} scenario can be realized in a
model-independent way due to operator mixing with $b\to s f\bar{f}$ four-fermion
operators. Such scenarios have been considered previously for $f =
\mbox{quarks}$ \cite{Datta:1998yq,Borzumati:1999qt, Hiller:2003js} as well as $f
=\tau$ \cite{Bobeth:2011st} and recently discussed in the present context of the
data \cite{Datta:2013kja} for $f = b$. In the case of $f= \mbox{quarks}$ also
hadronic charmless decays would be affected \cite{Beneke:2009eb}. Explicit
$Z'$-models have been discussed recently in
\cite{Gauld:2013qba,Buras:2013qja,Gauld:2013qja,Buras:2013dea}.

For nearly all combinations of scenarios and data sets, we find satisfactory $p$
values of about $5\%$ to $14\%$, and in the case of ``selection'', $p \ge 90\%$.
This indicates that it is possible to find a single parameter point (the
best-fit point), at which the data is adequately described. The Bayes factor, in
contrast, quantifies the ability to describe the data on average over the full
parameter space.  Concerning the \SM{} scenario, the Bayes factor indicates that
the additional parameters do not yield an advantage over the \SMnu{} scenario.
However, it is interesting to see that the Bayes factors are undecided in the
case of scenario \SMpNine{} for the ``full(+FF)'' data set. Taking lattice
results of $B\to K^*$ form factors into account, the Bayes factors increase
roughly by a factor $3$ for the ``full (+FF)'' data set in scenarios \SMp{} and
\SMpNine{}, but not in scenario \SM{}.

Evaluating the sensitivity to the prior shape, we repeated the ``full (+FF)''
fits replacing Gaussian priors with flat priors of the same range --- as in
\reffig{fig:subleading-btokstar} --- for all subleading parameters. The Wilson
coefficients are determined with the same or slightly higher precision as
before. Most of the subleading parameters play a minor role in the fit, so the
posterior is similar to the prior. Small shifts of a Gaussian prior become
larger shifts --- in the same direction --- of a flat prior. The two exceptions
are the large-recoil parameters $\zeta^{L\perp}_{K^*}$ and
$\zeta^{L0}_{K^*}$. Their posteriors are very similar to those of the fit with
Gaussian priors shown in \reffig{fig:subleading-btokstar} but marginally
wider. Due to the extra freedom, all models are able to better describe the
data. At the best-fit point, the $p$ values roughly double from $0.06$ to $0.13$
in the \SMnu, and from $0.08$ to $0.17$ in the \SMpNine.  The Bayes factor
slightly shifts in favor of \SMnu{} compared to \SMpNine{} from $1:1$ to
$1:1.6$, indicating that the models with variable $\wilson{i}$ on average do not
benefit as much from the added flexibility. In summary, changing the prior shape
of the subleading parameters does not entail any big surprise and corroborates
our main findings.

%
%
%
%--------+---------+---------+---------+---------+---------+---------+---------+
\section{Conclusions}

Our Bayesian analysis indicates that the standard model provides an adequate
description of the available measurements of rare leptonic, semileptonic, and
radiative $B$ decays.  Compared to our previous analysis~\cite{Beaujean:2012uj},
we determine the Wilson coefficients $\wilson{7,9,10}$ more accurately,
dominantly due to the reduction of the experimental uncertainties in the
exclusive decays and the addition of the inclusive decay $B\to X_s\gamma$.

Contrary to all similar analyses, our fits include the theory uncertainties
explicitly through nuisance parameters. We observe that tensions in the angular
and optimized observables in $B\to K^*\ell^+\ell^-$ decays can be lifted through
$(10$--$20)\%$ shifts in the transversity amplitudes at large recoil due to
subleading contributions. These shifts are present within the SM as well as the
model-independent extension of real-valued Wilson coefficients
$\wilson{7,9,10}$.  For the scenarios introducing additional chirality-flipped
coefficients $\wilson{7',9',10'}$, the shifts reduce to a few percent (except
$\zeta_{K^*}^{L\perp}$). We find $|\wilson{9',10'}| \lesssim 5$ at $95\%$
probability, see \reffig{fig:results-wilson-coeff-SMprime} and
\reftab{tab:wilson:coeff:1-dimCLs}, for the right-handed couplings, which holds
in the absence of scalar and tensor contributions. These constraints are
insensitive to the shape (Gaussian vs. flat) of the priors of subleading
corrections.

Among the information inferred from the data are constraints on the parameters
of the $B\to K^{(*)}$ form factors.  We have performed all fits with and without
the very recent lattice $B\to K^*$ form factor predictions
\cite{Horgan:2013hoa}. In both cases, the posterior ranges of the Wilson
coefficients $\wilson{7,7',10,10'}$ are essentially the same apart from minor
shifts in $\wilson{9,9'}$. Again in both cases, the posteriors of the $B\to
K^{*}$ form-factor parameters are very similar. This comes as a surprise given
the large difference in prior uncertainties but implies that the combination of
measurements supports the lattice input, even independently of the scenario.

The rough picture emerging from current data may be summarized as follows. The
low-$q^2$ $B\to K^*\ell^+\ell^-$ data prefer a negative new-physics contribution
to $\wilson{9}$~\cite{Descotes-Genon:2013wba}, which is not supported by $B\to K
\ell^+\ell^-$ data unless one allows a positive contribution to $\wilson{9'}$
(or alternatively $\wilson{10'}$)~\cite{Altmannshofer:2013foa}. Our Bayesian
analysis shows strong support for the standard model \SMnu{} compared to
additional new physics in Wilson coefficients $\wilson{7,9,10}$ in the
\SM{}-scenario and/or chirality-flipped $\wilson{7',9',10'}$ in the
\SMp{}-scenario in terms of Bayes factors. Only a reduced scenario \SMpNine{} of
the two Wilson coefficients $\wilson{9,9'}$ comes close to the standard model.
Including the $B\to K^*$ form-factor lattice predictions, the model comparison
suggests that scenario \SMpNine{} can provide an explanation of the data as
efficient as in the standard model with a Bayes factor of $1:1$.

A substantial reduction of uncertainties can be expected for LHCb, CMS, and
ATLAS measurements of $B^0\to K^{*0}\ell^+\ell^-$ and $B^+\to K^+\ell^+\ell^-$
once they publish the analysis of their 2012 data sets. It should also be
mentioned that $B\to K^*\gamma$ and $B\to K^{(*)} \ell^+\ell^-$ results from
Belle are not based on the final reprocessed data set and that BaBar's angular
analysis of $B\to K^*\ell^+\ell^-$ is still preliminary.  It remains to be seen
whether these improved analyses further substantiate the present hints of a $1$
to $2\sigma$ deviation from the SM prediction in $\wilson{9}$.

In our opinion, however, there remain two major challenges on the theory side.
The first is to improve our analytic knowledge of the $1/m_b$ corrections to the
exclusive decay amplitudes. The second is to reduce the uncertainty from
hadronic form factors, especially at low $q^2$. Without improvements on either,
there is little prospect to discern between small NP effects and large
subleading corrections. Another point of concern are potentially large
duality-violating effects that render the OPE at high $q^2$ invalid. They have
been estimated, though model-dependently, to be small \cite{Beylich:2011aq}. In
this regard, the experimental verification of certain relations
\cite{Bobeth:2012vn} among angular observables in $B\to K^*\ell^+\ell^-$ that
are predicted by the OPE would be very desirable. In the case that some of these
relations are not fulfilled, the analysis of the breaking pattern can provide
information on duality violation but also on additional new-physics scalar and
tensor interactions.

%
%
%
%--------+---------+---------+---------+---------+---------+---------+---------+
\begin{acknowledgments}
  D.v.D is grateful to Thorsten Feldmann, Tobias Huber, Soumitra Nandi and the
  late Nikolai Uraltsev for useful discussions and suggestions. C.B. thanks
  David Straub for discussions.  We are grateful to Christoph Langenbruch for
  pointing out a small numerical error within EOS. We acknowledge the support by
  the DFG Cluster of Excellence "Origin and Structure of the Universe". The
  analyses have been carried out on the computing facilities of the
  Computational Center for Particle and Astrophysics (C2PAP) and on the HORUS
  cluster at Siegen University to perform the fits. We further wish to thank
  Andrzej Buras, Thorsten Feldmann, Christian Hambrock, Gudrun Hiller, Alexander
  Khodjamirian and Stefan Schacht for comments on the manuscript.  This work is
  supported by the Deutsche Forschungsgemeinschaft (DFG) within research unit
  FOR 1873 (``QFET''). C.B. received partial support from the ERC Advanced Grant
  project ``FLAVOUR'' (267104).
\end{acknowledgments}

%--------+---------+---------+---------+---------+---------+---------+---------+
%
%  Appendix
%
%--------+---------+---------+---------+---------+---------+---------+---------+

\appendix

%
%
%
%--------+---------+---------+---------+---------+---------+---------+---------+
\section{Theoretical Treatment \label{app:theory-predictions}}

This appendix details the theoretical predictions for newly added observables,
and their respective nuisance parameters are discussed.  Further we explain
changes to the choice of priors for some of the nuisance parameters, as well as
a different parametrization of $B\to K^*$ form factors, always with respect to
our previous analysis~\cite{Beaujean:2012uj}. We also list the updated values of
input parameters whose uncertainty is neglected in
\reftab{tab:other:numeric:input}.

Concerning the set of common nuisance parameters --- the Wolfenstein parameters
of the CKM-quark-mixing matrix and the \msbar{} bottom- and charm-quark masses
entering the majority of observables --- we use the updated values given in
\reftab{tab:common-input} based on the most recent PDG combinations
\cite{Beringer:1900zz} and the tree-level result of the UTfit collaboration
\cite{Bona:2006ah}.

We model asymmetric uncertainty intervals for the priors with LogGamma
distributions (see \cite{Beaujean:2012uj} for details), while symmetric
intervals are implemented as Gaussian priors.

\begin{table}
\begin{center}
\renewcommand{\arraystretch}{1.3}
\begin{tabular}{cccc}
\hline
  Quantity & Unit & Value &  Reference
\\
\hline
  $\alpha_s(M_Z)$    &  $0.1184$   &  --   &  \cite{Beringer:1900zz}\\
\hline
  $M_Z$              &  $91.1876$  &  GeV  &  \cite{Beringer:1900zz}\\
  $M_W$              &  $80.385$   &  GeV  &  \cite{Beringer:1900zz}\\
  $m_t^{\rm pole}$   &  $173.5$    &  GeV  &  \cite{Beringer:1900zz}\\
\hline
  $M_{B^{+}}$        &  $5.27925$  &  GeV  &  \cite{Beringer:1900zz}\\
  $M_{B^{0}}$        &  $5.27958$  &  GeV  &  \cite{Beringer:1900zz}\\
  $\tau_{B^{+}}$     &  $1.641$    &  ps   &  \cite{Beringer:1900zz}\\
  $\tau_{B^{0}}$     &  $1.519$    &  ps   &  \cite{Beringer:1900zz}\\
  $\tau_{B^{+/0}}$   &  $1.580$    &  ps   &  $^\dagger$\\
  $f_{B^{+/0}}$      &  $190.6 \pm 4.7$  &  MeV  &  \cite{Laiho:2009eu}\\
\hline
  $M_{B_s}$          &  $5.36677$  &  GeV       &  \cite{Beringer:1900zz}\\
  $\tau_{B_s}$       &  $1.516$    &  ps        &  \cite{Amhis:2012bh,Beringer:1900zz}\\
  $\Delta \Gamma_s$  &  $0.081$    & ps$^{-1}$  &  \cite{Amhis:2012bh,Beringer:1900zz}\\
  $y_s$              &  $0.062$    &  --        &  \cite{Amhis:2012bh,Beringer:1900zz}\\
\hline
\end{tabular}
\renewcommand{\arraystretch}{1.0}
\caption{\label{tab:other:numeric:input} Numerical input that has been updated
  but is not used as nuisance parameters in the fit. $^\dagger$See text for
  additional details.
}
\end{center}
\end{table}

\begin{table}
\begin{center}
\renewcommand{\arraystretch}{1.3}
\begin{tabular}{cccc}
\hline
Quantity & Prior &  Unit &  Reference\\
\hline
  \multicolumn{4}{c}{\tabvsptop CKM}\\
\hline
$\lambda$             &  $0.22535 \pm 0.00065$  & -- & \cite{Bona:2006ah}\\
$A$                   &  $0.807 \pm 0.020$      & -- & \cite{Bona:2006ah}\\
$\bar{\rho}$          &  $0.128 \pm 0.055$      & -- & \cite{Bona:2006ah}\\
$\bar{\eta}$          &  $0.375 \pm 0.060$      & -- & \cite{Bona:2006ah}\\
\hline
  \multicolumn{4}{c}{\tabvsptop Quark Masses}\\
\hline
  $\overline{m}_c(m_c)$ &  $1.275 \pm 0.025$     & $\GeV$  & \cite{Beringer:1900zz}\\
  $\overline{m}_b(m_b)$ &  $4.18 \pm 0.03$       & $\GeV$  & \cite{Beringer:1900zz}\\
\hline
\end{tabular}
\renewcommand{\arraystretch}{1.0}
\caption{\label{tab:common-input} Prior distributions of common nuisance parameters.
}
\end{center}
\end{table}

%
%
%--------+---------+---------+---------+---------+---------+---------+---------+
\subsection{Inclusive decays $B\to X_s (\gamma,\, \ell^+\ell^-)$
  \label{sec:inclusive-decays}}

The branching ratio of the inclusive decay $B\to X_s \gamma$, $\mathcal{B}(B\to
X_s \gamma) \sim |\wilson{7}|^2 + |\wilson{7'}|^2$, represents the most
stringent constraint on the magnitude of the dipole Wilson coefficients
$\wilson{7,7'}$. In our analysis we include the known corrections to
next-to-leading order in $\alpha_s$ \cite{Chetyrkin:1996vx,Buras:2002tp} as well
as $\alpha_s \Lambda_{\rm QCD}^2/m_b^2$ corrections \cite{Ewerth:2009yr}.
Contrary to the common normalization to the semi-leptonic inclusive decay $B\to
X_c \ell \nu$, we express the branching ratio in terms of the averaged $B$ meson
life time for a 50:50 production ratio of $B^+B^-$ to $B^0 \bar{B}^0$ pairs at
the $\Upsilon(4S)$, $\tau_{B^{+/0}}$ given in \reftab{tab:other:numeric:input},
and the bottom-quark pole mass. In order to avoid renormalon ambiguities we
calculate the pole mass value from the \msbar{} mass $m_b(m_b)$ using the 3-loop
result \cite{Melnikov:2000qh}. The \msbar{} mass is part of our set of common
nuisance parameters and its uncertainty dominates the overall theory uncertainty
of inclusive decays. At order $\Lambda_{\rm QCD}^2/m_b^2$ in the heavy quark
expansion hadronic matrix elements of two dimension-five operators enter. They
are parametrized in terms of $\mu^2_{\pi}$ and $\mu^2_{G}$ for the expectation
values of the kinetic and the chromomagnetic operators, respectively.  The
parameters $\mu^2_\pi$ and $\mu^2_G$ enter the fit with priors according to
\cite{Uraltsev:2001ih}, which are listed in \reftab{tab:hadronic:nuisance}.  The
correlation between $\mu^2_{G}$ and the $b$-quark mass is accounted for by the
$B^*$--$B$ mass splitting. When confronting the theoretical prediction of the
mass splitting with the measurement \refeq{eq:Bmass:splitting:exp}, we include
also the effect of dimension-6 operators as described in \cite{Uraltsev:2001ih}.
Our SM prediction $\mathcal{B}(B\to X_s \gamma) = (3.14^{+0.22}_{-0.19})\cdot
10^{-4}$ is in good agreement with the NNLO result \cite{Misiak:2006zs}.  The
theory uncertainty of our prediction is determined from variation of the
Wolfenstein parameters, $m_b(m_b)$, $m_c(m_c)$, $\mu^2_{\pi}$, and $\mu^2_{G}$.

For the prediction of the branching ratio of the inclusive decay $B\to X_s
\ell^+\ell^-$ we work at NNLO in QCD and NLO in QED, including also
$\Lambda_{\rm QCD}^2/m_b^2$ subleading corrections, as described in
\cite{Bobeth:2003at,Huber:2005ig} except for the SM-SM$'$ interference terms. We
adopt the same normalization in terms of $\tau_{B^{+/0}}$ and bottom-quark pole
mass as described for $B\to X_s \gamma$.  The chirality-flipped operators are
included following \cite{Guetta:1997fw} and NLO QCD corrections to matrix
elements are accounted for in case they can be derived easily within the SM
operator basis. The overall theory uncertainty is determined as for $B\to
X_s\gamma$ from the variation of the same nuisance parameters. We obtain as the
SM prediction $\langle \mathcal{B}(B\to X_s \ell^+ \ell^-)\rangle_{[1,\,6]} =
(1.4\pm 0.1) \cdot 10^{-6}$.

\begin{table}
\begin{center}
\renewcommand{\arraystretch}{1.4}
\begin{tabular}{cccc}
\hline
  Quantity & Prior & Unit & Reference\\
\hline
  \multicolumn{4}{c}{Inclusive decays}\\
\hline
  $\mu^2_{\pi}(1\, \mbox{GeV})$  &  $0.45 \pm 0.10$        &  $\GeV^2$ &  \cite{Uraltsev:2001ih}\\
  $\mu^2_{G}(1\, \mbox{GeV})$    &  $0.35^{+0.03}_{-0.02}$ &  $\GeV^2$ &  \cite{Uraltsev:2001ih}\\
\hline
  \multicolumn{4}{c}{$B\to K$ form factors}\\
\hline
  $f_+(0)$    &  $0.34 \pm 0.05$         &  --    &  \cite{Khodjamirian:2010vf,Ball:2004ye}\\
  $b_1^+$     &  $-2.1^{+0.9}_{-1.6}$    &  --    &  \cite{Khodjamirian:2010vf}\\
\hline
  \multicolumn{4}{c}{$B\to K^{*}$ form factors}
\\
\hline
  $V(0)$      &  $0.36^{+0.23}_{-0.12}$  &  --  &  \cite{Khodjamirian:2010vf}\\
  $A_1(0)$    &  $0.25^{+0.16}_{-0.10}$  &  --  &  \cite{Khodjamirian:2010vf}\\
  $A_2(0)$    &  $0.23^{+0.19}_{-0.10}$  &  --  &  \cite{Khodjamirian:2010vf}\\
  $b_1^V$     &  $-4.8^{+0.8}_{-0.4}$    &  --  &  \cite{Khodjamirian:2010vf}\\
  $b_1^{A_1}$ &  $0.34^{+0.86}_{-0.80}$  &  --  &  \cite{Khodjamirian:2010vf}\\
  $b_1^{A_2}$ &  $-0.85^{+2.88}_{-1.35}$ &  --  &  \cite{Khodjamirian:2010vf}\\
\hline
  \multicolumn{4}{c}{$B_s$ decay constant}
\\
\hline
  $f_{B_s}$   &  $227.6 \pm 5.0$         &  MeV &  \cite{Laiho:2009eu,Bazavov:2011aa,
McNeile:2011ng, Na:2012kp}\\
\hline
\end{tabular}
\renewcommand{\arraystretch}{1.0}
\caption{\label{tab:hadronic:nuisance} Prior distributions of the nuisance parameters
   for hadronic quantities entering inclusive and exclusive decays.
}
\end{center}
\end{table}

%
%
%--------+---------+---------+---------+---------+---------+---------+---------+
\subsection{Form factors and decay constants \label{sec:form-factors}}

The most important change in the treatment of form factors is the consistent use
of the parametrization as in~\cite{Khodjamirian:2010vf}, for both $B\to K$ and
$B\to K^*$ transitions. It has the merits of a) a convenient expansion in a
small parameter $z$ that respects unitarity, b) correct behavior at the
$BK^{(*)}$ production threshold, c) correct asymptotic behavior for $q^2\to
\infty$, and d) a convenient parametrization at $q^2 = 0$.

\begin{table}
    \renewcommand{\arraystretch}{1.3}
    \begin{tabular}{c|ccc}
        \hline
        $q^2$ [\GeV] & $17$            & $20$            & $23$\\
        \hline
        $f_+(q^2)$   & $1.13 \pm 0.05$ & $1.63 \pm 0.07$ & $2.68 \pm 0.13$\\
        \hline
    \end{tabular}\\
    \renewcommand{\arraystretch}{1.0}
    \vspace{2\smallskipamount}
    \renewcommand{\arraystretch}{1.3}
    \begin{tabular}{c|ccc}
        \hline
        $q^2$ [\GeV]
             & $17$   & $20$    & $23$\\
        \hline
        $17$ & $1.00$ & $0.78$  & $0.30$\\
        $20$ & --     & $1.00$  & $0.71$\\
        $23$ & --     & --      & $1.00$\\
        \hline
    \end{tabular}
    \renewcommand{\arraystretch}{1.0}
    \caption{Reproduction of mean values, uncertainties (top) and correlation
        information (bottom) of lattice points \cite{Bouchard:2013eph} for the
        vector form factor $f_+(q^2)$ in $B\to K$ transitions.
        \label{tab:btok-lattice}
      }
\end{table}

\begin{table}
    \renewcommand{\arraystretch}{1.3}
    \begin{tabular}{c|ccc}
        \hline
        $q^2$ [\GeV]  & $15$            & $19.21$ \\
        \hline
        $V(q^2)$      & $1.14 \pm 0.11$ & $1.91 \pm 0.15$ \\
        $A_1(q^2)$    & $0.50 \pm 0.04$ & $0.62 \pm 0.04$ \\
        $A_{12}(q^2)$ & $0.37 \pm 0.04$ & $0.44 \pm 0.04$ \\
        \hline
    \end{tabular}\\
    \renewcommand{\arraystretch}{1.0}
    \vspace{2\smallskipamount}
    \renewcommand{\arraystretch}{1.3}
    \begin{tabular}{c|cc|cc|cc}
        \hline
           & \multicolumn{2}{c|}{$V$} & \multicolumn{2}{c|}{$A_1$} & \multicolumn{2}{c}{$A_{12}$}\\
        \hline
        $q^2$ [\GeV]
                & $15$   & $19.21$ & $15$   & $19.21$ & $15$   & $19.21$ \\
        \hline
        $15$    & $1.00$ & $0.39$  & $1.00$ & $0.50$  & $1.00$ & $0.21$ \\
        $19.21$ & --     & $1.00$  & --     & $1.00$  & --     & $1.00$ \\
        \hline
    \end{tabular}
    \renewcommand{\arraystretch}{1.0}
    \caption{Reproduction of mean values, uncertainties (top) and correlation
        information (bottom) of lattice points based on the results in the arXiv v1 of
        \cite{Horgan:2013hoa} for the vector form factors $V(q^2)$ and $A_{1,12}(q^2)$
        in $B\to K^*$ transitions. Note that the update of the results from the arXiv
        v1 to v2 introduce only minor changes to the mean values and the correlation
        coefficient for the $A_{12}$ lattice points.
        \label{tab:btokStar-lattice}
     }
\end{table}

For $B\to K$ we have modified the prior of the nuisance parameter $f_+(0)$ of
the $f_+$ form factor parametrization w.r.t our previous analysis but kept the
slope parameter $b_1^+$ as is. This change accounts for both LCSR results
\cite{Khodjamirian:2010vf,Ball:2004ye} that use the same approach of
$B$-interpolating currents and on-shell $K$-mesons. As a result the prior on
$f_+(0)$ is wider and the tension between the SM prediction of the $B\to K
\ell^+\ell^-$ branching ratio at $q^2 \in[1,
6]$~GeV$^2$~\cite{Khodjamirian:2012rm} and the LHCb measurement
\cite{Aaij:2012vr} is reduced. Moreover, the recent lattice predictions
\cite{Bouchard:2013eph} of the form factor $f_+$ at high $q^2$ are included in
our analysis as part of the likelihood for technical reasons. For this purpose
we reproduced lattice predictions at three values of $q^2 = 17,\, 20,\,
23$~GeV$^2$ as well as their correlation matrix based on the parametrization
given in \cite{Bouchard:2013eph}, see \reftab{tab:btok-lattice}.  (The $q^2$
values and number of points are chosen such that the correlation of neighboring
points does not exceed $80\%$). This constraint is included in the likelihood by
means of a multivariate Gaussian.

Due to the change of parametrization of the $B\to K^*$ form factors $V, A_1$ and
$A_2$, their three respective nuisance parameters are replaced by the three
form-factor normalizations at $q^2 = 0$ ($V(0), A_{1,2}(0)$) and three slope
parameters ($b_1^{V, A_{1,2}}$). The LCSR results of \cite{Khodjamirian:2010vf}
are chosen as the priors for normalizations and slopes.  We note that these
priors are less precise than the results of~\cite{Ball:2004rg} due to a novel
LCSR setup involving an on-shell $B$ meson and interpolation of the $K^*$ final
state. Beyond the informative priors we also include two additional constraints
on $B\to K^*$ form factors at $q^2 = 0$. First, the ratio $V(0)/A_1(0)$ is
constrained in the large energy limit as given by \cite{Hambrock:2013zya} (see
also references therein)
\begin{align}
  \label{eq:FF-constr:VoverA1}
  V(0)/A_1(0) & = 1.33\pm 0.40 \,,
\end{align}
where the uncertainty has been estimated based on power counting. Second,
we make use of the relation
\begin{equation}
  \label{eq:FF-constr:A0}
  A_0(0) =
    \frac{M_B + M_{K^*}}{2 M_{K^*}} A_1(0) - \frac{M_B - M_{K^*}}{2 M_{K^*}} A_2(0)
\end{equation}
where
\begin{equation}
  \label{eq:FF-constr:A0:LCSR}
  A_0(0) = 0.29^{+0.10}_{-0.07}
\end{equation}
is the LCSR result given in \cite{Khodjamirian:2010vf}.  We note that
\refeq{eq:FF-constr:VoverA1} and \refeq{eq:FF-constr:A0:LCSR} represent
additional constraints on the nuisance parameters $V(0)$ and $A_{1,2}(0)$.  The
motivation for this treatment is to tighten the constraints on $A_{1}(0)$, and
avoid unphysical (i.e. negative) values of the form-factor combination $A_0$.

Similar to our treatment of $B\to K$ lattice results, we reproduce lattice
predictions \cite{Horgan:2013hoa} for the $B\to K^*$ form factors $V$, $A_1$ and
$A_{12}$ at two kinematic points $q^2 = 15,\, 19.21$~GeV${}^2$. We obtain mean
values, variances, and the correlation coefficients between neighboring points
as given in \reftab{tab:btokStar-lattice}. Note that ref. \cite{Horgan:2013hoa}
does not provide information on cross-form factor correlation. We chose the
number and spacing of $q^2$ values so that the correlation of neighboring points
does not exceed $60\%$.

The updated prior of the $B_s$ decay constant $f_{B_s}$, entering the branching
ratio of $B_s\to \mu^+\mu^-$, takes into account recent lattice results, see
\reftab{tab:hadronic:nuisance}.

%
%
%--------+---------+---------+---------+---------+---------+---------+---------+
\subsection{Subleading $1/m_b$ \label{sec:subleading}}

\begin{table*}
    \centering
    \renewcommand{\arraystretch}{1.3}
%    \resizebox{\textwidth}{!}{
    \begin{tabular}{lccccccccc}
        \hline
        Source  & $\langle P'_4 \rangle_{[1,6]}$ & $\langle P'_4 \rangle_{[14.18,16]}$ & $\langle P'_4 \rangle_{[16,19]}$ & $\langle P'_5\rangle_{[1,6]}$ & $\langle P'_5 \rangle_{[14.18,16]}$ & $\langle P'_5 \rangle_{[16,19]}$ & $\langle P'_6\rangle_{[1,6]}$ & $\langle P'_6\rangle_{[14.18,16]}$ & $\langle P'_6 \rangle_{[16,19]}$ \\
        \hline
        \multicolumn{10}{c}{Measurement}\\
        \hline
        LHCb$^\dagger$ \cite{Aaij:2013qta}
        & $0.58^{+0.32}_{-0.36}$ & $-0.18^{+0.54}_{-0.70}$ & $0.70^{+0.44}_{-0.52}$ & $+0.21^{+0.20}_{-0.21}$ & $-0.79^{+0.27}_{-0.22}$ & $-0.60^{+0.21}_{-0.18}$ & $+0.18^{+0.21}_{-0.21}$ & $0.18^{+0.24}_{-0.25}$ & $-0.31^{+0.38}_{-0.39}$\\
        \hline
        \multicolumn{10}{c}{Predictions in \SMnu{}}\\
        \hline
        nominal priors
        & $0.47^{+0.07}_{-0.08}$ & $+1.21^{+0.08}_{-0.10}$ & $1.30^{+0.05}_{-0.05}$ & $-0.34^{+0.09}_{-0.08}$ & $-0.77^{+0.16}_{-0.14}$ & $-0.56^{+0.13}_{-0.13}$ & $-0.07^{+0.01}_{-0.02}$ & $\order{10^{-4}}$       & $\order{10^{-4}}$\\

        wide priors
        & $0.44^{+0.15}_{-0.15}$ & $+1.21^{+0.08}_{-0.10}$ & $1.31^{+0.04}_{-0.07}$ & $-0.32^{+0.18}_{-0.10}$ & $-0.77^{+0.16}_{-0.14}$ & $-0.54^{+0.13}_{-0.17}$ & $-0.07^{+0.02}_{-0.03}$ & $\order{10^{-4}}$       & $\order{10^{-4}}$\\
        \hline
        Ref. \cite{Jager:2012uw}
        & $0.46^{+0.16}_{-0.19}$ & --                      & --                      & $-0.28^{+0.30}_{-0.26}$ & --                      & --                      & $-0.07^{+0.08}_{-0.10}$ & --                      & -- \\
        Ref. \cite{Descotes-Genon:2013wba}
        & $0.56^{+0.07}_{-0.06}$ & $+1.16^{+0.19}_{-0.33}$ & $1.26^{+0.12}_{-0.25}$ & $-0.35^{+0.09}_{-0.10}$ & $-0.78^{+0.33}_{-0.36}$ & $-0.60^{+0.28}_{-0.37}$ & $-0.09^{+0.04}_{-0.05}$ & $0.00^{+0.00}_{-0.00}$ & $+0.00^{+0.00}_{-0.00}$\\
        Ref. \cite{Horgan:2013pva}
        & --                     & $+1.22^{+0.38}_{-0.38}$ & $1.30^{+0.02}_{-0.02}$ & --                              & $-0.71^{+0.07}_{-0.07}$ & $-0.54^{+0.04}_{-0.04}$ & --           & --                & -- \\
        \hline
        \multicolumn{10}{c}{Postdictions: this work, ``full'' data set}\\
        \hline
        \SMnu{}
        & $0.56^{+0.05}_{-0.05}$ & $+1.12^{+0.03}_{-0.03}$ & $1.24^{+0.02}_{-0.03}$ & $-0.27^{+0.02}_{-0.03}$ & $-0.84^{+0.04}_{-0.05}$ & $-0.66^{+0.04}_{-0.04}$ & $-0.054^{+0.005}_{-0.005}$ & $\order{10^{-4}}$       & $\order{10^{-4}}$\\
        \SM{}
        & $0.58^{+0.06}_{-0.05}$ & $+1.12^{+0.03}_{-0.03}$ & $1.23^{+0.03}_{-0.02}$ & $-0.28^{+0.03}_{-0.03}$ & $-0.86^{+0.04}_{-0.04}$ & $-0.67^{+0.04}_{-0.04}$ & $-0.054^{+0.005}_{-0.006}$ & $\order{10^{-4}}$       & $\order{10^{-4}}$\\
        \SMp{}
        & $0.54^{+0.07}_{-0.06}$ & $+1.19^{+0.02}_{-0.03}$ & $1.28^{+0.01}_{-0.02}$ & $-0.39^{+0.10}_{-0.04}$ & $-0.76^{+0.04}_{-0.04}$ & $-0.58^{+0.03}_{-0.03}$ & $-0.054^{+0.006}_{-0.006}$ & $\order{10^{-4}}$       & $\order{10^{-4}}$\\
        \hline
    \end{tabular}
%}
    \renewcommand{\arraystretch}{1.0}
    \caption{
        \label{tab:prediction-sl}
        Predictions based on \SMnu{} and wide (tripled uncertainty) priors,
        and postdictions after the fits, for the optimized observable $P'_{4,5,6}$
        in various $q^2$ bins. We compare our results with several sources.
        Note that for our predictions (postdictions) the uncertainties
        correspond to $68\%$ credibility intervals that arise from variation
        of only the nuisance parameters (all fit parameters).
        Note that the postdiction of $\langle P'_5\rangle_{[1,6]}$ in \SMp{}
        consists of two distinct regions around $-0.3$ and $-0.4$ from solutions
        $(A' + B')$ and $(C' + D')$, respectively.
        $\dagger$: Values have been adjusted to match the theory convention
        for the observable.
      }
\end{table*}

With increasing knowledge of the $B\to K^{(*)}$ form factors and measurement of
optimized --- i.e., form factor insensitive --- observables, the treatment of
subleading contributions to the amplitudes of both $B\to K\ell^+\ell^-$ and
$B\to K^*\ell^+\ell^-$ decays has increased in relevance. Especially their
analytic $q^2$ dependence is currently unknown, and their determination is not
within the scope of this work. However, we strive to infer the size of
contributions that go beyond the known QCDF and low recoil terms. In order to
achieve this goal we keep the parametrization of subleading terms as in our
previous work, except for the complex phases of the low recoil terms. Inference
of these phases is not possible in the absence of data on CP asymmetries in
$B\to K^{(*)}\ell^+\ell^-$ decays at low recoil \cite{Beaujean:2012uj}.  We
therefore remove these phases from the analysis.

The overall theory uncertainty of our predictions in the region of large recoil
differ substantially from those given in \cite{Jager:2012uw}, due to the
different treatment of subleading corrections to the form factor relations and
the contributions from $c\bar{c}$ resonances. We keep the parametrization as in
our previous work,
\begin{align}
    A^{L(R)}_\chi & \mapsto A^{L(R)}_\chi \zeta^{L(R)\chi}_{K^*}\,, &
    \chi & = \perp,\para, 0\,.
\end{align}
and obtain similar uncertainties in predictions of observables
as in \cite{Descotes-Genon:2013wba}.

We collect the experimental measurements as well as theoretical predictions from
the literature and this work in \reftab{tab:prediction-sl} for the optimized
observables $\langle P_{4,5,6}' \rangle$.  There we give predictions, i.e.,
before the fit, and postdictions that include experimental information, and
proceed for this purpose as described in \cite{Beaujean:2012uj}. The predictions
are restricted to the \SMnu{} scenario and are based on the prior distributions
of the nuisance parameters. Besides our nominal Gaussian prior choice with
$1\sigma$ ranges of $\pm 0.15$ ($15\%$ at amplitude level) for subleading
parameters at large recoil, $\zeta^{L(R)\chi}_{K^*,K}$, and also at low recoil,
we also show the results for wider $1\sigma$ range of $\pm 0.45$.  We do not
include the recent lattice results of $B\to K^*$ form factors
\cite{Horgan:2013hoa}.

The central values of our predictions agree within errors
with~\cite{Jager:2012uw}, \cite{Descotes-Genon:2013wba},
and~\cite{Horgan:2013pva}. At large recoil our theoretical uncertainties are of
the same size as in \cite{Descotes-Genon:2013wba} and much smaller than
in~\cite{Jager:2012uw}, who treat subleading corrections differently. Choosing
wider prior ranges leads to small shifts in the central value and can double the
theoretical uncertainty that is still smaller than the one in
\cite{Jager:2012uw}. At low recoil, subleading corrections are less important
and the wider prior ranges practically do not affect the overall uncertainties.

The postdictions are based on the posterior distributions of the fits for each
scenario with nominal prior distributions. This includes NP effects in the
Wilson coefficients in the scenarios \SM{} and \SMp{}.  The overall
uncertainties of the postdictions are smaller than the uncertainties of the
predictions. This can be attributed to the improved knowledge of form factors
and subleading nuisance parameters as witnessed by the prior-to-posterior
compression in \reftab{tab:FF:fit-results} and \reffig{fig:subleading-btokstar}.
The additional NP contributions in \SM{} compared to \SMnu{} do not change
postdictions of $\langle P_{4,5,6}' \rangle$ noticeably. On the other hand,
chirality-flipped operators in \SMp{} can induce shifts at large and low recoil
in $\langle P_{4,5}' \rangle$.

\begin{table*}
    \renewcommand{\arraystretch}{1.3}
    %\resizebox{\textwidth}{!}{
    \begin{tabular}{cc|c|cc|c}
        \hline
        \multicolumn{2}{c|}{Scenario} & \SM{} ``selection'' & \multicolumn{2}{c|}{\SM{} ``full''} & \SMp{} ``full''\\
        \multicolumn{2}{c|}{Solution} & $A$ & $A$ & $B$ & \\
        \hline
        \multirow{3}{*}{$\wilson{7}$}
        & $68\%$ & [$-0.37,\,-0.33$] & [$-0.37,\,-0.31$] & --                & [$-0.38,\,-0.30$]  $\cup$  [$+0.06,\,+0.19$]  $\cup$  [$+0.47,\,+0.51$]\\
        & $95\%$ & [$-0.39,\,-0.31$] & [$-0.39,\,-0.30$] & [$+0.47,\,+0.53$] & [$-0.38,\,-0.28$]  $\cup$  [$-0.12,\,+0.26$]  $\cup$  [$+0.44,\,+0.54$]\\
        & mode   & $-0.35$           & $-0.34$           & $+0.50$           & $-0.33$           $\wedge$ $+0.12$           $\wedge$ $+0.50$\\
        \hline
        \multirow{3}{*}{$\wilson{9}$}
        & $68\%$ & [$+1.85,\,+3.20$] & [$+3.25,\,+4.21$] & --                &                            [$-1.18,\,+0.75$]  $\cup$  [$+2.95,\,+4.01$]\\
        & $95\%$ & [$+1.06,\,+3.85$] & [$+3.03,\,+4.56$] & [$-5.30,\,-4.30$] & [$-5.32,\,-3.91$]  $\cup$  [$-1.89,\,+1.46$]  $\cup$  [$+2.69,\,+4.36$]\\
        & mode   & $+2.53$           & $+3.75$           & $-4.67$           & $-4.48$           $\wedge$ $+0.01$           $\wedge$ $+3.52$\\
        \hline
        \multirow{3}{*}{$\wilson{10}$}
        & $68\%$ & [$-4.66,\,-3.74$] & [$-4.80,\,-4.00$] & --                & [$-4.79,\,-4.05$] $\cup$   [$-0.90,\,+0.50$]\\
        & $95\%$ & [$-5.12,\,-3.28$] & [$-5.05,\,-3.71$] & [$+3.88,\,+4.72$] & [$-5.04,\,-3.80$] $\cup$   [$-1.23,\,+0.92$] $\cup$   [$+3.73,\,+4.81$]\\
        & mode   & $-4.19$           & $-4.36$           & $+4.32$           & $-4.42$           $\wedge$ $-0.53$           $\wedge$ $+4.44$\\
        \hline
        \multirow{3}{*}{$\wilson{7'}$}
        & $68\%$ & --                & --                & --                & [$-0.43,\,-0.39$]  $\cup$  [$-0.14,\,-0.01$]  $\cup$  [$+0.38,\,+0.44$]\\
        & $95\%$ & --                & --                & --                & [$-0.46,\,-0.36$]  $\cup$  [$-0.19,\,+0.17$]  $\cup$  [$+0.36,\,+0.46$]\\
        & mode   & --                & --                & --                & $-0.41$            $\wedge$ $-0.07$           $\wedge$ $+0.41$\\
        \hline
        \multirow{3}{*}{$\wilson{9'}$}
        & $68\%$ & --                & --                & --                & [$-4.60,\,-3.55$]  $\cup$  [$-1.10,\,+0.74$]  $\cup$  [$+3.72,\,+4.07$]\\
        & $95\%$ & --                & --                & --                & [$-5.04,\,-3.29$]  $\cup$  [$-1.97,\,+1.62$]  $\cup$  [$+3.28,\,+4.68$]\\
        & mode   & --                & --                & --                & $-4.03$            $\wedge$ $-0.09$           $\wedge$ $+3.85$\\
        \hline
        \multirow{3}{*}{$\wilson{10'}$}
        & $68\%$ & --                & --                & --                & [$-4.78,\,-4.11$]  $\cup$  [$-0.92,\,+0.34$]  $\cup$  [$+4.28,\,+4.45$]\\
        & $95\%$ & --                & --                & --                & [$-5.03,\,-3.86$]  $\cup$  [$-1.17,\,+0.92$]  $\cup$  [$+3.86,\,+4.87$]\\
        & mode   & --                & --                & --                & $-4.40$            $\wedge$ $-0.46$           $\wedge$ $+4.40$\\
        \hline
    \end{tabular}
%}
    \renewcommand{\arraystretch}{1.0}
    \caption{
        \label{tab:wilson:coeff:1-dimCLs}
        The 68\%- and 95\%-credibility intervals and the local modes of the
        marginalized 1D posterior distributions of the Wilson coefficients at
        $\mu = 4.2$ GeV, $P(\wilson{i}|D),\, i = 7,9,10,7',9',10'$, for nominal
        priors of nuisance parameters in the various scenarios. Note that for the
        \SMp{} scenario the individual solutions can not be disentangled within the
        1D posterior distributions, unlike for the SM scenario.
        For comparison, the SM values of the Wilson coefficients read
        $\wilson[SM]{7} = -0.34$,
        $\wilson[SM]{9} = +4.27$, $\wilson[SM]{10} = -4.17$, $\wilson[SM]{7'} = -0.01$,
        $\wilson[SM]{9'} = \wilson[SM]{10'} = 0$.
      }
\end{table*}

\begin{table*}
    \renewcommand{\arraystretch}{1.3}
    \newcommand{\la}{\langle}
    \newcommand{\ra}{\rangle}
    \let\bs\boldsymbol
    \begin{tabular}{ll|cccccc|cccccc}
        \hline
        & & \multicolumn{6}{c|}{\SM{} full, solution $A$} & \multicolumn{6}{c}{\SMp{} full, solution $A$'}\\
        \multicolumn{2}{c|}{Observable}                           & ATLAS       & BaBar       & Belle       & CDF         & CMS         & LHCb       & ATLAS       & BaBar       & Belle       & CDF         & CMS         & LHCb       \\
        \hline
        $B\to X_s\gamma$       & $\BR$                            & --          & $-0.1$      & $+0.4$      & --          & --          & --         & --          & $+0.2$      & $+0.8$      & --          & --          & --         \\
        \hline
        $B\to X_s\ell^+\ell^-$ & $\la\BR\ra_{[1,6]}$              & --          & $+0.5$      & $+0.3$      & --          & --          & --         & --          & $+0.2$      & $-0.2$      & --          & --          & --         \\
        \hline
        $B_s\to \mu^+\mu^-$    & $\BR$                            & --          & --          & --          & --          & $-0.7$      & $-0.7$     & --          & --          & --          & --          & $-0.4$      & $-0.4$     \\
        \hline
        \multirow{3}{*}{$B\to K^*\gamma$}
                               & $\BR$                            & --          & $+0.7$      & $-1.2$      & --          & --          & --         & --          & $+0.5$      & $-1.4$      & --          & --          & --         \\
                               & $S + C$                          & --          & $+0.4$      & $+0.7$      & --          & --          & --         & --          & $+0.8$      & $+0.4$      & --          & --          & --         \\
        \hline
        \multirow{5}{*}{$B\to K\ell^+\ell^-$}
                               & $\la\BR\ra_{[1,6]}$              & --          & $+0.3$      & $+0.3$      & $+0.3$      & --          & $-0.8$     & --          & $+0.2$      & $+0.3$      & $+0.2$      & --          & $-0.9$     \\
                               & $\la\BR\ra_{[14.18,16]}$         & --          & $+1.1$      & $+0.3$      & $+0.9$      & --          & $+0.8$     & --          & $+1.0$      & $+0.2$      & $+0.8$      & --          & $+0.6$     \\
                               & $\la\BR\ra_{[16,18]}$            & --          & --          & --          &  --         & --          & $+0.7$     & --          & --          & --          &  --         & --          & $+0.6$     \\
                               & $\la\BR\ra_{[16,23]}$            & --          & $+0.0$      & $+1.5$      & $-1.8$      & --          & --         & --          & $-0.1$      & $+1.4$      & $-1.9$      & --          & --         \\
                               & $\la\BR\ra_{[18,22]}$            & --          & --          & --          & --          & --          & $-0.8$     & --          & --          & --          & --          & --          & $-1.0$     \\
        \hline
        \multirow{24}{*}{$B\to K^*\ell^+\ell^-$}
                               & $\la\BR\ra_{[1,6]}$              & --          & $+0.5$      & $-0.7$      & $+0.3$      & $+0.9$      & $-0.4$     & --          & $+0.5$      & $-0.7$      & $+0.3$      & $+0.8$      & $-0.5$     \\
                               & $\la\BR\ra_{[14.18,16]}$         & --          & $+1.0$      & $-0.2$      & $+1.1$      & $-1.3$      & $-0.6$     & --          & $+1.1$      & $-0.1$      & $+1.2$      & $-1.1$      & $-0.4$     \\
                               & $\la\BR\ra_{[16,19]}$            & --          & $-0.6$      & $\bs{+2.6}$ & $-1.4$      & $+0.9$      & $-0.2$     & --          & $-0.5$      & $\bs{+2.7}$ & $-1.3$      & $+1.0$      & $-0.1$     \\
        \cline{2-14}
                               & $\la A_{\rm FB}\ra_{[1,6]}$      & $-0.9$      & $-1.9$      & $-1.2$      & $-1.9$      & $-0.4$      & --         & $-0.7$      & $-1.6$      & $-1.1$      & $-1.7$      & $-0.1$      & --         \\
                               & $\la A_{\rm FB}\ra_{[14.18,16]}$ & $-0.3$      & $+0.8$      & $-1.1$      & $-0.6$      & $+1.3$      & $-1.6$     & $-0.4$      & $+0.7$      & $-1.2$      & $-0.7$      & $+1.2$      & $-1.9$     \\
                               & $\la A_{\rm FB}\ra_{[16,19]}$    & $\bs{+2.1}$ & $+0.2$      & $-1.7$      & $-0.2$      & $-0.5$      & $+1.0$     & $+1.9$      & $+0.1$      & $-1.8$      & $-0.3$      & $-0.8$      & $+0.8$     \\
        \cline{2-14}
                               & $\la F_L\ra_{[1,6]}$             & $\bs{-2.5}$ & $\bs{-3.4}$ & $+0.4$      & $+1.2$      & $+1.1$      & $+1.1$     & $\bs{-2.5}$ & $\bs{-3.3}$ & $+0.4$      & $+1.3$      & $+1.1$      & $+1.1$     \\
                               & $\la F_L\ra_{[14.18,16]}$        & $-0.5$      & $+0.4$      & $-1.8$      & $+0.7$      & $+1.3$      & $-0.4$     & $-0.4$      & $+0.5$      & $-1.8$      & $+0.8$      & $+1.5$      & $-0.2$     \\
                               & $\la F_L\ra_{[16,19]}$           & $+0.1$      & $+1.2$      & $-1.5$      & $-1.5$      & $+1.3$      & $+0.5$     & $+0.2$      & $+1.3$      & $-1.4$      & $-1.5$      & $+1.4$      & $+0.6$     \\
        \cline{2-14}
                               & $\la A_T^{(2)}\ra_{[1,6]}$       & --          & --          & --          & $-0.2$      & --          & $+0.6$     & --          & --          & --          & $-0.3$      & --          & $-0.5$     \\
                               & $\la A_T^{(2)}\ra_{[14.18,16]}$  & --          & --          & --          & $+0.5$      & --          & $+1.1$     & --          & --          & --          & $+0.6$      & --          & $+1.3$     \\
                               & $\la A_T^{(2)}\ra_{[16,19]}$     & --          & --          & --          & $-0.1$      & --          & $-0.5$     & --          & --          & --          & $-0.1$      & --          & $-0.4$     \\
        \cline{2-14}
                               & $\la A_T^{({\rm re})}\ra_{[1,6]}$& --          & --          & --          & --          & --          & $+1.2$     & --          & --          & --          & --          & --          & $+1.7$     \\
        \cline{2-14}
                               & $\la P'_4\ra_{[1,6]}$            & --          & --          & --          & --          & --          & $-0.1$     & --          & --          & --          & --          & --          & $+0.0$     \\
                               & $\la P'_4\ra_{[14.18,16]}$       & --          & --          & --          & --          & --          & $\bs{-2.4}$& --          & --          & --          & --          & --          & $\bs{-2.4}$\\
                               & $\la P'_4\ra_{[16,19]}$          & --          & --          & --          & --          & --          & $-1.2$     & --          & --          & --          & --          & --          & $-1.2$     \\
        \cline{2-14}
                               & $\la P'_5\ra_{[1,6]}$            & --          & --          & --          & --          & --          & $+1.4$     & --          & --          & --          & --          & --          & $+1.7$     \\
                               & $\la P'_5\ra_{[14.18,16]}$       & --          & --          & --          & --          & --          & $+0.0$     & --          & --          & --          & --          & --          & $-0.2$     \\
                               & $\la P'_5\ra_{[16,19]}$          & --          & --          & --          & --          & --          & $+0.2$     & --          & --          & --          & --          & --          & $-0.1$     \\
        \cline{2-14}
                               & $\la P'_6\ra_{[1,6]}$            & --          & --          & --          & --          & --          & $+1.1$     & --          & --          & --          & --          & --          & $+1.0$     \\
                               & $\la P'_6\ra_{[14.18,16]}$       & --          & --          & --          & --          & --          & $+0.7$     & --          & --          & --          & --          & --          & $+0.7$     \\
                               & $\la P'_6\ra_{[16,19]}$          & --          & --          & --          & --          & --          & $-0.8$     & --          & --          & --          & --          & --          & $-0.8$     \\
        \hline
    \end{tabular}
    \renewcommand{\arraystretch}{1.0}
    \caption{Compilation of the pull values in units of Gaussian $\sigma$ at the SM-like best
      fit points $A$ in the \SM{} fit (left columns) and $A$' in the \SMp{} fit
      (right columns), listed per experiment and observable. Only pull values for
      fits with the ``full'' data set are listed. The single CLEO measurement
      of $\BR(B\to K^*\gamma)$ has a pull value $+0.3\sigma$ in both the \SM{} and
      \SMp{} fits. The pull values for the \SMnu{} fit deviate by less than
      $0.3\sigma$ from those of the \SM{} fit for the ``full'' data set, with the exception
      of $+2.3\sigma$ for the LHCb measurement of $\la P'_5\ra_{[1,6]}$.
      The pull values in scenario \SMp{} with the ``full (+FF)'' data set deviate by less
      than $0.4\sigma$ from those given for \SMp{} ``full'', except for the LHCb measurements of
      $\la A_T^{(2)} \ra_{[1,6]}$ with $+1.1\sigma$, $\la A_T^{({\rm re})}\ra_{[1,6]}$ with
      $+0.6\sigma$ and $\la P'_5\ra_{[1,6]}$ with $+1.2\sigma$.
      The pull values for \SM{} ``full (+FF)'' deviate by less then $0.3\sigma$ from those
      of \SM{} ``full'' except for $\la P'_5\ra_{[1,6]}$ with a slight increase of $+0.4\sigma$.
    }
    \label{tab:p-value-sm-full}
\end{table*}

%--------+---------+---------+---------+---------+---------+---------+---------+
%
%  References
%
%--------+---------+---------+---------+---------+---------+---------+---------+
\bibliographystyle{apsrev4-1}
\bibliography{references.bib}

\begin{thebibliography}{90}%
\makeatletter
\providecommand \@ifxundefined [1]{%
 \@ifx{#1\undefined}
}%
\providecommand \@ifnum [1]{%
 \ifnum #1\expandafter \@firstoftwo
 \else \expandafter \@secondoftwo
 \fi
}%
\providecommand \@ifx [1]{%
 \ifx #1\expandafter \@firstoftwo
 \else \expandafter \@secondoftwo
 \fi
}%
\providecommand \natexlab [1]{#1}%
\providecommand \enquote  [1]{``#1''}%
\providecommand \bibnamefont  [1]{#1}%
\providecommand \bibfnamefont [1]{#1}%
\providecommand \citenamefont [1]{#1}%
\providecommand \href@noop [0]{\@secondoftwo}%
\providecommand \href [0]{\begingroup \@sanitize@url \@href}%
\providecommand \@href[1]{\@@startlink{#1}\@@href}%
\providecommand \@@href[1]{\endgroup#1\@@endlink}%
\providecommand \@sanitize@url [0]{\catcode `\\12\catcode `\$12\catcode
  `\&12\catcode `\#12\catcode `\^12\catcode `\_12\catcode `\%12\relax}%
\providecommand \@@startlink[1]{}%
\providecommand \@@endlink[0]{}%
\providecommand \url  [0]{\begingroup\@sanitize@url \@url }%
\providecommand \@url [1]{\endgroup\@href {#1}{\urlprefix }}%
\providecommand \urlprefix  [0]{URL }%
\providecommand \Eprint [0]{\href }%
\providecommand \doibase [0]{http://dx.doi.org/}%
\providecommand \selectlanguage [0]{\@gobble}%
\providecommand \bibinfo  [0]{\@secondoftwo}%
\providecommand \bibfield  [0]{\@secondoftwo}%
\providecommand \translation [1]{[#1]}%
\providecommand \BibitemOpen [0]{}%
\providecommand \bibitemStop [0]{}%
\providecommand \bibitemNoStop [0]{.\EOS\space}%
\providecommand \EOS [0]{\spacefactor3000\relax}%
\providecommand \BibitemShut  [1]{\csname bibitem#1\endcsname}%
\let\auto@bib@innerbib\@empty
%</preamble>
\bibitem [{\citenamefont {Limosani}\ \emph {et~al.}(2009)\citenamefont
  {Limosani} \emph {et~al.}}]{Limosani:2009qg}%
  \BibitemOpen
  \bibfield  {author} {\bibinfo {author} {\bibfnamefont {A.}~\bibnamefont
  {Limosani}} \emph {et~al.} (\bibinfo {collaboration} {Belle Collaboration}),\
  }\href {\doibase 10.1103/PhysRevLett.103.241801} {\bibfield  {journal}
  {\bibinfo  {journal} {Phys.Rev.Lett.}\ }\textbf {\bibinfo {volume} {103}},\
  \bibinfo {pages} {241801} (\bibinfo {year} {2009})},\ \Eprint
  {http://arxiv.org/abs/0907.1384} {arXiv:0907.1384 [hep-ex]} \BibitemShut
  {NoStop}%
%%CITATION = ARXIV:0907.1384;%%
\bibitem [{\citenamefont {Lees}\ \emph
  {et~al.}(2012{\natexlab{a}})\citenamefont {Lees} \emph
  {et~al.}}]{Lees:2012ufa}%
  \BibitemOpen
  \bibfield  {author} {\bibinfo {author} {\bibfnamefont {J.}~\bibnamefont
  {Lees}} \emph {et~al.} (\bibinfo {collaboration} {BaBar Collaboration}),\
  }\href {\doibase 10.1103/PhysRevD.86.112008} {\bibfield  {journal} {\bibinfo
  {journal} {Phys.Rev.}\ }\textbf {\bibinfo {volume} {D86}},\ \bibinfo {pages}
  {112008} (\bibinfo {year} {2012}{\natexlab{a}})},\ \Eprint
  {http://arxiv.org/abs/1207.5772} {arXiv:1207.5772 [hep-ex]} \BibitemShut
  {NoStop}%
%%CITATION = ARXIV:1207.5772;%%
\bibitem [{\citenamefont {Aubert}\ \emph {et~al.}(2004)\citenamefont {Aubert}
  \emph {et~al.}}]{Aubert:2004it}%
  \BibitemOpen
  \bibfield  {author} {\bibinfo {author} {\bibfnamefont {B.}~\bibnamefont
  {Aubert}} \emph {et~al.} (\bibinfo {collaboration} {BaBar Collaboration}),\
  }\href {\doibase 10.1103/PhysRevLett.93.081802} {\bibfield  {journal}
  {\bibinfo  {journal} {Phys.Rev.Lett.}\ }\textbf {\bibinfo {volume} {93}},\
  \bibinfo {pages} {081802} (\bibinfo {year} {2004})},\ \Eprint
  {http://arxiv.org/abs/hep-ex/0404006} {arXiv:hep-ex/0404006 [hep-ex]}
  \BibitemShut {NoStop}%
%%CITATION = HEP-EX/0404006;%%
\bibitem [{\citenamefont {Iwasaki}\ \emph {et~al.}(2005)\citenamefont {Iwasaki}
  \emph {et~al.}}]{Iwasaki:2005sy}%
  \BibitemOpen
  \bibfield  {author} {\bibinfo {author} {\bibfnamefont {M.}~\bibnamefont
  {Iwasaki}} \emph {et~al.} (\bibinfo {collaboration} {Belle Collaboration}),\
  }\href {\doibase 10.1103/PhysRevD.72.092005} {\bibfield  {journal} {\bibinfo
  {journal} {Phys.Rev.}\ }\textbf {\bibinfo {volume} {D72}},\ \bibinfo {pages}
  {092005} (\bibinfo {year} {2005})},\ \Eprint
  {http://arxiv.org/abs/hep-ex/0503044} {arXiv:hep-ex/0503044 [hep-ex]}
  \BibitemShut {NoStop}%
%%CITATION = HEP-EX/0503044;%%
\bibitem [{\citenamefont {Aaij}\ \emph
  {et~al.}(2013{\natexlab{a}})\citenamefont {Aaij} \emph
  {et~al.}}]{Aaij:2013aka}%
  \BibitemOpen
  \bibfield  {author} {\bibinfo {author} {\bibfnamefont {R.}~\bibnamefont
  {Aaij}} \emph {et~al.} (\bibinfo {collaboration} {LHCb collaboration}),\
  }\href {\doibase 10.1103/PhysRevLett.111.101805} {\bibfield  {journal}
  {\bibinfo  {journal} {Phys.Rev.Lett.}\ }\textbf {\bibinfo {volume} {111}},\
  \bibinfo {pages} {101805} (\bibinfo {year} {2013}{\natexlab{a}})},\ \Eprint
  {http://arxiv.org/abs/1307.5024} {arXiv:1307.5024 [hep-ex]} \BibitemShut
  {NoStop}%
%%CITATION = ARXIV:1307.5024;%%
\bibitem [{\citenamefont {Chatrchyan}\ \emph
  {et~al.}(2013{\natexlab{a}})\citenamefont {Chatrchyan} \emph
  {et~al.}}]{Chatrchyan:2013bka}%
  \BibitemOpen
  \bibfield  {author} {\bibinfo {author} {\bibfnamefont {S.}~\bibnamefont
  {Chatrchyan}} \emph {et~al.} (\bibinfo {collaboration} {CMS Collaboration}),\
  }\href {\doibase 10.1103/PhysRevLett.111.101804} {\bibfield  {journal}
  {\bibinfo  {journal} {Phys.Rev.Lett.}\ }\textbf {\bibinfo {volume} {111}},\
  \bibinfo {pages} {101804} (\bibinfo {year} {2013}{\natexlab{a}})},\ \Eprint
  {http://arxiv.org/abs/1307.5025} {arXiv:1307.5025 [hep-ex]} \BibitemShut
  {NoStop}%
%%CITATION = ARXIV:1307.5025;%%
\bibitem [{\citenamefont {Coan}\ \emph {et~al.}(2000)\citenamefont {Coan} \emph
  {et~al.}}]{Coan:1999kh}%
  \BibitemOpen
  \bibfield  {author} {\bibinfo {author} {\bibfnamefont {T.}~\bibnamefont
  {Coan}} \emph {et~al.} (\bibinfo {collaboration} {CLEO Collaboration}),\
  }\href {\doibase 10.1103/PhysRevLett.84.5283} {\bibfield  {journal} {\bibinfo
   {journal} {Phys.Rev.Lett.}\ }\textbf {\bibinfo {volume} {84}},\ \bibinfo
  {pages} {5283} (\bibinfo {year} {2000})},\ \Eprint
  {http://arxiv.org/abs/hep-ex/9912057} {arXiv:hep-ex/9912057 [hep-ex]}
  \BibitemShut {NoStop}%
%%CITATION = HEP-EX/9912057;%%
\bibitem [{\citenamefont {Aubert}\ \emph {et~al.}(2008)\citenamefont {Aubert}
  \emph {et~al.}}]{Aubert:2008gy}%
  \BibitemOpen
  \bibfield  {author} {\bibinfo {author} {\bibfnamefont {B.}~\bibnamefont
  {Aubert}} \emph {et~al.} (\bibinfo {collaboration} {BaBar Collaboration}),\
  }\href {\doibase 10.1103/PhysRevD.78.071102} {\bibfield  {journal} {\bibinfo
  {journal} {Phys.Rev.}\ }\textbf {\bibinfo {volume} {D78}},\ \bibinfo {pages}
  {071102} (\bibinfo {year} {2008})},\ \Eprint {http://arxiv.org/abs/0807.3103}
  {arXiv:0807.3103 [hep-ex]} \BibitemShut {NoStop}%
%%CITATION = ARXIV:0807.3103;%%
\bibitem [{\citenamefont {Aubert}\ \emph {et~al.}(2009)\citenamefont {Aubert}
  \emph {et~al.}}]{Aubert:2009ak}%
  \BibitemOpen
  \bibfield  {author} {\bibinfo {author} {\bibfnamefont {B.}~\bibnamefont
  {Aubert}} \emph {et~al.} (\bibinfo {collaboration} {BaBar Collaboration}),\
  }\href {\doibase 10.1103/PhysRevLett.103.211802} {\bibfield  {journal}
  {\bibinfo  {journal} {Phys.Rev.Lett.}\ }\textbf {\bibinfo {volume} {103}},\
  \bibinfo {pages} {211802} (\bibinfo {year} {2009})},\ \Eprint
  {http://arxiv.org/abs/0906.2177} {arXiv:0906.2177 [hep-ex]} \BibitemShut
  {NoStop}%
%%CITATION = ARXIV:0906.2177;%%
\bibitem [{\citenamefont {Nakao}\ \emph {et~al.}(2004)\citenamefont {Nakao}
  \emph {et~al.}}]{Nakao:2004th}%
  \BibitemOpen
  \bibfield  {author} {\bibinfo {author} {\bibfnamefont {M.}~\bibnamefont
  {Nakao}} \emph {et~al.} (\bibinfo {collaboration} {BELLE Collaboration}),\
  }\href {\doibase 10.1103/PhysRevD.69.112001} {\bibfield  {journal} {\bibinfo
  {journal} {Phys.Rev.}\ }\textbf {\bibinfo {volume} {D69}},\ \bibinfo {pages}
  {112001} (\bibinfo {year} {2004})},\ \Eprint
  {http://arxiv.org/abs/hep-ex/0402042} {arXiv:hep-ex/0402042 [hep-ex]}
  \BibitemShut {NoStop}%
%%CITATION = HEP-EX/0402042;%%
\bibitem [{\citenamefont {Ushiroda}\ \emph {et~al.}(2006)\citenamefont
  {Ushiroda} \emph {et~al.}}]{Ushiroda:2006fi}%
  \BibitemOpen
  \bibfield  {author} {\bibinfo {author} {\bibfnamefont {Y.}~\bibnamefont
  {Ushiroda}} \emph {et~al.} (\bibinfo {collaboration} {Belle Collaboration}),\
  }\href {\doibase 10.1103/PhysRevD.74.111104} {\bibfield  {journal} {\bibinfo
  {journal} {Phys.Rev.}\ }\textbf {\bibinfo {volume} {D74}},\ \bibinfo {pages}
  {111104} (\bibinfo {year} {2006})},\ \Eprint
  {http://arxiv.org/abs/hep-ex/0608017} {arXiv:hep-ex/0608017 [hep-ex]}
  \BibitemShut {NoStop}%
%%CITATION = HEP-EX/0608017;%%
\bibitem [{\citenamefont {Lees}\ \emph
  {et~al.}(2012{\natexlab{b}})\citenamefont {Lees} \emph
  {et~al.}}]{Lees:2012tva}%
  \BibitemOpen
  \bibfield  {author} {\bibinfo {author} {\bibfnamefont {J.}~\bibnamefont
  {Lees}} \emph {et~al.} (\bibinfo {collaboration} {BaBar Collaboration}),\
  }\href {\doibase 10.1103/PhysRevD.86.032012} {\bibfield  {journal} {\bibinfo
  {journal} {Phys.Rev.}\ }\textbf {\bibinfo {volume} {D86}},\ \bibinfo {pages}
  {032012} (\bibinfo {year} {2012}{\natexlab{b}})},\ \Eprint
  {http://arxiv.org/abs/1204.3933} {arXiv:1204.3933 [hep-ex]} \BibitemShut
  {NoStop}%
%%CITATION = ARXIV:1204.3933;%%
\bibitem [{\citenamefont {Wei}\ \emph {et~al.}(2009)\citenamefont {Wei} \emph
  {et~al.}}]{Wei:2009zv}%
  \BibitemOpen
  \bibfield  {author} {\bibinfo {author} {\bibfnamefont {J.-T.}\ \bibnamefont
  {Wei}} \emph {et~al.} (\bibinfo {collaboration} {BELLE Collaboration}),\
  }\href {\doibase 10.1103/PhysRevLett.103.171801} {\bibfield  {journal}
  {\bibinfo  {journal} {Phys.Rev.Lett.}\ }\textbf {\bibinfo {volume} {103}},\
  \bibinfo {pages} {171801} (\bibinfo {year} {2009})},\ \Eprint
  {http://arxiv.org/abs/0904.0770} {arXiv:0904.0770 [hep-ex]} \BibitemShut
  {NoStop}%
%%CITATION = ARXIV:0904.0770;%%
\bibitem [{\citenamefont {{[CDF Collaboration]}}(2012)}]{CDF:2012:BKstarll}%
  \BibitemOpen
  \bibfield  {author} {\bibinfo {author} {\bibnamefont {{[CDF
  Collaboration]}}},\ }\href@noop {} {\  (\bibinfo {year} {2012})},\ \Eprint
  {http://arxiv.org/abs/Public Note 10894} {Public Note 10894} \BibitemShut
  {NoStop}%
\bibitem [{\citenamefont {Aaij}\ \emph
  {et~al.}(2013{\natexlab{b}})\citenamefont {Aaij} \emph
  {et~al.}}]{Aaij:2012vr}%
  \BibitemOpen
  \bibfield  {author} {\bibinfo {author} {\bibfnamefont {R.}~\bibnamefont
  {Aaij}} \emph {et~al.} (\bibinfo {collaboration} {LHCb Collaboration}),\
  }\href {\doibase 10.1007/JHEP02(2013)105} {\bibfield  {journal} {\bibinfo
  {journal} {JHEP}\ }\textbf {\bibinfo {volume} {1302}},\ \bibinfo {pages}
  {105} (\bibinfo {year} {2013}{\natexlab{b}})},\ \Eprint
  {http://arxiv.org/abs/1209.4284} {arXiv:1209.4284 [hep-ex]} \BibitemShut
  {NoStop}%
%%CITATION = ARXIV:1209.4284;%%
\bibitem [{\citenamefont {Aaij}\ \emph
  {et~al.}(2013{\natexlab{c}})\citenamefont {Aaij} \emph
  {et~al.}}]{Aaij:2013iag}%
  \BibitemOpen
  \bibfield  {author} {\bibinfo {author} {\bibfnamefont {R.}~\bibnamefont
  {Aaij}} \emph {et~al.} (\bibinfo {collaboration} {LHCb Collaboration}),\
  }\href {\doibase 10.1007/JHEP08(2013)131} {\bibfield  {journal} {\bibinfo
  {journal} {JHEP}\ }\textbf {\bibinfo {volume} {1308}},\ \bibinfo {pages}
  {131} (\bibinfo {year} {2013}{\natexlab{c}})},\ \Eprint
  {http://arxiv.org/abs/1304.6325} {arXiv:1304.6325 [hep-ex]} \BibitemShut
  {NoStop}%
%%CITATION = ARXIV:1304.6325;%%
\bibitem [{\citenamefont {Chatrchyan}\ \emph
  {et~al.}(2013{\natexlab{b}})\citenamefont {Chatrchyan} \emph
  {et~al.}}]{Chatrchyan:2013cda}%
  \BibitemOpen
  \bibfield  {author} {\bibinfo {author} {\bibfnamefont {S.}~\bibnamefont
  {Chatrchyan}} \emph {et~al.} (\bibinfo {collaboration} {CMS Collaboration}),\
  }\href {\doibase 10.1016/j.physletb.2013.10.017} {\bibfield  {journal}
  {\bibinfo  {journal} {Phys.Lett.}\ }\textbf {\bibinfo {volume} {B727}},\
  \bibinfo {pages} {77} (\bibinfo {year} {2013}{\natexlab{b}})},\ \Eprint
  {http://arxiv.org/abs/1308.3409} {arXiv:1308.3409 [hep-ex]} \BibitemShut
  {NoStop}%
%%CITATION = ARXIV:1308.3409;%%
\bibitem [{\citenamefont {Aad}\ \emph {et~al.}(2013)\citenamefont {Aad} \emph
  {et~al.}}]{ATLAS:2013ola}%
  \BibitemOpen
  \bibfield  {author} {\bibinfo {author} {\bibfnamefont {G.}~\bibnamefont
  {Aad}} \emph {et~al.} (\bibinfo {collaboration} {ATLAS Collaboration}),\
  }\href@noop {} {\  (\bibinfo {year} {2013})},\ \Eprint
  {http://arxiv.org/abs/ATLAS-CONF-2013-038, ATLAS-COM-CONF-2013-043}
  {ATLAS-CONF-2013-038, ATLAS-COM-CONF-2013-043} \BibitemShut {NoStop}%
%%CITATION = ATLAS-CONF-2013-038 ETC.;%%
\bibitem [{\citenamefont {Aaij}\ \emph
  {et~al.}(2013{\natexlab{d}})\citenamefont {Aaij} \emph
  {et~al.}}]{Aaij:2013qta}%
  \BibitemOpen
  \bibfield  {author} {\bibinfo {author} {\bibfnamefont {R.}~\bibnamefont
  {Aaij}} \emph {et~al.} (\bibinfo {collaboration} {LHCb collaboration}),\
  }\href {\doibase 10.1103/PhysRevLett.111.191801} {\bibfield  {journal}
  {\bibinfo  {journal} {Phys.Rev.Lett.}\ }\textbf {\bibinfo {volume} {111}},\
  \bibinfo {pages} {191801} (\bibinfo {year} {2013}{\natexlab{d}})},\ \Eprint
  {http://arxiv.org/abs/1308.1707} {arXiv:1308.1707 [hep-ex]} \BibitemShut
  {NoStop}%
%%CITATION = ARXIV:1308.1707;%%
\bibitem [{\citenamefont {Kr{\"u}ger}\ and\ \citenamefont
  {Matias}(2005)}]{Kruger:2005ep}%
  \BibitemOpen
  \bibfield  {author} {\bibinfo {author} {\bibfnamefont {F.}~\bibnamefont
  {Kr{\"u}ger}}\ and\ \bibinfo {author} {\bibfnamefont {J.}~\bibnamefont
  {Matias}},\ }\href {\doibase 10.1103/PhysRevD.71.094009} {\bibfield
  {journal} {\bibinfo  {journal} {Phys.Rev.}\ }\textbf {\bibinfo {volume}
  {D71}},\ \bibinfo {pages} {094009} (\bibinfo {year} {2005})},\ \Eprint
  {http://arxiv.org/abs/hep-ph/0502060} {arXiv:hep-ph/0502060 [hep-ph]}
  \BibitemShut {NoStop}%
%%CITATION = HEP-PH/0502060;%%
\bibitem [{\citenamefont {Egede}\ \emph {et~al.}(2008)\citenamefont {Egede},
  \citenamefont {Hurth}, \citenamefont {Matias}, \citenamefont {Ramon},\ and\
  \citenamefont {Reece}}]{Egede:2008uy}%
  \BibitemOpen
  \bibfield  {author} {\bibinfo {author} {\bibfnamefont {U.}~\bibnamefont
  {Egede}}, \bibinfo {author} {\bibfnamefont {T.}~\bibnamefont {Hurth}},
  \bibinfo {author} {\bibfnamefont {J.}~\bibnamefont {Matias}}, \bibinfo
  {author} {\bibfnamefont {M.}~\bibnamefont {Ramon}}, \ and\ \bibinfo {author}
  {\bibfnamefont {W.}~\bibnamefont {Reece}},\ }\href {\doibase
  10.1088/1126-6708/2008/11/032} {\bibfield  {journal} {\bibinfo  {journal}
  {JHEP}\ }\textbf {\bibinfo {volume} {0811}},\ \bibinfo {pages} {032}
  (\bibinfo {year} {2008})},\ \Eprint {http://arxiv.org/abs/0807.2589}
  {arXiv:0807.2589 [hep-ph]} \BibitemShut {NoStop}%
%%CITATION = ARXIV:0807.2589;%%
\bibitem [{\citenamefont {Bobeth}\ \emph {et~al.}(2010)\citenamefont {Bobeth},
  \citenamefont {Hiller},\ and\ \citenamefont {van Dyk}}]{Bobeth:2010wg}%
  \BibitemOpen
  \bibfield  {author} {\bibinfo {author} {\bibfnamefont {C.}~\bibnamefont
  {Bobeth}}, \bibinfo {author} {\bibfnamefont {G.}~\bibnamefont {Hiller}}, \
  and\ \bibinfo {author} {\bibfnamefont {D.}~\bibnamefont {van Dyk}},\ }\href
  {\doibase 10.1007/JHEP07(2010)098} {\bibfield  {journal} {\bibinfo  {journal}
  {JHEP}\ }\textbf {\bibinfo {volume} {1007}},\ \bibinfo {pages} {098}
  (\bibinfo {year} {2010})},\ \Eprint {http://arxiv.org/abs/1006.5013}
  {arXiv:1006.5013 [hep-ph]} \BibitemShut {NoStop}%
%%CITATION = ARXIV:1006.5013;%%
\bibitem [{\citenamefont {Bobeth}\ \emph {et~al.}(2011)\citenamefont {Bobeth},
  \citenamefont {Hiller},\ and\ \citenamefont {van Dyk}}]{Bobeth:2011gi}%
  \BibitemOpen
  \bibfield  {author} {\bibinfo {author} {\bibfnamefont {C.}~\bibnamefont
  {Bobeth}}, \bibinfo {author} {\bibfnamefont {G.}~\bibnamefont {Hiller}}, \
  and\ \bibinfo {author} {\bibfnamefont {D.}~\bibnamefont {van Dyk}},\ }\href
  {\doibase 10.1007/JHEP07(2011)067} {\bibfield  {journal} {\bibinfo  {journal}
  {JHEP}\ }\textbf {\bibinfo {volume} {1107}},\ \bibinfo {pages} {067}
  (\bibinfo {year} {2011})},\ \Eprint {http://arxiv.org/abs/1105.0376}
  {arXiv:1105.0376 [hep-ph]} \BibitemShut {NoStop}%
%%CITATION = ARXIV:1105.0376;%%
\bibitem [{\citenamefont {Becirevic}\ and\ \citenamefont
  {Schneider}(2012)}]{Becirevic:2011bp}%
  \BibitemOpen
  \bibfield  {author} {\bibinfo {author} {\bibfnamefont {D.}~\bibnamefont
  {Becirevic}}\ and\ \bibinfo {author} {\bibfnamefont {E.}~\bibnamefont
  {Schneider}},\ }\href {\doibase 10.1016/j.nuclphysb.2011.09.004} {\bibfield
  {journal} {\bibinfo  {journal} {Nucl.Phys.}\ }\textbf {\bibinfo {volume}
  {B854}},\ \bibinfo {pages} {321} (\bibinfo {year} {2012})},\ \Eprint
  {http://arxiv.org/abs/1106.3283} {arXiv:1106.3283 [hep-ph]} \BibitemShut
  {NoStop}%
%%CITATION = ARXIV:1106.3283;%%
\bibitem [{\citenamefont {Matias}\ \emph {et~al.}(2012)\citenamefont {Matias},
  \citenamefont {Mescia}, \citenamefont {Ramon},\ and\ \citenamefont
  {Virto}}]{Matias:2012xw}%
  \BibitemOpen
  \bibfield  {author} {\bibinfo {author} {\bibfnamefont {J.}~\bibnamefont
  {Matias}}, \bibinfo {author} {\bibfnamefont {F.}~\bibnamefont {Mescia}},
  \bibinfo {author} {\bibfnamefont {M.}~\bibnamefont {Ramon}}, \ and\ \bibinfo
  {author} {\bibfnamefont {J.}~\bibnamefont {Virto}},\ }\href {\doibase
  10.1007/JHEP04(2012)104} {\bibfield  {journal} {\bibinfo  {journal} {JHEP}\
  }\textbf {\bibinfo {volume} {1204}},\ \bibinfo {pages} {104} (\bibinfo {year}
  {2012})},\ \Eprint {http://arxiv.org/abs/1202.4266} {arXiv:1202.4266
  [hep-ph]} \BibitemShut {NoStop}%
%%CITATION = ARXIV:1202.4266;%%
\bibitem [{\citenamefont {Das}\ and\ \citenamefont {Sinha}(2012)}]{Das:2012kz}%
  \BibitemOpen
  \bibfield  {author} {\bibinfo {author} {\bibfnamefont {D.}~\bibnamefont
  {Das}}\ and\ \bibinfo {author} {\bibfnamefont {R.}~\bibnamefont {Sinha}},\
  }\href {\doibase 10.1103/PhysRevD.86.056006} {\bibfield  {journal} {\bibinfo
  {journal} {Phys.Rev.}\ }\textbf {\bibinfo {volume} {D86}},\ \bibinfo {pages}
  {056006} (\bibinfo {year} {2012})},\ \Eprint {http://arxiv.org/abs/1205.1438}
  {arXiv:1205.1438 [hep-ph]} \BibitemShut {NoStop}%
%%CITATION = ARXIV:1205.1438;%%
\bibitem [{\citenamefont {Descotes-Genon}\ \emph
  {et~al.}(2013{\natexlab{a}})\citenamefont {Descotes-Genon}, \citenamefont
  {Matias}, \citenamefont {Ramon},\ and\ \citenamefont
  {Virto}}]{DescotesGenon:2012zf}%
  \BibitemOpen
  \bibfield  {author} {\bibinfo {author} {\bibfnamefont {S.}~\bibnamefont
  {Descotes-Genon}}, \bibinfo {author} {\bibfnamefont {J.}~\bibnamefont
  {Matias}}, \bibinfo {author} {\bibfnamefont {M.}~\bibnamefont {Ramon}}, \
  and\ \bibinfo {author} {\bibfnamefont {J.}~\bibnamefont {Virto}},\ }\href
  {\doibase 10.1007/JHEP01(2013)048} {\bibfield  {journal} {\bibinfo  {journal}
  {JHEP}\ }\textbf {\bibinfo {volume} {1301}},\ \bibinfo {pages} {048}
  (\bibinfo {year} {2013}{\natexlab{a}})},\ \Eprint
  {http://arxiv.org/abs/1207.2753} {arXiv:1207.2753 [hep-ph]} \BibitemShut
  {NoStop}%
%%CITATION = ARXIV:1207.2753;%%
\bibitem [{\citenamefont {Bobeth}\ \emph {et~al.}(2013)\citenamefont {Bobeth},
  \citenamefont {Hiller},\ and\ \citenamefont {van Dyk}}]{Bobeth:2012vn}%
  \BibitemOpen
  \bibfield  {author} {\bibinfo {author} {\bibfnamefont {C.}~\bibnamefont
  {Bobeth}}, \bibinfo {author} {\bibfnamefont {G.}~\bibnamefont {Hiller}}, \
  and\ \bibinfo {author} {\bibfnamefont {D.}~\bibnamefont {van Dyk}},\ }\href
  {\doibase 10.1103/PhysRevD.87.034016} {\bibfield  {journal} {\bibinfo
  {journal} {Phys.Rev.}\ }\textbf {\bibinfo {volume} {D87}},\ \bibinfo {pages}
  {034016} (\bibinfo {year} {2013})},\ \Eprint {http://arxiv.org/abs/1212.2321}
  {arXiv:1212.2321 [hep-ph]} \BibitemShut {NoStop}%
%%CITATION = ARXIV:1212.2321;%%
\bibitem [{\citenamefont {Hambrock}\ and\ \citenamefont
  {Hiller}(2012)}]{Hambrock:2012dg}%
  \BibitemOpen
  \bibfield  {author} {\bibinfo {author} {\bibfnamefont {C.}~\bibnamefont
  {Hambrock}}\ and\ \bibinfo {author} {\bibfnamefont {G.}~\bibnamefont
  {Hiller}},\ }\href {\doibase 10.1103/PhysRevLett.109.091802} {\bibfield
  {journal} {\bibinfo  {journal} {Phys.Rev.Lett.}\ }\textbf {\bibinfo {volume}
  {109}},\ \bibinfo {pages} {091802} (\bibinfo {year} {2012})},\ \Eprint
  {http://arxiv.org/abs/1204.4444} {arXiv:1204.4444 [hep-ph]} \BibitemShut
  {NoStop}%
%%CITATION = ARXIV:1204.4444;%%
\bibitem [{\citenamefont {Beaujean}\ \emph {et~al.}(2012)\citenamefont
  {Beaujean}, \citenamefont {Bobeth}, \citenamefont {van Dyk},\ and\
  \citenamefont {Wacker}}]{Beaujean:2012uj}%
  \BibitemOpen
  \bibfield  {author} {\bibinfo {author} {\bibfnamefont {F.}~\bibnamefont
  {Beaujean}}, \bibinfo {author} {\bibfnamefont {C.}~\bibnamefont {Bobeth}},
  \bibinfo {author} {\bibfnamefont {D.}~\bibnamefont {van Dyk}}, \ and\
  \bibinfo {author} {\bibfnamefont {C.}~\bibnamefont {Wacker}},\ }\href
  {\doibase 10.1007/JHEP08(2012)030} {\bibfield  {journal} {\bibinfo  {journal}
  {JHEP}\ }\textbf {\bibinfo {volume} {1208}},\ \bibinfo {pages} {030}
  (\bibinfo {year} {2012})},\ \Eprint {http://arxiv.org/abs/1205.1838}
  {arXiv:1205.1838 [hep-ph]} \BibitemShut {NoStop}%
%%CITATION = ARXIV:1205.1838;%%
\bibitem [{\citenamefont {Hambrock}\ \emph {et~al.}(2013)\citenamefont
  {Hambrock}, \citenamefont {Hiller}, \citenamefont {Schacht},\ and\
  \citenamefont {Zwicky}}]{Hambrock:2013zya}%
  \BibitemOpen
  \bibfield  {author} {\bibinfo {author} {\bibfnamefont {C.}~\bibnamefont
  {Hambrock}}, \bibinfo {author} {\bibfnamefont {G.}~\bibnamefont {Hiller}},
  \bibinfo {author} {\bibfnamefont {S.}~\bibnamefont {Schacht}}, \ and\
  \bibinfo {author} {\bibfnamefont {R.}~\bibnamefont {Zwicky}},\ }\href@noop {}
  {\  (\bibinfo {year} {2013})},\ \Eprint {http://arxiv.org/abs/1308.4379}
  {arXiv:1308.4379 [hep-ph]} \BibitemShut {NoStop}%
%%CITATION = ARXIV:1308.4379;%%
\bibitem [{\citenamefont {Beneke}\ \emph {et~al.}(2001)\citenamefont {Beneke},
  \citenamefont {Feldmann},\ and\ \citenamefont {Seidel}}]{Beneke:2001at}%
  \BibitemOpen
  \bibfield  {author} {\bibinfo {author} {\bibfnamefont {M.}~\bibnamefont
  {Beneke}}, \bibinfo {author} {\bibfnamefont {T.}~\bibnamefont {Feldmann}}, \
  and\ \bibinfo {author} {\bibfnamefont {D.}~\bibnamefont {Seidel}},\ }\href
  {\doibase 10.1016/S0550-3213(01)00366-2} {\bibfield  {journal} {\bibinfo
  {journal} {Nucl.Phys.}\ }\textbf {\bibinfo {volume} {B612}},\ \bibinfo
  {pages} {25} (\bibinfo {year} {2001})},\ \Eprint
  {http://arxiv.org/abs/hep-ph/0106067} {arXiv:hep-ph/0106067 [hep-ph]}
  \BibitemShut {NoStop}%
%%CITATION = HEP-PH/0106067;%%
\bibitem [{\citenamefont {Beneke}\ \emph {et~al.}(2005)\citenamefont {Beneke},
  \citenamefont {Feldmann},\ and\ \citenamefont {Seidel}}]{Beneke:2004dp}%
  \BibitemOpen
  \bibfield  {author} {\bibinfo {author} {\bibfnamefont {M.}~\bibnamefont
  {Beneke}}, \bibinfo {author} {\bibfnamefont {T.}~\bibnamefont {Feldmann}}, \
  and\ \bibinfo {author} {\bibfnamefont {D.}~\bibnamefont {Seidel}},\ }\href
  {\doibase 10.1140/epjc/s2005-02181-5} {\bibfield  {journal} {\bibinfo
  {journal} {Eur.Phys.J.}\ }\textbf {\bibinfo {volume} {C41}},\ \bibinfo
  {pages} {173} (\bibinfo {year} {2005})},\ \Eprint
  {http://arxiv.org/abs/hep-ph/0412400} {arXiv:hep-ph/0412400 [hep-ph]}
  \BibitemShut {NoStop}%
%%CITATION = HEP-PH/0412400;%%
\bibitem [{\citenamefont {Ball}\ \emph {et~al.}(2007)\citenamefont {Ball},
  \citenamefont {Jones},\ and\ \citenamefont {Zwicky}}]{Ball:2006eu}%
  \BibitemOpen
  \bibfield  {author} {\bibinfo {author} {\bibfnamefont {P.}~\bibnamefont
  {Ball}}, \bibinfo {author} {\bibfnamefont {G.~W.}\ \bibnamefont {Jones}}, \
  and\ \bibinfo {author} {\bibfnamefont {R.}~\bibnamefont {Zwicky}},\ }\href
  {\doibase 10.1103/PhysRevD.75.054004} {\bibfield  {journal} {\bibinfo
  {journal} {Phys.Rev.}\ }\textbf {\bibinfo {volume} {D75}},\ \bibinfo {pages}
  {054004} (\bibinfo {year} {2007})},\ \Eprint
  {http://arxiv.org/abs/hep-ph/0612081} {arXiv:hep-ph/0612081 [hep-ph]}
  \BibitemShut {NoStop}%
%%CITATION = HEP-PH/0612081;%%
\bibitem [{\citenamefont {Khodjamirian}\ \emph {et~al.}(2010)\citenamefont
  {Khodjamirian}, \citenamefont {Mannel}, \citenamefont {Pivovarov},\ and\
  \citenamefont {Wang}}]{Khodjamirian:2010vf}%
  \BibitemOpen
  \bibfield  {author} {\bibinfo {author} {\bibfnamefont {A.}~\bibnamefont
  {Khodjamirian}}, \bibinfo {author} {\bibfnamefont {T.}~\bibnamefont
  {Mannel}}, \bibinfo {author} {\bibfnamefont {A.}~\bibnamefont {Pivovarov}}, \
  and\ \bibinfo {author} {\bibfnamefont {Y.-M.}\ \bibnamefont {Wang}},\ }\href
  {\doibase 10.1007/JHEP09(2010)089} {\bibfield  {journal} {\bibinfo  {journal}
  {JHEP}\ }\textbf {\bibinfo {volume} {1009}},\ \bibinfo {pages} {089}
  (\bibinfo {year} {2010})},\ \Eprint {http://arxiv.org/abs/1006.4945}
  {arXiv:1006.4945 [hep-ph]} \BibitemShut {NoStop}%
%%CITATION = ARXIV:1006.4945;%%
\bibitem [{\citenamefont {Khodjamirian}\ \emph {et~al.}(2013)\citenamefont
  {Khodjamirian}, \citenamefont {Mannel},\ and\ \citenamefont
  {Wang}}]{Khodjamirian:2012rm}%
  \BibitemOpen
  \bibfield  {author} {\bibinfo {author} {\bibfnamefont {A.}~\bibnamefont
  {Khodjamirian}}, \bibinfo {author} {\bibfnamefont {T.}~\bibnamefont
  {Mannel}}, \ and\ \bibinfo {author} {\bibfnamefont {Y.}~\bibnamefont
  {Wang}},\ }\href {\doibase 10.1007/JHEP02(2013)010} {\bibfield  {journal}
  {\bibinfo  {journal} {JHEP}\ }\textbf {\bibinfo {volume} {1302}},\ \bibinfo
  {pages} {010} (\bibinfo {year} {2013})},\ \Eprint
  {http://arxiv.org/abs/1211.0234} {arXiv:1211.0234 [hep-ph]} \BibitemShut
  {NoStop}%
%%CITATION = ARXIV:1211.0234;%%
\bibitem [{\citenamefont {Dimou}\ \emph {et~al.}(2013)\citenamefont {Dimou},
  \citenamefont {Lyon},\ and\ \citenamefont {Zwicky}}]{Dimou:2012un}%
  \BibitemOpen
  \bibfield  {author} {\bibinfo {author} {\bibfnamefont {M.}~\bibnamefont
  {Dimou}}, \bibinfo {author} {\bibfnamefont {J.}~\bibnamefont {Lyon}}, \ and\
  \bibinfo {author} {\bibfnamefont {R.}~\bibnamefont {Zwicky}},\ }\href
  {\doibase 10.1103/PhysRevD.87.074008} {\bibfield  {journal} {\bibinfo
  {journal} {Phys.Rev.}\ }\textbf {\bibinfo {volume} {D87}},\ \bibinfo {pages}
  {074008} (\bibinfo {year} {2013})},\ \Eprint {http://arxiv.org/abs/1212.2242}
  {arXiv:1212.2242 [hep-ph]} \BibitemShut {NoStop}%
%%CITATION = ARXIV:1212.2242;%%
\bibitem [{\citenamefont {J{\"a}ger}\ and\ \citenamefont
  {Martin~Camalich}(2013)}]{Jager:2012uw}%
  \BibitemOpen
  \bibfield  {author} {\bibinfo {author} {\bibfnamefont {S.}~\bibnamefont
  {J{\"a}ger}}\ and\ \bibinfo {author} {\bibfnamefont {J.}~\bibnamefont
  {Martin~Camalich}},\ }\href {\doibase 10.1007/JHEP05(2013)043} {\bibfield
  {journal} {\bibinfo  {journal} {JHEP}\ }\textbf {\bibinfo {volume} {1305}},\
  \bibinfo {pages} {043} (\bibinfo {year} {2013})},\ \Eprint
  {http://arxiv.org/abs/1212.2263} {arXiv:1212.2263 [hep-ph]} \BibitemShut
  {NoStop}%
%%CITATION = ARXIV:1212.2263;%%
\bibitem [{\citenamefont {Grinstein}\ and\ \citenamefont
  {Pirjol}(2004)}]{Grinstein:2004vb}%
  \BibitemOpen
  \bibfield  {author} {\bibinfo {author} {\bibfnamefont {B.}~\bibnamefont
  {Grinstein}}\ and\ \bibinfo {author} {\bibfnamefont {D.}~\bibnamefont
  {Pirjol}},\ }\href {\doibase 10.1103/PhysRevD.70.114005} {\bibfield
  {journal} {\bibinfo  {journal} {Phys.Rev.}\ }\textbf {\bibinfo {volume}
  {D70}},\ \bibinfo {pages} {114005} (\bibinfo {year} {2004})},\ \Eprint
  {http://arxiv.org/abs/hep-ph/0404250} {arXiv:hep-ph/0404250 [hep-ph]}
  \BibitemShut {NoStop}%
%%CITATION = HEP-PH/0404250;%%
\bibitem [{\citenamefont {Beylich}\ \emph {et~al.}(2011)\citenamefont
  {Beylich}, \citenamefont {Buchalla},\ and\ \citenamefont
  {Feldmann}}]{Beylich:2011aq}%
  \BibitemOpen
  \bibfield  {author} {\bibinfo {author} {\bibfnamefont {M.}~\bibnamefont
  {Beylich}}, \bibinfo {author} {\bibfnamefont {G.}~\bibnamefont {Buchalla}}, \
  and\ \bibinfo {author} {\bibfnamefont {T.}~\bibnamefont {Feldmann}},\ }\href
  {\doibase 10.1140/epjc/s10052-011-1635-0} {\bibfield  {journal} {\bibinfo
  {journal} {Eur.Phys.J.}\ }\textbf {\bibinfo {volume} {C71}},\ \bibinfo
  {pages} {1635} (\bibinfo {year} {2011})},\ \Eprint
  {http://arxiv.org/abs/1101.5118} {arXiv:1101.5118 [hep-ph]} \BibitemShut
  {NoStop}%
%%CITATION = ARXIV:1101.5118;%%
\bibitem [{\citenamefont {Descotes-Genon}\ \emph {et~al.}(2011)\citenamefont
  {Descotes-Genon}, \citenamefont {Ghosh}, \citenamefont {Matias},\ and\
  \citenamefont {Ramon}}]{DescotesGenon:2011yn}%
  \BibitemOpen
  \bibfield  {author} {\bibinfo {author} {\bibfnamefont {S.}~\bibnamefont
  {Descotes-Genon}}, \bibinfo {author} {\bibfnamefont {D.}~\bibnamefont
  {Ghosh}}, \bibinfo {author} {\bibfnamefont {J.}~\bibnamefont {Matias}}, \
  and\ \bibinfo {author} {\bibfnamefont {M.}~\bibnamefont {Ramon}},\ }\href
  {\doibase 10.1007/JHEP06(2011)099} {\bibfield  {journal} {\bibinfo  {journal}
  {JHEP}\ }\textbf {\bibinfo {volume} {1106}},\ \bibinfo {pages} {099}
  (\bibinfo {year} {2011})},\ \Eprint {http://arxiv.org/abs/1104.3342}
  {arXiv:1104.3342 [hep-ph]} \BibitemShut {NoStop}%
%%CITATION = ARXIV:1104.3342;%%
\bibitem [{\citenamefont {Altmannshofer}\ \emph {et~al.}(2012)\citenamefont
  {Altmannshofer}, \citenamefont {Paradisi},\ and\ \citenamefont
  {Straub}}]{Altmannshofer:2011gn}%
  \BibitemOpen
  \bibfield  {author} {\bibinfo {author} {\bibfnamefont {W.}~\bibnamefont
  {Altmannshofer}}, \bibinfo {author} {\bibfnamefont {P.}~\bibnamefont
  {Paradisi}}, \ and\ \bibinfo {author} {\bibfnamefont {D.~M.}\ \bibnamefont
  {Straub}},\ }\href {\doibase 10.1007/JHEP04(2012)008} {\bibfield  {journal}
  {\bibinfo  {journal} {JHEP}\ }\textbf {\bibinfo {volume} {1204}},\ \bibinfo
  {pages} {008} (\bibinfo {year} {2012})},\ \Eprint
  {http://arxiv.org/abs/1111.1257} {arXiv:1111.1257 [hep-ph]} \BibitemShut
  {NoStop}%
%%CITATION = ARXIV:1111.1257;%%
\bibitem [{\citenamefont {Bobeth}\ \emph {et~al.}(2012)\citenamefont {Bobeth},
  \citenamefont {Hiller}, \citenamefont {van Dyk},\ and\ \citenamefont
  {Wacker}}]{Bobeth:2011nj}%
  \BibitemOpen
  \bibfield  {author} {\bibinfo {author} {\bibfnamefont {C.}~\bibnamefont
  {Bobeth}}, \bibinfo {author} {\bibfnamefont {G.}~\bibnamefont {Hiller}},
  \bibinfo {author} {\bibfnamefont {D.}~\bibnamefont {van Dyk}}, \ and\
  \bibinfo {author} {\bibfnamefont {C.}~\bibnamefont {Wacker}},\ }\href
  {\doibase 10.1007/JHEP01(2012)107} {\bibfield  {journal} {\bibinfo  {journal}
  {JHEP}\ }\textbf {\bibinfo {volume} {1201}},\ \bibinfo {pages} {107}
  (\bibinfo {year} {2012})},\ \Eprint {http://arxiv.org/abs/1111.2558}
  {arXiv:1111.2558 [hep-ph]} \BibitemShut {NoStop}%
%%CITATION = ARXIV:1111.2558;%%
\bibitem [{\citenamefont {Altmannshofer}\ and\ \citenamefont
  {Straub}(2012)}]{Altmannshofer:2012az}%
  \BibitemOpen
  \bibfield  {author} {\bibinfo {author} {\bibfnamefont {W.}~\bibnamefont
  {Altmannshofer}}\ and\ \bibinfo {author} {\bibfnamefont {D.~M.}\ \bibnamefont
  {Straub}},\ }\href {\doibase 10.1007/JHEP08(2012)121} {\bibfield  {journal}
  {\bibinfo  {journal} {JHEP}\ }\textbf {\bibinfo {volume} {1208}},\ \bibinfo
  {pages} {121} (\bibinfo {year} {2012})},\ \Eprint
  {http://arxiv.org/abs/1206.0273} {arXiv:1206.0273 [hep-ph]} \BibitemShut
  {NoStop}%
%%CITATION = ARXIV:1206.0273;%%
\bibitem [{\citenamefont {Aaij}\ \emph
  {et~al.}(2013{\natexlab{e}})\citenamefont {Aaij} \emph
  {et~al.}}]{LHCb:2012kz}%
  \BibitemOpen
  \bibfield  {author} {\bibinfo {author} {\bibfnamefont {R.}~\bibnamefont
  {Aaij}} \emph {et~al.} (\bibinfo {collaboration} {LHCb Collaboration}),\
  }\href {\doibase 10.1103/PhysRevLett.110.031801} {\bibfield  {journal}
  {\bibinfo  {journal} {Phys.Rev.Lett.}\ }\textbf {\bibinfo {volume} {110}},\
  \bibinfo {pages} {031801} (\bibinfo {year} {2013}{\natexlab{e}})},\ \Eprint
  {http://arxiv.org/abs/1210.4492} {arXiv:1210.4492 [hep-ex]} \BibitemShut
  {NoStop}%
%%CITATION = ARXIV:1210.4492;%%
\bibitem [{\citenamefont {Aaij}\ \emph
  {et~al.}(2013{\natexlab{f}})\citenamefont {Aaij} \emph
  {et~al.}}]{Aaij:2013dgw}%
  \BibitemOpen
  \bibfield  {author} {\bibinfo {author} {\bibfnamefont {R.}~\bibnamefont
  {Aaij}} \emph {et~al.} (\bibinfo {collaboration} {LHCb collaboration}),\
  }\href {\doibase 10.1103/PhysRevLett.111.151801} {\bibfield  {journal}
  {\bibinfo  {journal} {Phys.Rev.Lett.}\ }\textbf {\bibinfo {volume} {111}},\
  \bibinfo {pages} {151801} (\bibinfo {year} {2013}{\natexlab{f}})},\ \Eprint
  {http://arxiv.org/abs/1308.1340} {arXiv:1308.1340 [hep-ex]} \BibitemShut
  {NoStop}%
%%CITATION = ARXIV:1308.1340;%%
\bibitem [{\citenamefont {Descotes-Genon}\ \emph
  {et~al.}(2013{\natexlab{b}})\citenamefont {Descotes-Genon}, \citenamefont
  {Matias},\ and\ \citenamefont {Virto}}]{Descotes-Genon:2013wba}%
  \BibitemOpen
  \bibfield  {author} {\bibinfo {author} {\bibfnamefont {S.}~\bibnamefont
  {Descotes-Genon}}, \bibinfo {author} {\bibfnamefont {J.}~\bibnamefont
  {Matias}}, \ and\ \bibinfo {author} {\bibfnamefont {J.}~\bibnamefont
  {Virto}},\ }\href {\doibase 10.1103/PhysRevD.88.074002} {\bibfield  {journal}
  {\bibinfo  {journal} {Phys.Rev.}\ }\textbf {\bibinfo {volume} {D88}},\
  \bibinfo {pages} {074002} (\bibinfo {year} {2013}{\natexlab{b}})},\ \Eprint
  {http://arxiv.org/abs/1307.5683} {arXiv:1307.5683 [hep-ph]} \BibitemShut
  {NoStop}%
%%CITATION = ARXIV:1307.5683;%%
\bibitem [{\citenamefont {Altmannshofer}\ and\ \citenamefont
  {Straub}(2013)}]{Altmannshofer:2013foa}%
  \BibitemOpen
  \bibfield  {author} {\bibinfo {author} {\bibfnamefont {W.}~\bibnamefont
  {Altmannshofer}}\ and\ \bibinfo {author} {\bibfnamefont {D.~M.}\ \bibnamefont
  {Straub}},\ }\href {\doibase 10.1140/epjc/s10052-013-2646-9} {\bibfield
  {journal} {\bibinfo  {journal} {Eur.Phys.J.}\ }\textbf {\bibinfo {volume}
  {C73}},\ \bibinfo {pages} {2646} (\bibinfo {year} {2013})},\ \Eprint
  {http://arxiv.org/abs/1308.1501} {arXiv:1308.1501 [hep-ph]} \BibitemShut
  {NoStop}%
%%CITATION = ARXIV:1308.1501;%%
\bibitem [{\citenamefont {Horgan}\ \emph
  {et~al.}(2013{\natexlab{a}})\citenamefont {Horgan}, \citenamefont {Liu},
  \citenamefont {Meinel},\ and\ \citenamefont {Wingate}}]{Horgan:2013pva}%
  \BibitemOpen
  \bibfield  {author} {\bibinfo {author} {\bibfnamefont {R.~R.}\ \bibnamefont
  {Horgan}}, \bibinfo {author} {\bibfnamefont {Z.}~\bibnamefont {Liu}},
  \bibinfo {author} {\bibfnamefont {S.}~\bibnamefont {Meinel}}, \ and\ \bibinfo
  {author} {\bibfnamefont {M.}~\bibnamefont {Wingate}},\ }\href@noop {} {\
  (\bibinfo {year} {2013}{\natexlab{a}})},\ \Eprint
  {http://arxiv.org/abs/1310.3887} {arXiv:1310.3887 [hep-ph]} \BibitemShut
  {NoStop}%
%%CITATION = ARXIV:1310.3887;%%
\bibitem [{\citenamefont {Bouchard}\ \emph {et~al.}(2013)\citenamefont
  {Bouchard}, \citenamefont {Lepage}, \citenamefont {Monahan}, \citenamefont
  {Na},\ and\ \citenamefont {Shigemitsu}}]{Bouchard:2013eph}%
  \BibitemOpen
  \bibfield  {author} {\bibinfo {author} {\bibfnamefont {C.}~\bibnamefont
  {Bouchard}}, \bibinfo {author} {\bibfnamefont {G.~P.}\ \bibnamefont
  {Lepage}}, \bibinfo {author} {\bibfnamefont {C.}~\bibnamefont {Monahan}},
  \bibinfo {author} {\bibfnamefont {H.}~\bibnamefont {Na}}, \ and\ \bibinfo
  {author} {\bibfnamefont {J.}~\bibnamefont {Shigemitsu}},\ }\href {\doibase
  10.1103/PhysRevD.88.054509} {\bibfield  {journal} {\bibinfo  {journal}
  {Phys.Rev.}\ }\textbf {\bibinfo {volume} {D88}},\ \bibinfo {pages} {054509}
  (\bibinfo {year} {2013})},\ \Eprint {http://arxiv.org/abs/1306.2384}
  {arXiv:1306.2384 [hep-lat]} \BibitemShut {NoStop}%
%%CITATION = ARXIV:1306.2384;%%
\bibitem [{\citenamefont {Horgan}\ \emph
  {et~al.}(2013{\natexlab{b}})\citenamefont {Horgan}, \citenamefont {Liu},
  \citenamefont {Meinel},\ and\ \citenamefont {Wingate}}]{Horgan:2013hoa}%
  \BibitemOpen
  \bibfield  {author} {\bibinfo {author} {\bibfnamefont {R.~R.}\ \bibnamefont
  {Horgan}}, \bibinfo {author} {\bibfnamefont {Z.}~\bibnamefont {Liu}},
  \bibinfo {author} {\bibfnamefont {S.}~\bibnamefont {Meinel}}, \ and\ \bibinfo
  {author} {\bibfnamefont {M.}~\bibnamefont {Wingate}},\ }\href@noop {} {\
  (\bibinfo {year} {2013}{\natexlab{b}})},\ \Eprint
  {http://arxiv.org/abs/1310.3722} {arXiv:1310.3722 [hep-lat]} \BibitemShut
  {NoStop}%
%%CITATION = ARXIV:1310.3722;%%
\bibitem [{\citenamefont {Bobeth}\ \emph {et~al.}(2004)\citenamefont {Bobeth},
  \citenamefont {Gambino}, \citenamefont {Gorbahn},\ and\ \citenamefont
  {Haisch}}]{Bobeth:2003at}%
  \BibitemOpen
  \bibfield  {author} {\bibinfo {author} {\bibfnamefont {C.}~\bibnamefont
  {Bobeth}}, \bibinfo {author} {\bibfnamefont {P.}~\bibnamefont {Gambino}},
  \bibinfo {author} {\bibfnamefont {M.}~\bibnamefont {Gorbahn}}, \ and\
  \bibinfo {author} {\bibfnamefont {U.}~\bibnamefont {Haisch}},\ }\href
  {\doibase 10.1088/1126-6708/2004/04/071} {\bibfield  {journal} {\bibinfo
  {journal} {JHEP}\ }\textbf {\bibinfo {volume} {0404}},\ \bibinfo {pages}
  {071} (\bibinfo {year} {2004})},\ \Eprint
  {http://arxiv.org/abs/hep-ph/0312090} {arXiv:hep-ph/0312090 [hep-ph]}
  \BibitemShut {NoStop}%
%%CITATION = HEP-PH/0312090;%%
\bibitem [{\citenamefont {Huber}\ \emph {et~al.}(2006)\citenamefont {Huber},
  \citenamefont {Lunghi}, \citenamefont {Misiak},\ and\ \citenamefont
  {Wyler}}]{Huber:2005ig}%
  \BibitemOpen
  \bibfield  {author} {\bibinfo {author} {\bibfnamefont {T.}~\bibnamefont
  {Huber}}, \bibinfo {author} {\bibfnamefont {E.}~\bibnamefont {Lunghi}},
  \bibinfo {author} {\bibfnamefont {M.}~\bibnamefont {Misiak}}, \ and\ \bibinfo
  {author} {\bibfnamefont {D.}~\bibnamefont {Wyler}},\ }\href {\doibase
  10.1016/j.nuclphysb.2006.01.037} {\bibfield  {journal} {\bibinfo  {journal}
  {Nucl.Phys.}\ }\textbf {\bibinfo {volume} {B740}},\ \bibinfo {pages} {105}
  (\bibinfo {year} {2006})},\ \Eprint {http://arxiv.org/abs/hep-ph/0512066}
  {arXiv:hep-ph/0512066 [hep-ph]} \BibitemShut {NoStop}%
%%CITATION = HEP-PH/0512066;%%
\bibitem [{\citenamefont {Beringer}\ \emph {et~al.}(2012)\citenamefont
  {Beringer} \emph {et~al.}}]{Beringer:1900zz}%
  \BibitemOpen
  \bibfield  {author} {\bibinfo {author} {\bibfnamefont {J.}~\bibnamefont
  {Beringer}} \emph {et~al.} (\bibinfo {collaboration} {Particle Data Group}),\
  }\href {\doibase 10.1103/PhysRevD.86.010001} {\bibfield  {journal} {\bibinfo
  {journal} {Phys.Rev.}\ }\textbf {\bibinfo {volume} {D86}},\ \bibinfo {pages}
  {010001} (\bibinfo {year} {2012})}\BibitemShut {NoStop}%
%%CITATION = PHRVA,D86,010001;%%
\bibitem [{\citenamefont {Aaltonen}\ \emph {et~al.}(2012)\citenamefont
  {Aaltonen} \emph {et~al.}}]{Aaltonen:2011ja}%
  \BibitemOpen
  \bibfield  {author} {\bibinfo {author} {\bibfnamefont {T.}~\bibnamefont
  {Aaltonen}} \emph {et~al.} (\bibinfo {collaboration} {CDF Collaboration}),\
  }\href {\doibase 10.1103/PhysRevLett.108.081807} {\bibfield  {journal}
  {\bibinfo  {journal} {Phys.Rev.Lett.}\ }\textbf {\bibinfo {volume} {108}},\
  \bibinfo {pages} {081807} (\bibinfo {year} {2012})},\ \Eprint
  {http://arxiv.org/abs/1108.0695} {arXiv:1108.0695 [hep-ex]} \BibitemShut
  {NoStop}%
%%CITATION = ARXIV:1108.0695;%%
\bibitem [{\citenamefont {Altmannshofer}\ \emph {et~al.}(2009)\citenamefont
  {Altmannshofer}, \citenamefont {Ball}, \citenamefont {Bharucha},
  \citenamefont {Buras}, \citenamefont {Straub} \emph
  {et~al.}}]{Altmannshofer:2008dz}%
  \BibitemOpen
  \bibfield  {author} {\bibinfo {author} {\bibfnamefont {W.}~\bibnamefont
  {Altmannshofer}}, \bibinfo {author} {\bibfnamefont {P.}~\bibnamefont {Ball}},
  \bibinfo {author} {\bibfnamefont {A.}~\bibnamefont {Bharucha}}, \bibinfo
  {author} {\bibfnamefont {A.~J.}\ \bibnamefont {Buras}}, \bibinfo {author}
  {\bibfnamefont {D.~M.}\ \bibnamefont {Straub}},  \emph {et~al.},\ }\href
  {\doibase 10.1088/1126-6708/2009/01/019} {\bibfield  {journal} {\bibinfo
  {journal} {JHEP}\ }\textbf {\bibinfo {volume} {0901}},\ \bibinfo {pages}
  {019} (\bibinfo {year} {2009})},\ \Eprint {http://arxiv.org/abs/0811.1214}
  {arXiv:0811.1214 [hep-ph]} \BibitemShut {NoStop}%
%%CITATION = ARXIV:0811.1214;%%
\bibitem [{\citenamefont {Aaltonen}\ \emph {et~al.}(2011)\citenamefont
  {Aaltonen} \emph {et~al.}}]{Aaltonen:2011qs}%
  \BibitemOpen
  \bibfield  {author} {\bibinfo {author} {\bibfnamefont {T.}~\bibnamefont
  {Aaltonen}} \emph {et~al.} (\bibinfo {collaboration} {CDF Collaboration}),\
  }\href {\doibase 10.1103/PhysRevLett.107.201802} {\bibfield  {journal}
  {\bibinfo  {journal} {Phys.Rev.Lett.}\ }\textbf {\bibinfo {volume} {107}},\
  \bibinfo {pages} {201802} (\bibinfo {year} {2011})},\ \Eprint
  {http://arxiv.org/abs/1107.3753} {arXiv:1107.3753 [hep-ex]} \BibitemShut
  {NoStop}%
%%CITATION = ARXIV:1107.3753;%%
\bibitem [{\citenamefont {Aaij}\ \emph
  {et~al.}(2013{\natexlab{g}})\citenamefont {Aaij} \emph
  {et~al.}}]{Aaij:2013pta}%
  \BibitemOpen
  \bibfield  {author} {\bibinfo {author} {\bibfnamefont {R.}~\bibnamefont
  {Aaij}} \emph {et~al.} (\bibinfo {collaboration} {LHCb collaboration}),\
  }\href {\doibase 10.1103/PhysRevLett.111.112003} {\bibfield  {journal}
  {\bibinfo  {journal} {Phys. Rev. Lett. 111,}\ }\textbf {\bibinfo {volume}
  {112003}} (\bibinfo {year} {2013}{\natexlab{g}}),\
  10.1103/PhysRevLett.111.112003},\ \Eprint {http://arxiv.org/abs/1307.7595}
  {arXiv:1307.7595 [hep-ex]} \BibitemShut {NoStop}%
%%CITATION = ARXIV:1307.7595;%%
\bibitem [{\citenamefont {{[LHCb Collaboration]}}(2012)}]{LHCb:2012:BKstarll}%
  \BibitemOpen
  \bibfield  {author} {\bibinfo {author} {\bibnamefont {{[LHCb
  Collaboration]}}},\ }\href@noop {} {\  (\bibinfo {year} {2012})},\ \Eprint
  {http://arxiv.org/abs/LHCb-CONF-2012-008} {LHCb-CONF-2012-008} \BibitemShut
  {NoStop}%
\bibitem [{\citenamefont {Poireau}(2012)}]{Poireau:2012by}%
  \BibitemOpen
  \bibfield  {author} {\bibinfo {author} {\bibfnamefont {V.}~\bibnamefont
  {Poireau}} (\bibinfo {collaboration} {BaBar collaboration}),\ }\href@noop {}
  {\  (\bibinfo {year} {2012})},\ \Eprint {http://arxiv.org/abs/1205.2201}
  {arXiv:1205.2201 [hep-ex]} \BibitemShut {NoStop}%
%%CITATION = ARXIV:1205.2201;%%
\bibitem [{\citenamefont {van Dyk}\ \emph {et~al.}()\citenamefont {van Dyk}
  \emph {et~al.}}]{EOS}%
  \BibitemOpen
  \bibfield  {author} {\bibinfo {author} {\bibfnamefont {D.}~\bibnamefont {van
  Dyk}} \emph {et~al.},\ }\href@noop {} {\emph {\bibinfo {title} {{EOS -- A HEP
  Program for Flavor Physics}}}},\ \bibinfo {note}
  {\url{http://project.het.physik.tu-dortmund.de/eos}}\BibitemShut {NoStop}%
\bibitem [{\citenamefont {Beaujean}\ and\ \citenamefont
  {Caldwell}(2013)}]{Beaujean:2013:PMC}%
  \BibitemOpen
  \bibfield  {author} {\bibinfo {author} {\bibfnamefont {F.}~\bibnamefont
  {Beaujean}}\ and\ \bibinfo {author} {\bibfnamefont {A.}~\bibnamefont
  {Caldwell}},\ }\href@noop {} {\  (\bibinfo {year} {2013})},\ \Eprint
  {http://arxiv.org/abs/1304.7808} {arXiv:1304.7808 [stat.CO]} \BibitemShut
  {NoStop}%
\bibitem [{\citenamefont {Powell}(2009)}]{Powell:2009}%
  \BibitemOpen
  \bibfield  {author} {\bibinfo {author} {\bibfnamefont {M.~J.}\ \bibnamefont
  {Powell}},\ }\href@noop {} {\bibfield  {journal} {\bibinfo  {journal}
  {Cambridge NA Report NA2009/06, University of Cambridge, Cambridge}\ }
  (\bibinfo {year} {2009})}\BibitemShut {NoStop}%
\bibitem [{\citenamefont {Powell}(1994)}]{Powell:1994}%
  \BibitemOpen
  \bibfield  {author} {\bibinfo {author} {\bibfnamefont {M.~J.}\ \bibnamefont
  {Powell}},\ }in\ \href@noop {} {\emph {\bibinfo {booktitle} {Advances in
  optimization and numerical analysis}}}\ (\bibinfo  {publisher} {Springer},\
  \bibinfo {year} {1994})\ pp.\ \bibinfo {pages} {51--67}\BibitemShut {NoStop}%
\bibitem [{\citenamefont {Johnson}()}]{NLopt}%
  \BibitemOpen
  \bibfield  {author} {\bibinfo {author} {\bibfnamefont {S.~G.}\ \bibnamefont
  {Johnson}},\ }\href@noop {} {\emph {\bibinfo {title} {{The NLopt
  nonlinear-optimization package}}}},\ \bibinfo {note}
  {\url{http://ab-initio.mit.edu/nlopt}}\BibitemShut {NoStop}%
\bibitem [{\citenamefont {Datta}\ \emph {et~al.}(1999)\citenamefont {Datta},
  \citenamefont {O'Donnell}, \citenamefont {Pakvasa},\ and\ \citenamefont
  {Zhang}}]{Datta:1998yq}%
  \BibitemOpen
  \bibfield  {author} {\bibinfo {author} {\bibfnamefont {A.}~\bibnamefont
  {Datta}}, \bibinfo {author} {\bibfnamefont {P.}~\bibnamefont {O'Donnell}},
  \bibinfo {author} {\bibfnamefont {S.}~\bibnamefont {Pakvasa}}, \ and\
  \bibinfo {author} {\bibfnamefont {X.}~\bibnamefont {Zhang}},\ }\href
  {\doibase 10.1103/PhysRevD.60.014011} {\bibfield  {journal} {\bibinfo
  {journal} {Phys.Rev.}\ }\textbf {\bibinfo {volume} {D60}},\ \bibinfo {pages}
  {014011} (\bibinfo {year} {1999})},\ \Eprint
  {http://arxiv.org/abs/hep-ph/9812325} {arXiv:hep-ph/9812325 [hep-ph]}
  \BibitemShut {NoStop}%
%%CITATION = HEP-PH/9812325;%%
\bibitem [{\citenamefont {Borzumati}\ \emph {et~al.}(2000)\citenamefont
  {Borzumati}, \citenamefont {Greub}, \citenamefont {Hurth},\ and\
  \citenamefont {Wyler}}]{Borzumati:1999qt}%
  \BibitemOpen
  \bibfield  {author} {\bibinfo {author} {\bibfnamefont {F.}~\bibnamefont
  {Borzumati}}, \bibinfo {author} {\bibfnamefont {C.}~\bibnamefont {Greub}},
  \bibinfo {author} {\bibfnamefont {T.}~\bibnamefont {Hurth}}, \ and\ \bibinfo
  {author} {\bibfnamefont {D.}~\bibnamefont {Wyler}},\ }\href {\doibase
  10.1103/PhysRevD.62.075005} {\bibfield  {journal} {\bibinfo  {journal}
  {Phys.Rev.}\ }\textbf {\bibinfo {volume} {D62}},\ \bibinfo {pages} {075005}
  (\bibinfo {year} {2000})},\ \Eprint {http://arxiv.org/abs/hep-ph/9911245}
  {arXiv:hep-ph/9911245 [hep-ph]} \BibitemShut {NoStop}%
%%CITATION = HEP-PH/9911245;%%
\bibitem [{\citenamefont {Hiller}\ and\ \citenamefont
  {Kr{\"u}ger}(2004)}]{Hiller:2003js}%
  \BibitemOpen
  \bibfield  {author} {\bibinfo {author} {\bibfnamefont {G.}~\bibnamefont
  {Hiller}}\ and\ \bibinfo {author} {\bibfnamefont {F.}~\bibnamefont
  {Kr{\"u}ger}},\ }\href {\doibase 10.1103/PhysRevD.69.074020} {\bibfield
  {journal} {\bibinfo  {journal} {Phys.Rev.}\ }\textbf {\bibinfo {volume}
  {D69}},\ \bibinfo {pages} {074020} (\bibinfo {year} {2004})},\ \Eprint
  {http://arxiv.org/abs/hep-ph/0310219} {arXiv:hep-ph/0310219 [hep-ph]}
  \BibitemShut {NoStop}%
%%CITATION = HEP-PH/0310219;%%
\bibitem [{\citenamefont {Bobeth}\ and\ \citenamefont
  {Haisch}(2013)}]{Bobeth:2011st}%
  \BibitemOpen
  \bibfield  {author} {\bibinfo {author} {\bibfnamefont {C.}~\bibnamefont
  {Bobeth}}\ and\ \bibinfo {author} {\bibfnamefont {U.}~\bibnamefont
  {Haisch}},\ }\href {\doibase 10.5506/APhysPolB.44.127} {\bibfield  {journal}
  {\bibinfo  {journal} {Acta Phys.Polon.}\ }\textbf {\bibinfo {volume} {B44}},\
  \bibinfo {pages} {127} (\bibinfo {year} {2013})},\ \Eprint
  {http://arxiv.org/abs/1109.1826} {arXiv:1109.1826 [hep-ph]} \BibitemShut
  {NoStop}%
%%CITATION = ARXIV:1109.1826;%%
\bibitem [{\citenamefont {Datta}\ \emph {et~al.}(2013)\citenamefont {Datta},
  \citenamefont {Duraisamy},\ and\ \citenamefont {Ghosh}}]{Datta:2013kja}%
  \BibitemOpen
  \bibfield  {author} {\bibinfo {author} {\bibfnamefont {A.}~\bibnamefont
  {Datta}}, \bibinfo {author} {\bibfnamefont {M.}~\bibnamefont {Duraisamy}}, \
  and\ \bibinfo {author} {\bibfnamefont {D.}~\bibnamefont {Ghosh}},\
  }\href@noop {} {\  (\bibinfo {year} {2013})},\ \Eprint
  {http://arxiv.org/abs/1310.1937} {arXiv:1310.1937 [hep-ph]} \BibitemShut
  {NoStop}%
%%CITATION = ARXIV:1310.1937;%%
\bibitem [{\citenamefont {Beneke}\ \emph {et~al.}(2009)\citenamefont {Beneke},
  \citenamefont {Li},\ and\ \citenamefont {Vernazza}}]{Beneke:2009eb}%
  \BibitemOpen
  \bibfield  {author} {\bibinfo {author} {\bibfnamefont {M.}~\bibnamefont
  {Beneke}}, \bibinfo {author} {\bibfnamefont {X.-Q.}\ \bibnamefont {Li}}, \
  and\ \bibinfo {author} {\bibfnamefont {L.}~\bibnamefont {Vernazza}},\ }\href
  {\doibase 10.1140/epjc/s10052-009-0989-z} {\bibfield  {journal} {\bibinfo
  {journal} {Eur.Phys.J.}\ }\textbf {\bibinfo {volume} {C61}},\ \bibinfo
  {pages} {429} (\bibinfo {year} {2009})},\ \Eprint
  {http://arxiv.org/abs/0901.4841} {arXiv:0901.4841 [hep-ph]} \BibitemShut
  {NoStop}%
%%CITATION = ARXIV:0901.4841;%%
\bibitem [{\citenamefont {Gauld}\ \emph
  {et~al.}(2014{\natexlab{a}})\citenamefont {Gauld}, \citenamefont {Goertz},\
  and\ \citenamefont {Haisch}}]{Gauld:2013qba}%
  \BibitemOpen
  \bibfield  {author} {\bibinfo {author} {\bibfnamefont {R.}~\bibnamefont
  {Gauld}}, \bibinfo {author} {\bibfnamefont {F.}~\bibnamefont {Goertz}}, \
  and\ \bibinfo {author} {\bibfnamefont {U.}~\bibnamefont {Haisch}},\ }\href
  {\doibase 10.1103/PhysRevD.89.015005} {\bibfield  {journal} {\bibinfo
  {journal} {Phys.Rev.}\ }\textbf {\bibinfo {volume} {D89}},\ \bibinfo {pages}
  {015005} (\bibinfo {year} {2014}{\natexlab{a}})},\ \Eprint
  {http://arxiv.org/abs/1308.1959} {arXiv:1308.1959 [hep-ph]} \BibitemShut
  {NoStop}%
%%CITATION = ARXIV:1308.1959;%%
\bibitem [{\citenamefont {Buras}\ and\ \citenamefont
  {Girrbach}(2013)}]{Buras:2013qja}%
  \BibitemOpen
  \bibfield  {author} {\bibinfo {author} {\bibfnamefont {A.~J.}\ \bibnamefont
  {Buras}}\ and\ \bibinfo {author} {\bibfnamefont {J.}~\bibnamefont
  {Girrbach}},\ }\href {\doibase 10.1007/JHEP12(2013)009} {\bibfield  {journal}
  {\bibinfo  {journal} {JHEP}\ }\textbf {\bibinfo {volume} {1312}},\ \bibinfo
  {pages} {009} (\bibinfo {year} {2013})},\ \Eprint
  {http://arxiv.org/abs/1309.2466} {arXiv:1309.2466 [hep-ph]} \BibitemShut
  {NoStop}%
%%CITATION = ARXIV:1309.2466;%%
\bibitem [{\citenamefont {Gauld}\ \emph
  {et~al.}(2014{\natexlab{b}})\citenamefont {Gauld}, \citenamefont {Goertz},\
  and\ \citenamefont {Haisch}}]{Gauld:2013qja}%
  \BibitemOpen
  \bibfield  {author} {\bibinfo {author} {\bibfnamefont {R.}~\bibnamefont
  {Gauld}}, \bibinfo {author} {\bibfnamefont {F.}~\bibnamefont {Goertz}}, \
  and\ \bibinfo {author} {\bibfnamefont {U.}~\bibnamefont {Haisch}},\ }\href
  {\doibase 10.1007/JHEP01(2014)069} {\bibfield  {journal} {\bibinfo  {journal}
  {JHEP}\ }\textbf {\bibinfo {volume} {1401}},\ \bibinfo {pages} {069}
  (\bibinfo {year} {2014}{\natexlab{b}})},\ \Eprint
  {http://arxiv.org/abs/1310.1082} {arXiv:1310.1082 [hep-ph]} \BibitemShut
  {NoStop}%
%%CITATION = ARXIV:1310.1082;%%
\bibitem [{\citenamefont {Buras}\ \emph {et~al.}(2013)\citenamefont {Buras},
  \citenamefont {De~Fazio},\ and\ \citenamefont {Girrbach}}]{Buras:2013dea}%
  \BibitemOpen
  \bibfield  {author} {\bibinfo {author} {\bibfnamefont {A.~J.}\ \bibnamefont
  {Buras}}, \bibinfo {author} {\bibfnamefont {F.}~\bibnamefont {De~Fazio}}, \
  and\ \bibinfo {author} {\bibfnamefont {J.}~\bibnamefont {Girrbach}},\
  }\href@noop {} {\  (\bibinfo {year} {2013})},\ \Eprint
  {http://arxiv.org/abs/1311.6729} {arXiv:1311.6729 [hep-ph]} \BibitemShut
  {NoStop}%
%%CITATION = ARXIV:1311.6729;%%
\bibitem [{\citenamefont {Bona}\ \emph {et~al.}(2006)\citenamefont {Bona} \emph
  {et~al.}}]{Bona:2006ah}%
  \BibitemOpen
  \bibfield  {author} {\bibinfo {author} {\bibfnamefont {M.}~\bibnamefont
  {Bona}} \emph {et~al.} (\bibinfo {collaboration} {UTfit Collaboration}),\
  }\href {\doibase 10.1088/1126-6708/2006/10/081} {\bibfield  {journal}
  {\bibinfo  {journal} {JHEP}\ }\textbf {\bibinfo {volume} {0610}},\ \bibinfo
  {pages} {081} (\bibinfo {year} {2006})},\ \bibinfo {note} {we use the updated
  data from Winter 2013 (pre-Moriond 13)},\ \Eprint
  {http://arxiv.org/abs/hep-ph/0606167} {arXiv:hep-ph/0606167 [hep-ph]}
  \BibitemShut {NoStop}%
%%CITATION = HEP-PH/0606167;%%
\bibitem [{\citenamefont {Laiho}\ \emph {et~al.}(2010)\citenamefont {Laiho},
  \citenamefont {Lunghi},\ and\ \citenamefont {Van~de Water}}]{Laiho:2009eu}%
  \BibitemOpen
  \bibfield  {author} {\bibinfo {author} {\bibfnamefont {J.}~\bibnamefont
  {Laiho}}, \bibinfo {author} {\bibfnamefont {E.}~\bibnamefont {Lunghi}}, \
  and\ \bibinfo {author} {\bibfnamefont {R.~S.}\ \bibnamefont {Van~de Water}},\
  }\href {\doibase 10.1103/PhysRevD.81.034503} {\bibfield  {journal} {\bibinfo
  {journal} {Phys.Rev.}\ }\textbf {\bibinfo {volume} {D81}},\ \bibinfo {pages}
  {034503} (\bibinfo {year} {2010})},\ \bibinfo {note} {we use the update as
  presented on \url{http://latticeaverages.org} in June 2013.},\ \Eprint
  {http://arxiv.org/abs/0910.2928} {arXiv:0910.2928 [hep-ph]} \BibitemShut
  {NoStop}%
%%CITATION = ARXIV:0910.2928;%%
\bibitem [{\citenamefont {Amhis}\ \emph {et~al.}(2012)\citenamefont {Amhis}
  \emph {et~al.}}]{Amhis:2012bh}%
  \BibitemOpen
  \bibfield  {author} {\bibinfo {author} {\bibfnamefont {Y.}~\bibnamefont
  {Amhis}} \emph {et~al.} (\bibinfo {collaboration} {Heavy Flavor Averaging
  Group}),\ }\href@noop {} {\  (\bibinfo {year} {2012})},\ \Eprint
  {http://arxiv.org/abs/1207.1158} {arXiv:1207.1158 [hep-ex]} \BibitemShut
  {NoStop}%
%%CITATION = ARXIV:1207.1158;%%
\bibitem [{\citenamefont {Chetyrkin}\ \emph {et~al.}(1997)\citenamefont
  {Chetyrkin}, \citenamefont {Misiak},\ and\ \citenamefont
  {Munz}}]{Chetyrkin:1996vx}%
  \BibitemOpen
  \bibfield  {author} {\bibinfo {author} {\bibfnamefont {K.~G.}\ \bibnamefont
  {Chetyrkin}}, \bibinfo {author} {\bibfnamefont {M.}~\bibnamefont {Misiak}}, \
  and\ \bibinfo {author} {\bibfnamefont {M.}~\bibnamefont {Munz}},\ }\href
  {\doibase 10.1016/S0370-2693(97)00324-9} {\bibfield  {journal} {\bibinfo
  {journal} {Phys.Lett.}\ }\textbf {\bibinfo {volume} {B400}},\ \bibinfo
  {pages} {206} (\bibinfo {year} {1997})},\ \Eprint
  {http://arxiv.org/abs/hep-ph/9612313} {arXiv:hep-ph/9612313 [hep-ph]}
  \BibitemShut {NoStop}%
%%CITATION = HEP-PH/9612313;%%
\bibitem [{\citenamefont {Buras}\ \emph {et~al.}(2002)\citenamefont {Buras},
  \citenamefont {Czarnecki}, \citenamefont {Misiak},\ and\ \citenamefont
  {Urban}}]{Buras:2002tp}%
  \BibitemOpen
  \bibfield  {author} {\bibinfo {author} {\bibfnamefont {A.~J.}\ \bibnamefont
  {Buras}}, \bibinfo {author} {\bibfnamefont {A.}~\bibnamefont {Czarnecki}},
  \bibinfo {author} {\bibfnamefont {M.}~\bibnamefont {Misiak}}, \ and\ \bibinfo
  {author} {\bibfnamefont {J.}~\bibnamefont {Urban}},\ }\href {\doibase
  10.1016/S0550-3213(02)00261-4} {\bibfield  {journal} {\bibinfo  {journal}
  {Nucl.Phys.}\ }\textbf {\bibinfo {volume} {B631}},\ \bibinfo {pages} {219}
  (\bibinfo {year} {2002})},\ \Eprint {http://arxiv.org/abs/hep-ph/0203135}
  {arXiv:hep-ph/0203135 [hep-ph]} \BibitemShut {NoStop}%
%%CITATION = HEP-PH/0203135;%%
\bibitem [{\citenamefont {Ewerth}\ \emph {et~al.}(2010)\citenamefont {Ewerth},
  \citenamefont {Gambino},\ and\ \citenamefont {Nandi}}]{Ewerth:2009yr}%
  \BibitemOpen
  \bibfield  {author} {\bibinfo {author} {\bibfnamefont {T.}~\bibnamefont
  {Ewerth}}, \bibinfo {author} {\bibfnamefont {P.}~\bibnamefont {Gambino}}, \
  and\ \bibinfo {author} {\bibfnamefont {S.}~\bibnamefont {Nandi}},\ }\href
  {\doibase 10.1016/j.nuclphysb.2009.12.035} {\bibfield  {journal} {\bibinfo
  {journal} {Nucl.Phys.}\ }\textbf {\bibinfo {volume} {B830}},\ \bibinfo
  {pages} {278} (\bibinfo {year} {2010})},\ \Eprint
  {http://arxiv.org/abs/0911.2175} {arXiv:0911.2175 [hep-ph]} \BibitemShut
  {NoStop}%
%%CITATION = ARXIV:0911.2175;%%
\bibitem [{\citenamefont {Melnikov}\ and\ \citenamefont
  {Ritbergen}(2000)}]{Melnikov:2000qh}%
  \BibitemOpen
  \bibfield  {author} {\bibinfo {author} {\bibfnamefont {K.}~\bibnamefont
  {Melnikov}}\ and\ \bibinfo {author} {\bibfnamefont {T.~v.}\ \bibnamefont
  {Ritbergen}},\ }\href {\doibase 10.1016/S0370-2693(00)00507-4} {\bibfield
  {journal} {\bibinfo  {journal} {Phys.Lett.}\ }\textbf {\bibinfo {volume}
  {B482}},\ \bibinfo {pages} {99} (\bibinfo {year} {2000})},\ \Eprint
  {http://arxiv.org/abs/hep-ph/9912391} {arXiv:hep-ph/9912391 [hep-ph]}
  \BibitemShut {NoStop}%
%%CITATION = HEP-PH/9912391;%%
\bibitem [{\citenamefont {Uraltsev}(2002)}]{Uraltsev:2001ih}%
  \BibitemOpen
  \bibfield  {author} {\bibinfo {author} {\bibfnamefont {N.}~\bibnamefont
  {Uraltsev}},\ }\href {\doibase 10.1016/S0370-2693(02)02616-3} {\bibfield
  {journal} {\bibinfo  {journal} {Phys.Lett.}\ }\textbf {\bibinfo {volume}
  {B545}},\ \bibinfo {pages} {337} (\bibinfo {year} {2002})},\ \Eprint
  {http://arxiv.org/abs/hep-ph/0111166} {arXiv:hep-ph/0111166 [hep-ph]}
  \BibitemShut {NoStop}%
%%CITATION = HEP-PH/0111166;%%
\bibitem [{\citenamefont {Misiak}\ \emph {et~al.}(2007)\citenamefont {Misiak},
  \citenamefont {Asatrian}, \citenamefont {Bieri}, \citenamefont {Czakon},
  \citenamefont {Czarnecki} \emph {et~al.}}]{Misiak:2006zs}%
  \BibitemOpen
  \bibfield  {author} {\bibinfo {author} {\bibfnamefont {M.}~\bibnamefont
  {Misiak}}, \bibinfo {author} {\bibfnamefont {H.}~\bibnamefont {Asatrian}},
  \bibinfo {author} {\bibfnamefont {K.}~\bibnamefont {Bieri}}, \bibinfo
  {author} {\bibfnamefont {M.}~\bibnamefont {Czakon}}, \bibinfo {author}
  {\bibfnamefont {A.}~\bibnamefont {Czarnecki}},  \emph {et~al.},\ }\href
  {\doibase 10.1103/PhysRevLett.98.022002} {\bibfield  {journal} {\bibinfo
  {journal} {Phys.Rev.Lett.}\ }\textbf {\bibinfo {volume} {98}},\ \bibinfo
  {pages} {022002} (\bibinfo {year} {2007})},\ \Eprint
  {http://arxiv.org/abs/hep-ph/0609232} {arXiv:hep-ph/0609232 [hep-ph]}
  \BibitemShut {NoStop}%
%%CITATION = HEP-PH/0609232;%%
\bibitem [{\citenamefont {Guetta}\ and\ \citenamefont
  {Nardi}(1998)}]{Guetta:1997fw}%
  \BibitemOpen
  \bibfield  {author} {\bibinfo {author} {\bibfnamefont {D.}~\bibnamefont
  {Guetta}}\ and\ \bibinfo {author} {\bibfnamefont {E.}~\bibnamefont {Nardi}},\
  }\href {\doibase 10.1103/PhysRevD.58.012001} {\bibfield  {journal} {\bibinfo
  {journal} {Phys.Rev.}\ }\textbf {\bibinfo {volume} {D58}},\ \bibinfo {pages}
  {012001} (\bibinfo {year} {1998})},\ \Eprint
  {http://arxiv.org/abs/hep-ph/9707371} {arXiv:hep-ph/9707371 [hep-ph]}
  \BibitemShut {NoStop}%
%%CITATION = HEP-PH/9707371;%%
\bibitem [{\citenamefont {Ball}\ and\ \citenamefont
  {Zwicky}(2005{\natexlab{a}})}]{Ball:2004ye}%
  \BibitemOpen
  \bibfield  {author} {\bibinfo {author} {\bibfnamefont {P.}~\bibnamefont
  {Ball}}\ and\ \bibinfo {author} {\bibfnamefont {R.}~\bibnamefont {Zwicky}},\
  }\href {\doibase 10.1103/PhysRevD.71.014015} {\bibfield  {journal} {\bibinfo
  {journal} {Phys.Rev.}\ }\textbf {\bibinfo {volume} {D71}},\ \bibinfo {pages}
  {014015} (\bibinfo {year} {2005}{\natexlab{a}})},\ \Eprint
  {http://arxiv.org/abs/hep-ph/0406232} {arXiv:hep-ph/0406232 [hep-ph]}
  \BibitemShut {NoStop}%
%%CITATION = HEP-PH/0406232;%%
\bibitem [{\citenamefont {Bazavov}\ \emph {et~al.}(2012)\citenamefont {Bazavov}
  \emph {et~al.}}]{Bazavov:2011aa}%
  \BibitemOpen
  \bibfield  {author} {\bibinfo {author} {\bibfnamefont {A.}~\bibnamefont
  {Bazavov}} \emph {et~al.} (\bibinfo {collaboration} {Fermilab Lattice
  Collaboration, MILC Collaboration}),\ }\href {\doibase
  10.1103/PhysRevD.85.114506} {\bibfield  {journal} {\bibinfo  {journal}
  {Phys.Rev.}\ }\textbf {\bibinfo {volume} {D85}},\ \bibinfo {pages} {114506}
  (\bibinfo {year} {2012})},\ \Eprint {http://arxiv.org/abs/1112.3051}
  {arXiv:1112.3051 [hep-lat]} \BibitemShut {NoStop}%
%%CITATION = ARXIV:1112.3051;%%
\bibitem [{\citenamefont {McNeile}\ \emph {et~al.}(2012)\citenamefont
  {McNeile}, \citenamefont {Davies}, \citenamefont {Follana}, \citenamefont
  {Hornbostel},\ and\ \citenamefont {Lepage}}]{McNeile:2011ng}%
  \BibitemOpen
  \bibfield  {author} {\bibinfo {author} {\bibfnamefont {C.}~\bibnamefont
  {McNeile}}, \bibinfo {author} {\bibfnamefont {C.}~\bibnamefont {Davies}},
  \bibinfo {author} {\bibfnamefont {E.}~\bibnamefont {Follana}}, \bibinfo
  {author} {\bibfnamefont {K.}~\bibnamefont {Hornbostel}}, \ and\ \bibinfo
  {author} {\bibfnamefont {G.}~\bibnamefont {Lepage}},\ }\href {\doibase
  10.1103/PhysRevD.85.031503} {\bibfield  {journal} {\bibinfo  {journal}
  {Phys.Rev.}\ }\textbf {\bibinfo {volume} {D85}},\ \bibinfo {pages} {031503}
  (\bibinfo {year} {2012})},\ \Eprint {http://arxiv.org/abs/1110.4510}
  {arXiv:1110.4510 [hep-lat]} \BibitemShut {NoStop}%
%%CITATION = ARXIV:1110.4510;%%
\bibitem [{\citenamefont {Na}\ \emph {et~al.}(2012)\citenamefont {Na},
  \citenamefont {Monahan}, \citenamefont {Davies}, \citenamefont {Horgan},
  \citenamefont {Lepage} \emph {et~al.}}]{Na:2012kp}%
  \BibitemOpen
  \bibfield  {author} {\bibinfo {author} {\bibfnamefont {H.}~\bibnamefont
  {Na}}, \bibinfo {author} {\bibfnamefont {C.~J.}\ \bibnamefont {Monahan}},
  \bibinfo {author} {\bibfnamefont {C.~T.}\ \bibnamefont {Davies}}, \bibinfo
  {author} {\bibfnamefont {R.}~\bibnamefont {Horgan}}, \bibinfo {author}
  {\bibfnamefont {G.~P.}\ \bibnamefont {Lepage}},  \emph {et~al.},\ }\href
  {\doibase 10.1103/PhysRevD.86.034506} {\bibfield  {journal} {\bibinfo
  {journal} {Phys.Rev.}\ }\textbf {\bibinfo {volume} {D86}},\ \bibinfo {pages}
  {034506} (\bibinfo {year} {2012})},\ \Eprint {http://arxiv.org/abs/1202.4914}
  {arXiv:1202.4914 [hep-lat]} \BibitemShut {NoStop}%
%%CITATION = ARXIV:1202.4914;%%
\bibitem [{\citenamefont {Ball}\ and\ \citenamefont
  {Zwicky}(2005{\natexlab{b}})}]{Ball:2004rg}%
  \BibitemOpen
  \bibfield  {author} {\bibinfo {author} {\bibfnamefont {P.}~\bibnamefont
  {Ball}}\ and\ \bibinfo {author} {\bibfnamefont {R.}~\bibnamefont {Zwicky}},\
  }\href {\doibase 10.1103/PhysRevD.71.014029} {\bibfield  {journal} {\bibinfo
  {journal} {Phys.Rev.}\ }\textbf {\bibinfo {volume} {D71}},\ \bibinfo {pages}
  {014029} (\bibinfo {year} {2005}{\natexlab{b}})},\ \Eprint
  {http://arxiv.org/abs/hep-ph/0412079} {arXiv:hep-ph/0412079 [hep-ph]}
  \BibitemShut {NoStop}%
%%CITATION = HEP-PH/0412079;%%
\end{thebibliography}%

\end{document}